\documentclass[aps,
		prd,
		reprint,
		onecolumn,
		superscriptaddress,
		shortbibliography,
		nofootinbib,
		floatfix,
		notitlepage
		]{revtex4-1}

\usepackage{amsmath,amssymb,amsfonts}
\usepackage[dvipsnames]{xcolor}
\usepackage{hyperref}
\hypersetup{colorlinks=true,citecolor=Cyan,urlcolor=Red}
\usepackage{graphicx}

\usepackage{slashed}

\newcommand{\be}{\begin{equation}}
\newcommand{\ee}{\end{equation}}
\newcommand{\bi}{\begin{itemize}}
\newcommand{\ei}{\end{itemize}}
\newcommand{\bea}{\begin{eqnarray}}
\newcommand{\eea}{\end{eqnarray}}
\newcommand{\ud}{\mathrm{d}}		

\newcommand{\LCm}{{\scriptscriptstyle -}} 
\newcommand{\LCp}{{\scriptscriptstyle +}}

\newcommand{\LCperp}{{\scriptscriptstyle \perp}}

\newcommand\redsout{\bgroup\markoverwith{\textcolor{red}{\rule[0.5ex]{2pt}{0.4pt}}}\ULon}
\usepackage[normalem]{ulem}
\makeatletter
\newcommand\wave{\bgroup \markoverwith{\textcolor{red}{\lower3.5\p@\hbox{\sixly \char58}}}\ULon}
\makeatother

\begin{document}

\title{Pair production in strong electric fields}
\author{Zhiyu Lei}
\email{u201810186@hust.edu.cn}
\affiliation{Institute of Modern Physics, Chinese Academy of Sciences, Lanzhou 730000, China}
\affiliation{School of Physics, Huazhong University of Science and Technology, Wuhan 430074, China}
\author{Bolun Hu}
\email{hubolun@impcas.ac.cn}
\affiliation{School of Physical Science and Technology, Lanzhou University, Lanzhou 730000, China}
\affiliation{Institute of Modern Physics, Chinese Academy of Sciences, Lanzhou 730000, China}
\affiliation{School of Nuclear Science and Technology, University of Chinese Academy of Sciences, Beijing 100049, China}

\author{Xingbo Zhao}
\email{xbzhao@impcas.ac.cn}
\affiliation{Institute of Modern Physics, Chinese Academy of Sciences, Lanzhou 730000, China}
\affiliation{School of Nuclear Science and Technology, University of Chinese Academy of Sciences, Beijing 100049, China}
\begin{abstract}	
   
 We study the electron-positron pair production process in strong background electric fields with spatio-temporal inhomogeneity in the time-dependent Basis Light-front Quantization approach. We calculate the observables such as the invariant mass and the longitudinal momentum distribution of the produced pairs as functions of evolution time and compare the vacuum decay rate with that of the Schwinger effect. We observe a critical intensity of background field, above which the vacuum decay rate is no longer oscillating with time periodically as in perturbation theory. This work may provide the foundation for the study of the pair production process in strong fields with realistic spacetime structures. 
\end{abstract}

\maketitle
\section{Introduction}
Nonperturbative processes have been a topic of considerable interest since the inception of quantum field theory. Light-front quantization, with simple vacuum structure and kinematic Lorentz boosts, provides the Hamiltonian formalism as a convenient and efficient approach to quantum field theory. Basis Light-front Quantization (BLFQ)~\cite{Vary:2009gt} has been constructed as a nonperturbative method for solving the structure of relativistic bound states in the light-front Hamiltonian formalism. BLFQ has been applied to various hadron systems~\cite{Li:2015iaw,Li:2015zda,Karmanov:2016yzu,Chen:2016dlk,Li:2017mlw,Adhikari:2018umb,Tang:2018myz,Jia:2018ary,Jia:2018hxd,Lan:2019vui,Lan:2019rba,Du:2019qsz,Mondal:2019yph,Lan:2019img,Mondal:2019jdg,Du:2019ips,Xu:2020xbt,Xu:2021wwj} and to positronium~\cite{Wiecki:2015xxa}. Recently BLFQ has been extended to time-dependent Basis Light-front Quantization (tBLFQ)~\cite{Zhao:2013cma}. The tBLFQ approach simulates the time evolution of quantum field configurations in terms of a time-dependent and fully relativistic light-front Schrödinger equation. This approach is thus suitable for studying nonperturbative processes in the presence of strong and possibly time-dependent background fields. The tBLFQ approach has been applied to nonlinear Compton scattering~\cite{Zhao:2013cma,Hu:2019hjx} in strong electromagnetic fields, to scattering of an electron off a heavy-ion~\cite{Chen:2018vdw} and to scattering of a quark in a color-glass-condensate field modeling a nucleus~\cite{Li:2020uhl,Li:2020bti}. In this work, we employ tBLFQ to study the pair production process in strong electric fields. The electron and positron pair produced from the vacuum in the presence of a uniform and constant electric field is the famous Schwinger effect; for recent reviews see~\cite{Gelis:2015kya,Hu:2019dij}. 

High-intensity laser facilities are being constructed with the goal of studying strong-field QED including the pair production process. The major modern high-intensity laser systems include the European High Power laser Energy Research facility (HiPER, Rutherford Laboratory, England), the Extreme Light Infrastructure (ELI), the Exawatt Center for Extreme Light Studies (XCELS, Russia), the X-ray Free Electron Laser (XFEL)~\cite{Abramowicz:2021zja}, and the Shanghai Synchrotron Radiation Facility (SSRF, China). The typical strength of the electric field for the Schwinger effect is around $10^{18}\rm \,V/m$, which is still beyond the reach of these contemporary facilities, so various mechanisms in stimulating the pair production process have been proposed~\cite{Bamber:1999zt,Schutzhold:2008pz,Dunne:2009gi,Baier:2009it,Bulanov:2010ei}. In modern laser facilities the electromagnetic fields are produced in the form of highly focused pulses. In order to analyze and understand the experimental results, nonperturbative theoretical approaches with the capability of treating complex time-dependent background fields are required, which is an ideal application scenario for tBLFQ. Compared to e.g. particle-in-cell (PIC) approaches~\cite{Gonoskov:2014mda} though, BLFQ cannot yet handle as many particles or collective effects.

This paper is organized as follows. We introduce the formalism of tBLFQ in Sec.~\ref{tBLFQ}, where we detail the procedure of constructing the basis in tBLFQ. In Sec.~\ref{Background} we demonstrate the profiles of the background electric fields in this work. In Sec.~\ref{Results} we provide the numerical results for the observables of the produced pairs. In Sec.~\ref{Conclusions} we present our conclusions and outlook. The appendices include the conventions, the matrix elements of the background field interaction in the tBLFQ basis, and a comparison of the production probability in the weak field limit with time-dependent perturbation theory.
 
\section{Time-dependent basis light-front quantization}\label{tBLFQ}
We define the light-front time and the longitudinal coordinate as $x^{\LCp}:=x^{0}+x^{3}$ and $x^{\LCm}:=x^{0}-x^{3}$, respectively; the remaining $2$ spatial coordinates $x^{\LCperp}:=\left( x^{1},x^{2} \right)$ are the transverse directions~\cite{Hu:2019hjx,Zhao:2013cma}. In the light-front formalism of quantum field theory, the time evolution of the system is governed by the light-front Schr\"odinger equation
\begin{equation}
   i \frac{\partial}{\partial x^{\LCp}}\left|\,\psi ; x^{\LCp}\,\right\rangle=\frac{1}{2} P^{\LCm}\left(x^{\LCp}\right)\left|\,\psi ; x^{\LCp}\,\right\rangle\;,
   \label{eqn:Schrodinger}
\end{equation}
in which $P^{\LCm}$ is the light-front QED Hamiltonian whose explicit form will be given in the following text, and $\left|\,\psi ; x^{\LCp}\,\right\rangle$ is the state vector which includes physically accessible states. For numerical calculations, truncation is necessary to render the Hilbert space finite-dimensional. tBLFQ works in a finite-dimensional discretized Fock space that is truncated both at the level of the Fock sector, and at the level of each particle for appropriate combinations of quantum numbers. In this paper, we keep the Fock space with the vacuum $|\,0\,\rangle$ and through the sector including $4$ pairs of electrons and positrons, so the Fock space now contains the following sectors
\begin{equation}
   |\,0\,\rangle, \;|\,e^{+}e^{-}\,\rangle,\;|\,e^{+}e^{-}e^{+}e^{-}\,\rangle,\;|\,e^{+}e^{-}e^{+}e^{-}e^{+}e^{-}\,\rangle,\;|\,e^{+}e^{-}e^{+}e^{-}e^{+}e^{-}e^{+}e^{-}\,\rangle \;.
   \label{}
\end{equation}
By excluding the dynamical photons, we essentially ignore the back-reaction~\cite{Hebenstreit:2013qxa,Kasper:2014uaa,Taya:2017pdp,Otto:2018hya} of the electron-positron pairs on the background field. We follow the light-front QED Hamiltonian derived from the QED Lagrangian in~\cite{Zhao:2013cma}, in which the terms containing gauge boson fields vanish in our Fock-space truncation, so we also drop the instantaneous photon interaction for the lack of dynamical photon interaction, due to gauge invariance~\cite{Tang:1991rc}. Thus the Hamiltonian in our case is reduced to
\begin{equation} 
	P^{\LCm}=\int \mathrm{d}^{2} x^{\LCperp} \mathrm{d} x^{\LCm} \frac{1}{2} \bar{\Psi} \gamma^{\LCp} \frac{m_{e}^{2}+\left(i \partial^{\LCperp}\right)^{2}}{i \partial^{\LCp}} \Psi +\frac{e^{2}}{2} \bar{\Psi} \gamma^{\mu} \mathcal{A}_{\mu} \frac{\gamma^{\LCp}}{i \partial^{\LCp}} \gamma^{\nu} \mathcal{A}_{\nu} \Psi+e \bar{\Psi} \gamma^{\mu} \Psi \mathcal{A}_{\mu}\;,
	\label{eqn:fullhamiltonian}
\end{equation}
in which $e=\sqrt{4 \pi / 137}$ is the electron charge and $m_{e}=0.511\rm \,MeV$ is the electron mass; $\Psi$ denotes the field operator of the fermions and $\mathcal A$ denotes the background field which is treated as a classical field. The first term of Eq.~\eqref{eqn:fullhamiltonian} is the kinetic energy term we label as $P_{0}^{\LCm}$; 
the second term is an instantaneous fermion interaction that will vanish since the external field we will adopt contains only a nonzero longitudinal component [see Eq.~\eqref{BG} in Sec.~\ref{Background}]; the third term we label as $V$ is the vertex interaction that creates electron-positron pairs and accelerates them. 
In this work, for simplicity we adopt the background field independent of the transverse coordinates; see Eq.~\eqref{BG}. We thus neglect the excitation of the transverse motion,\footnote{In this work, we still work with $\rm{QED}_{3+1}$, which is distinct from $\rm{QED}_{1+1}$, where the fermions have no spin.} and approximate it by a Gaussian wave packet for each fermion
\begin{equation}
   \Phi^{b}\left( p^{\LCperp} \right)=\frac{2\sqrt{\pi}}{b}\exp\left( -\frac{ p_{1}^{2}+p_{2}^{2} }{2b^{2}} \right)\;,
   \label{}
\end{equation}
where $b$ is the width of the Gaussian profile in momentum space. This profile corresponds to the lowest state among the $2$-dimensional harmonic oscillator ($2$DHO) eigenstates, which are adopted in our previous tBLFQ studies~\cite{Zhao:2013cma,Hu:2019hjx}. We plan to expand our transverse basis by including the excited eigenstates of the $2$DHO and study the dynamics in the transverse directions in a future study.

We compactify the longitudinal direction in a circle of length $x^{\LCm}=2L$ and impose the anti-periodic boundary condition for each fermion in the longitudinal direction; consequently the longitudinal momentum $p^{\LCp}$ takes half-integers multiplied by $2\pi/L$, 
\begin{equation}
   p^{\LCp}=\frac{2 \pi}{L} k \; \text { for } \; k=\frac{1}{2}, \frac{3}{2}, \frac{5}{2}, \ldots\;.
\end{equation}
Here we take $L=2\pi\rm \,MeV^{-1}$ so $k$ can be interpreted as momentum in units of $\rm \,MeV$. Note that the zero modes (states with zero longitudinal momentum) are absent in this basis. 

In order to keep the basis finite, we retain the basis states with the total momentum of all the particles not larger than a maximum value. Thereby we introduce a truncation parameter $K_{\max}$ in the longitudinal direction such that 
\begin{equation}
   \sum_{i} k_{i}\le K_{\max}\;,
   \label{}
\end{equation}
where $k_{i}$ is the longitudinal momentum of the electrons (denoted by even $i$) and the positrons (denoted by odd $i$) in the Fock states. In our case the background field~\eqref{BG} has only a longitudinal component and thus does not change the total helicity of the system through the interaction $V$ in Eq.~\eqref{eqn:fullhamiltonian}, so we only retain the basis states with zero total helicity 
$$\sum_{i}\lambda_{i}=0\;,$$
in which $\lambda_{i}$ is the helicity of the electrons and the positrons in the basis states. We further reduce the basis size by taking the advantage of the antisymmetry property of the many-fermion system whereby we just keep one copy of the basis states representing the same physical state. For example, the basis states with two identical fermions labeled ``1'' and ``3'' having $k_{1}=1/2,\lambda_{1}=\uparrow, k_{3}=3/2,\lambda_{3}=\downarrow$ and $k_{1}=3/2,\lambda_{1}=\downarrow, k_{3}=1/2,\lambda_{3}=\uparrow$ represent the same physical state and we only keep the latter.

Now we have a finite basis, and it is then straightforward to express the Hamiltonian of the system in this basis and evolve the system according to the Schr\"odinger equation~\eqref{eqn:Schrodinger}, which has the following formal solution 
\begin{equation}
   \left|\,\psi ; x^{\LCp}\,\right\rangle=\mathcal{T}_{\LCp} \exp \left(-\frac{i}{2} \int_{0}^{x^{\LCp}} P^{\LCm}\left( x^{\LCp} \right)\right)|\,\psi ; 0\,\rangle\;,
   \label{eqn:discrete}
\end{equation}
where $\mathcal{T}_{\LCp}$ is light-front time ordering. The matrix elements of $P^{\LCm}$ in our basis can be obtained by using the mode expansion of the field operator and the anti-commutation relations; for details see Appendix~\ref{melements}.
In the numerical implementation of Eq.~\ref{eqn:discrete}, we adopt the second-order difference scheme MSD2~\cite{askar1978askar,Hu:2019hjx,Zhao:2013cma} rather than the naive ``Euler scheme'', by relating the state at $x^{\LCp}+\delta x^{\LCp}$ to those at both $x^{\LCp}$ and $x^{\LCp}-\delta x^{\LCp}$ 
\begin{equation}
\left|\,\psi ; x^{\LCp}+\delta x^{\LCp}\,\right\rangle=\left|\,\psi ; x^{\LCp}-\delta x^{\LCp}\,\right\rangle+\left(\mathrm e^{-i P^{\LCm} \delta x^{\LCp} / 2}-\mathrm e^{i P^{\LCm} \delta x^{\LCp} / 2}\right)\left|\,\psi ; x^{\LCp}\,\right\rangle 
\approx\left|\,\psi ; x^{\LCp}-\delta x^{\LCp}\,\right\rangle-i P^{\LCm}\delta x^{\LCp}\left|\,\psi ; x^{\LCp}\,\right\rangle\;.
\end{equation}
It is then straightforward to calculate the time evolution of observables by sandwiching the corresponding operators with the state vectors
\begin{equation}
   \langle\, \hat{O}\,\rangle(x^{\LCp})=\left\langle\,\psi ; x^{\LCp}\,\right|\hat{O}\,\left|\,\psi ; x^{\LCp}\,\right\rangle\;.    
   \label{}
\end{equation}
\section{Background field}\label{Background}
\begin{figure*}[t!]
\centering
   \begin{center}
      \begin{tabular}{@{}cccc@{}}
         \includegraphics[width=.47\textwidth]{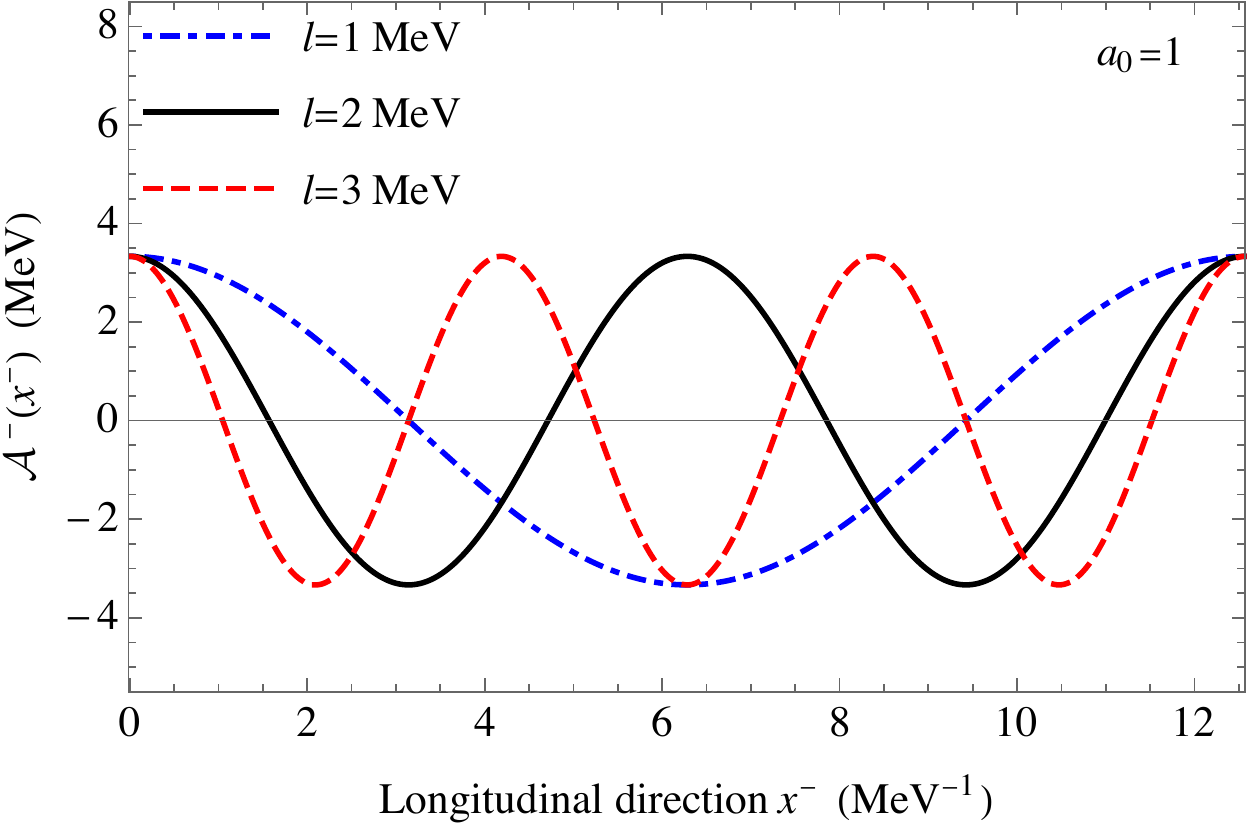} 
         \includegraphics[width=.47\textwidth]{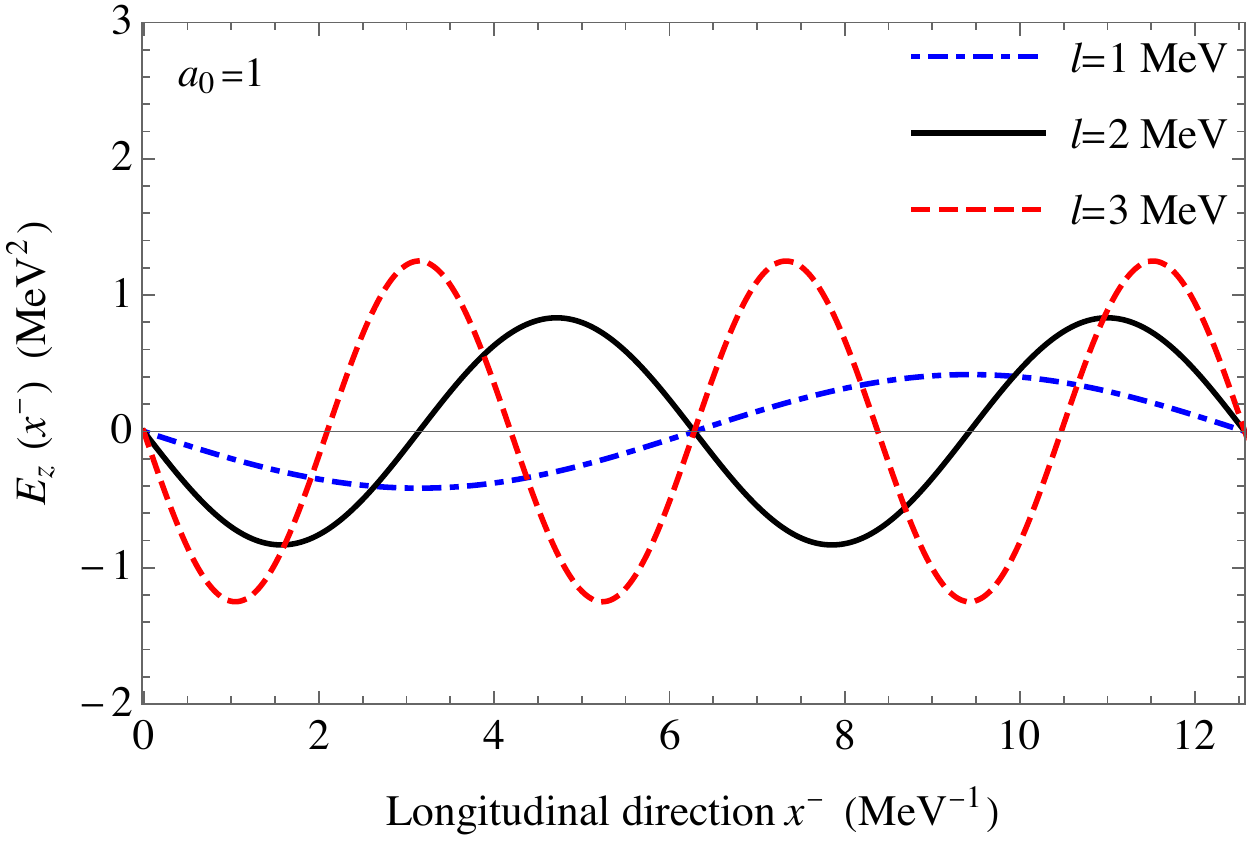}
  \end{tabular}
  \caption{Left: the longitudinal component of the background field $\mathcal A$ with different momenta $l=1\rm \,MeV$, $2\rm \,MeV$ and $3\rm \,MeV$, at the intensity $a_{0}=1$. Right: the corresponding electric field $E_{z}$ with different momenta $l=1\rm \,MeV$, $2\rm \,MeV$ and $3\rm \,MeV$, at the intensity $a_{0}=1$.}
   \label{fig:BG}
   \end{center}
\end{figure*}

\begin{figure*}[t!]
	\centering
	\begin{center}
		\begin{tabular}{@{}cccc@{}}
			\includegraphics[width=.478\textwidth]{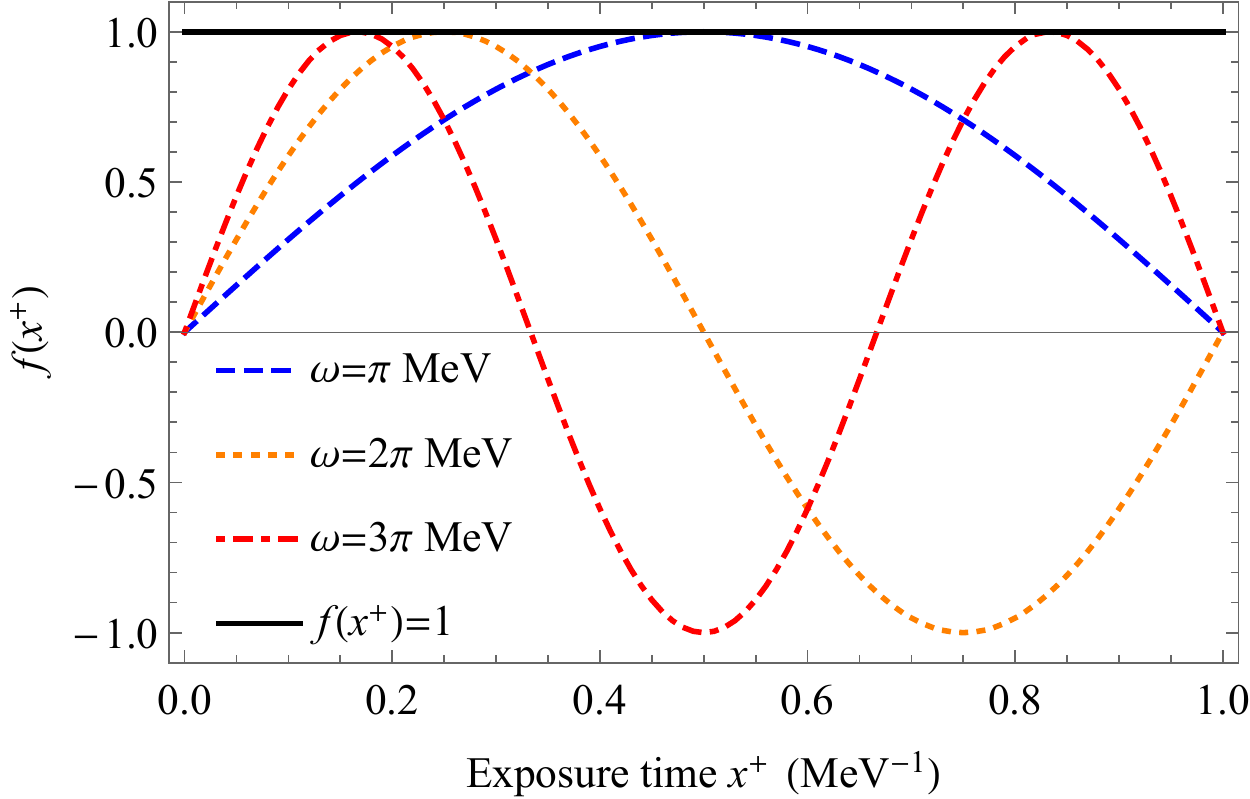} 
		\end{tabular}
                \caption{The time profiles $f\left( x^{\LCp} \right)=1$ and $f\left( x^{\LCp} \right)=\sin\left( \omega x^{\LCp} \right)$ with $\omega=\pi\mathrm{\,MeV}, 2\pi\mathrm{\,MeV}$ and $3\pi\mathrm{\,MeV}$, respectively.}
		\label{fig:TP}
	\end{center}
\end{figure*}

In this work, for simplicity we adopt the following background field containing only a longitudinal nonzero component
\begin{equation}
   e \mathcal{A}^{\mu}(x^{\LCp},x^{\LCm})=2 \delta_{\LCm}^{\mu} m_{e} a_{0} \cos \left(\frac{lx^{\LCm}}{2} \right)f\left(x^{\LCp}\right)\;,
   \label{BG}
\end{equation}
in which $a_{0}$ is a dimensionless overall factor for the intensity of the background field, and $l$ is the frequency, or momentum, of the background in the longitudinal direction, with the $1/2$ in front of $l$ coming from the light-front conventions; see Appendix~\ref{fields} for details. This background field is inhomogeneous in the longitudinal direction and is homogeneous in the transverse directions. The time profile $f\left( x^{\LCp} \right)$ is chosen to simulate the temporally focused lasers in modern facilities, and also provides an unambiguous definition of the particle number in the final state~\cite{Parker:1968mv,Tanji:2008ku,Hebenstreit:2009km,Kim:2011jw,Dabrowski:2016tsx,Ilderton:2021zej}, which basically means that we count the particle number only when the backgrounds are turned off. This background field corresponds to an electric field in the $z$-direction
\begin{equation}
   E_{z}=F_{03}=\frac{1}{2}\left( \frac{1}{2} F_{\LCm\LCp} -\frac{1}{2}F_{\LCp\LCm}  \right)=\frac{1}{2}\partial_{\LCm} \mathcal{A}_{\LCp}\;,
   \label{electric}
\end{equation}
with all the other components of the electromagnetic field vanishing. In this work, we will study the background~\eqref{BG} with $3$ different momenta $l=1\rm \, MeV$, $2\rm \, MeV$ and $3\rm \, MeV$. For $a_{0}=1$ and $l=1\rm \,MeV$, the peak value of the electric field~\eqref{electric} is approximately half of the Schwinger limit $E_{c}=m_{e}^{2}/e\approx0.87\rm \,MeV^{2}\approx1.32\times10^{18}\rm\, V/m$. The $\mathcal A$ fields and their resulting electric fields are shown in Fig.~\ref{fig:BG}. The fields with larger $l$ have shorter periods, and lead to larger magnitudes of the electric field. In the light-front time direction we will study several time profiles: $f\left( x^{\LCp} \right)=1$ and $f\left( x^{\LCp} \right)=\sin\left( \omega x^{\LCp} \right)$, with frequencies $\omega=\pi\mathrm{\,MeV}, 2\pi\mathrm{\,MeV}$, $3\pi\mathrm{\,MeV}$, and $5\pi\mathrm{\,MeV}$. The sinusoidal time profiles are employed to simulate the asymptotic switch-on-and-off behavior of experimentally generated laser pulses. We compare the time profiles in Fig.~\ref{fig:TP}. Note that the parameters $\omega$ and $l$ here are still far from reach for the present or near-future facilities, but are chosen for the convenience of our calculation.

\section{Observables}\label{Results}
In this section, we present the time evolution of observables obtained from the tBLFQ calculation. 
One of the advantages of our approach is that we can track each particle exclusively in the Fock space up to $4$ electron-positron pairs, which is difficult in other methods~\cite{Gelis:2015kya}. It should be noted here and later that whenever we mention ``$n$-pairs'' of electrons and positrons, we are anticipating the exclusive measurements so that ``one pair'' represents ``one pair and only one pair''. The invariant mass of the system can be obtained by $M^{2}=P^{\LCp}P^{\LCm}-\left(P^{\LCperp}  \right)^{2}$, namely,
\begin{equation}
   \begin{aligned}
      \left\langle\, M^{2}\,\right\rangle(x^{\LCp})&=\left\langle\, P^{\LCp}P_{0}^{\LCm}-\left(  P^{\LCperp}\right)^{2}\,\right\rangle(x^{\LCp})=\left\langle\,\psi ; x^{\LCp}\,\right|\,P^{\LCp}P_{0}^{\LCm}-\left(  P^{\LCperp}\right)^{2}\,\left|\,\psi ; x^{\LCp}\,\right\rangle\;,    
   \label{eqn:invmass}
   \end{aligned}
\end{equation}
in which  $P^{\LCp}$ is the total longitudinal momentum of all the particles
\begin{equation}
   P^{\LCp}=\int \mathrm{d}^{2} x^{\LCperp} \mathrm{d} x^{\LCm} \frac{1}{2} \bar{\Psi} \gamma^{\LCp}i \partial^{\LCp} \Psi\;,
   \label{eqn:total_momentum}
\end{equation}
and $P^{\LCm}_{0}$ is the total kinetic energy of them, which is given by
\begin{equation}
   P_{0}^{\LCm}=\int \mathrm{d}^{2} x^{\LCperp} \mathrm{d} x^{\LCm} \frac{1}{2} \bar{\Psi} \gamma^{\LCp} \frac{m_{e}^{2}+\left(i\partial^{\LCperp}  \right)^{2}}{i \partial^{\LCp}} \Psi\;,
   \label{eqn:free_hamiltonian}
\end{equation}
and $P^{\LCperp}$ is their total transverse momentum
\begin{equation}
   P^{\LCperp}=\int \mathrm{d}^{2} x^{\LCperp} \mathrm{d} x^{\LCm} \frac{1}{2} \bar{\Psi} \gamma^{\LCp}i \partial^{\LCperp} \Psi\;.
   \label{eqn:total_momentum}
\end{equation}
Note that in~\eqref{eqn:invmass} we take into account only the kinetic energy of the pairs and ignore the interaction of the background field.
We also study the longitudinal momentum distribution of the produced particles, which represents the probability of finding an electron (or a positron) with momentum $p^{\LCp}$ and helicity $\lambda$. It can be given by the expectation value of the fermion number operator 
\begin{equation}
   \rho(p^{\LCp},\lambda,x^{\LCp})=\left\langle\,\psi ; x^{\LCp}\,\right|\,b^{\dagger}\left( p^{\LCp},\lambda \right)b\left( p^{\LCp} ,\lambda\right)\,\left|\,\psi ; x^{\LCp}\,\right\rangle\;,\\
   \label{eqn:md}
\end{equation}
where $b^{\dagger}$ and $b$ are the creation and annihilation operators of the electron respectively; for details, see Appendix~\ref{fields}. We note that $\rho$ is based on the free number operator, and measures the number of pairs when the background field $\mathcal{A}$ is turned off.

In the remainder of this section, we present our results according to the time profile $f(x^{\LCp})$ of the background field. We mainly focus on two cases of the time profile: $f\left( x^{\LCp} \right)=1$, $f\left( x^{\LCp} \right)=\sin\left( \omega x^{\LCp} \right)$ with $\omega=\pi\rm\,MeV$ and present their results in Sec.~\ref{1} and Sec.~\ref{pi} respectively. The results obtained in fields with frequencies $\omega=2\pi\mathrm{\,MeV}, 3\pi\mathrm{\,MeV}$, and $5\pi\mathrm{\,MeV}$ are presented in Sec.~\ref{other}. 

\subsection{Background fields with time profile $f\left( x^{\LCp} \right)=1$}\label{1}
In this subsection, we study the constant time profile, i.e.~$f\left( x^{\LCp} \right)=1$. We first study the dependences of the observables on the truncation parameter $K_{\max}$ and then we consider their dependences on the field intensity parameter $a_{0}$ and momentum $l$. The observables we study include the probability of finding electron-positron pairs, the pair-production rate, the invariant mass, and the momentum distributions.

\begin{figure*}[t!]
	\centering
	\begin{center}
		\begin{tabular}{@{}cccc@{}}
			\includegraphics[width=.50\textwidth]{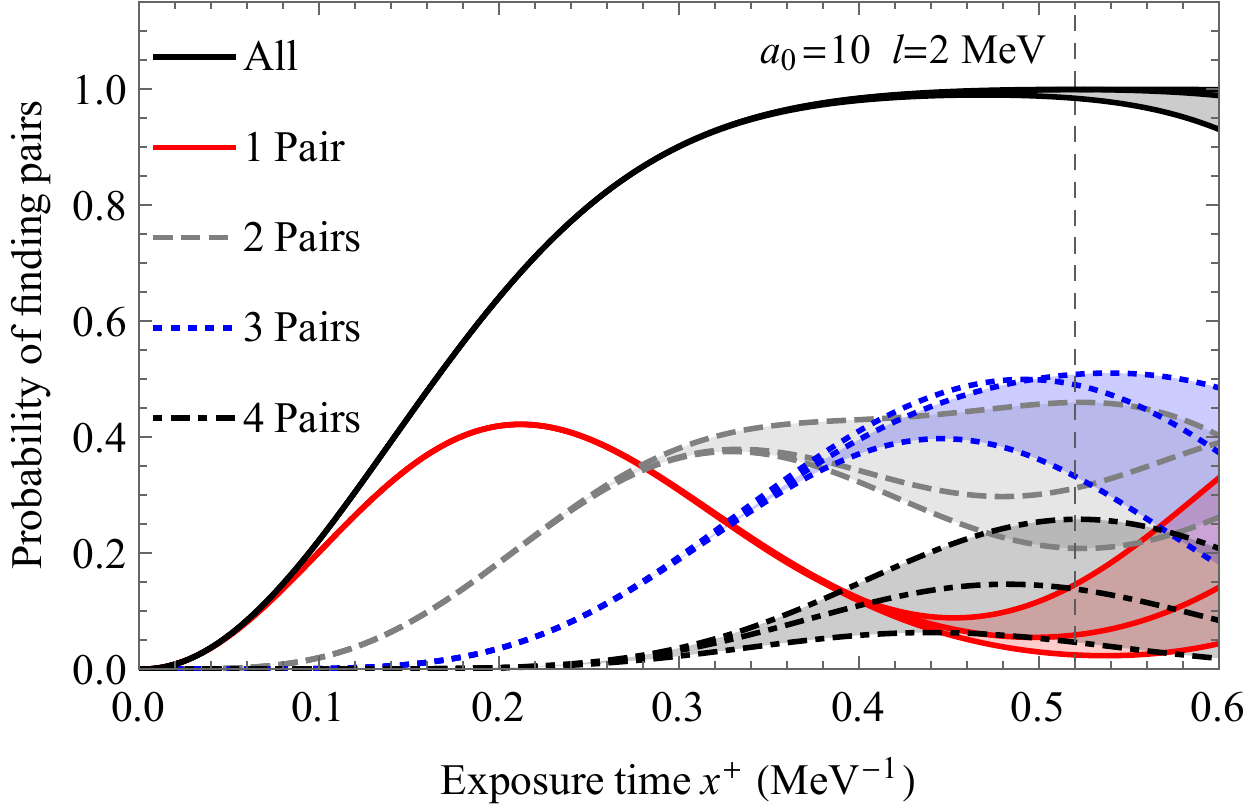} 
		\end{tabular}
                \caption{Time evolution of the probabilities of finding $n$ pairs of electrons and positrons (and their total) in the background field~\eqref{BG} with $f\left( x^{\LCp} \right)=1$, $a_{0}=10$ and $l=2\rm \,MeV$. Results in bases with $K_{\max}=8$, $10$ and $12$ are presented in different curves with the same plot style, corresponding to the top, middle and bottom curves in each shaded area. One exception is the curve corresponding to the probability of finding $3$ pairs for $K_{\max}=10$, which is at the top before $x^{\LCp}\approx0.5 {\rm \,MeV}$.
                Other parameter: $b=m_{e}$.}
		\label{fig:const_kmax_dependence}
	\end{center}
\end{figure*}

We show the time evolution of the probabilities of finding $n$ pairs of electrons and positrons, as well as their total, obtained in bases with $K_{\max}=8$, $10$ and $12$ in Fig.\footnote{$K_{\max}=8$ is the minimal $K_{\max}$ for $4$ pairs of electrons and positrons to be present in the basis.}\ref{fig:const_kmax_dependence}.
 The probability of finding each of $n$ pairs forms a band. Before $x^{\LCp}=0.3\rm \,MeV^{-1}$ the width of each band is small, implying good convergence with $K_{\max}$, but as time passes the curves in each band show increasing discrepancies.
The band representing the probability of finding $1$ pair of electron and positron increases first between $x^{\LCp}=0\rm \,MeV^{-1}$ and  $x^{\LCp}\approx0.2\rm \,MeV^{-1}$ and then decreases afterwards. The probabilities of finding more pairs of electrons and positrons sequentially increase. 
We find that when the probability of finding $2$ pairs becomes significant at $x^{\LCp}\approx0.2\rm \,MeV^{-1}$ the probability of finding $1$ pair of electron and positron decreases, and that the same pattern also applies to the probabilities of finding $2$ and $3$ pairs of electrons and positrons. 
This pattern is due to the fact that the background field can only create $1$ electron-positron pair at one time (see Appendix~\ref{melements}), so it is only when the probabilities of finding a smaller number of pairs become significant, the production of additional pairs can be efficient. Once the states with larger number of pairs have been abundantly populated, the probabilities of the smaller number of pairs are depleted, leading to the slowdown in producing additional pairs. 
We observe broadening in bands starting around $x^{\LCp}=0.3\rm \,MeV^{-1}$, and attribute them to the limited basis space corresponding to the smaller $K_{\max}$, where the truncation error starts to appear. 
As suggested by the trends in the curves, we expect there would be a significant probability of finding $5$ pairs of electrons and positrons at the time when the probability of finding $4$ pairs start to decrease. However, the $5$-pairs sector is absent in our basis and we thus consider the results after $x^{\LCp}\approx0.52\rm \,MeV^{-1}$ may be subject to severe truncation artifacts even for $K_{\max}=12$. Compared to the probabilities of finding individual pairs, the total probability is less sensitive to $K_{\max}$. In the remainder of this subsection, we will take $K_{\max}=12$ and study the dependences of observables on other parameters.

\begin{figure*}[t!]
	\centering
	\begin{center}
		\begin{tabular}{@{}cccc@{}}
			\includegraphics[width=.47\textwidth]{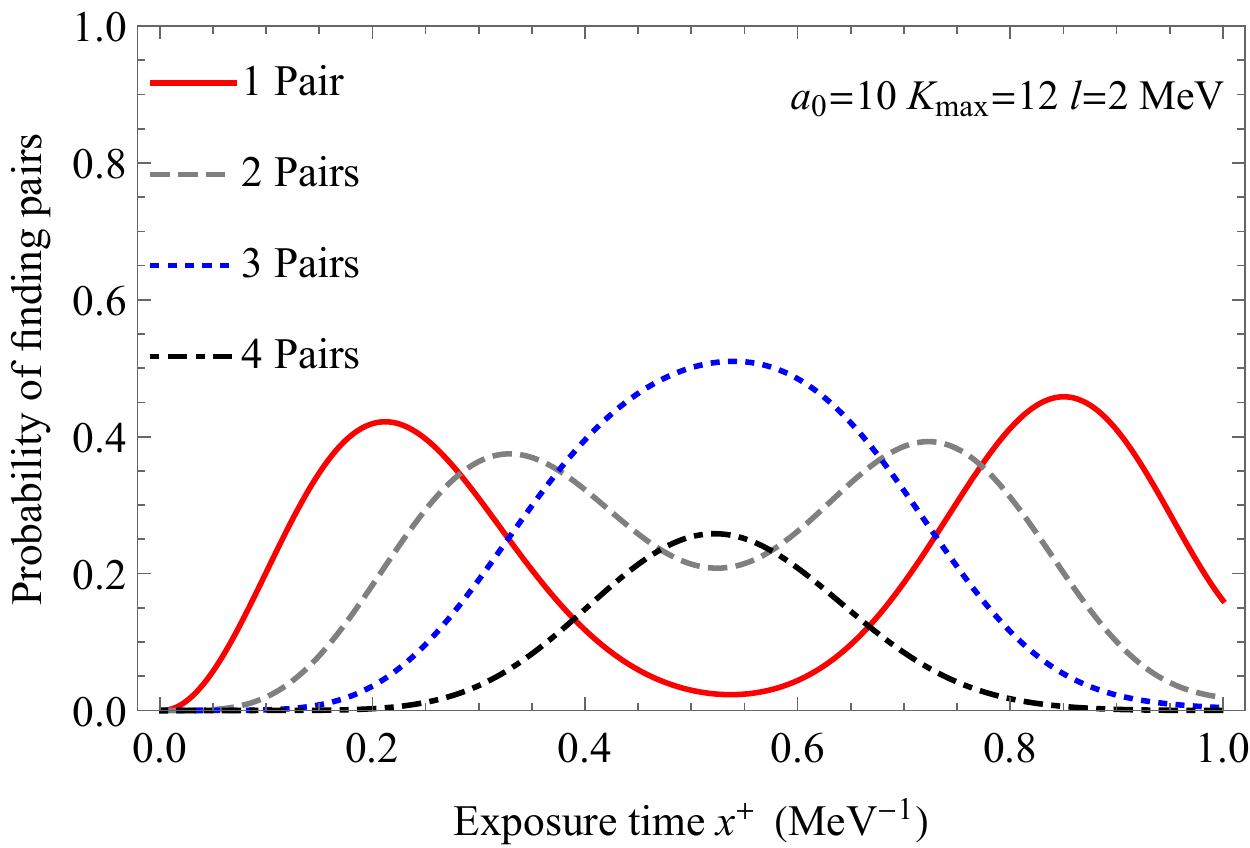} 
			\includegraphics[width=.47\textwidth]{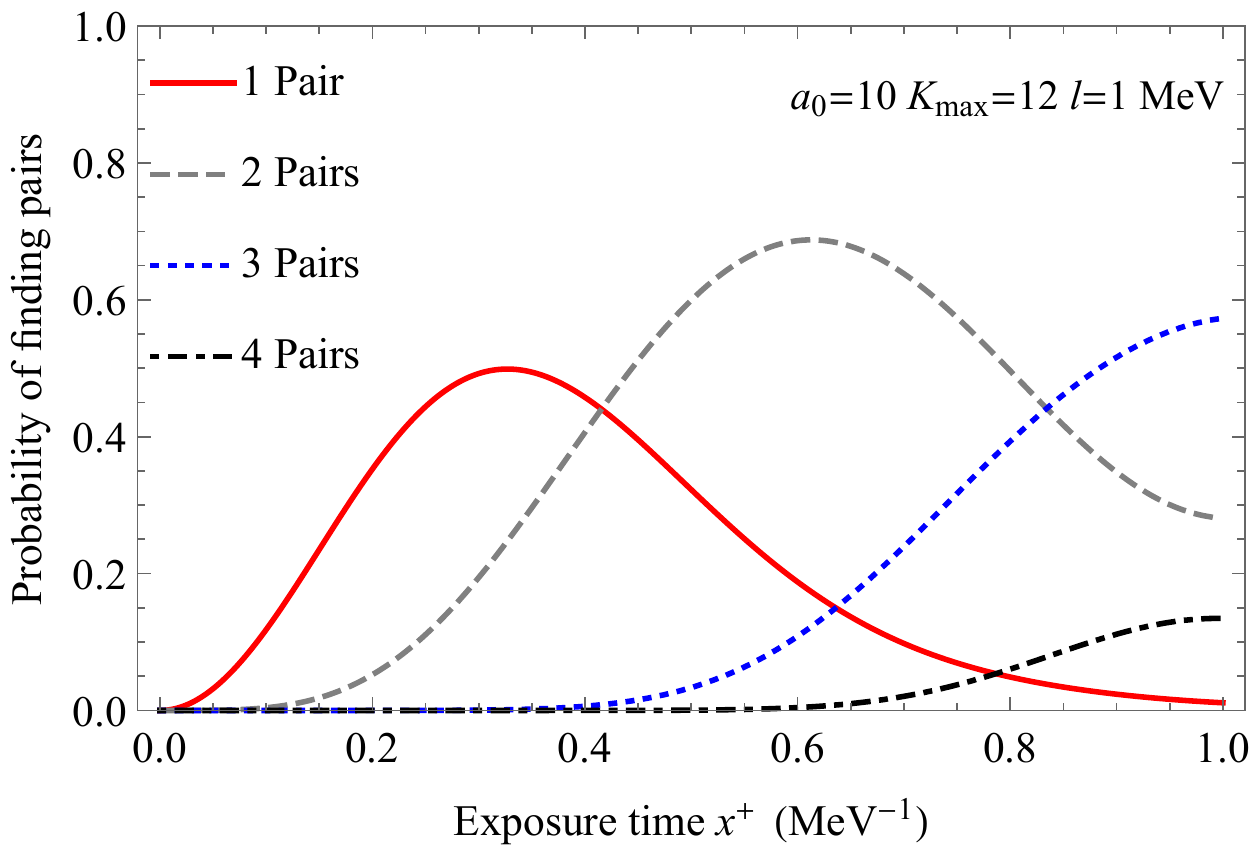} 
		\end{tabular}
                \caption{Time evolution of the probabilities of finding $n$ pairs of electrons and positrons in the background field~\eqref{BG} with $f\left( x^{\LCp} \right)=1$ and $a_{0}=10$. In the left panel the longitudinal momentum of the background field $l=2\rm \,MeV$ and in the right panel $l=1\rm \,MeV$. Other parameters: $K_{\max}=12$, $b=m_{e}$.}
		\label{fig:const_sector_l}
	\end{center}
\end{figure*}

We show the dependence of the pair production probability on the background field momentum $l$ in Fig.~\ref{fig:const_sector_l}.  We represent the time evolution of the probabilities of finding $n$ pairs of electrons and positrons in the background fields with longitudinal momentum $l=2\rm \,MeV$ and $l=1\rm \,MeV$, in the left and the right panel, respectively. At the initial stage of the time evolution, the probabilities of finding $1$ electron-positron pair are comparable. However, the probabilities of finding multiple pairs increase at a smaller rate in the $l=1\rm \,MeV$ case. This is because background fields with larger momentum accelerate particles to wider ranges of momenta and therefore the productions of multiple pairs are less affected by the Pauli exclusion principle. 

\begin{figure*}[t!]
	\centering
	\begin{center}
		\begin{tabular}{@{}cccc@{}}
			\includegraphics[width=.47\textwidth]{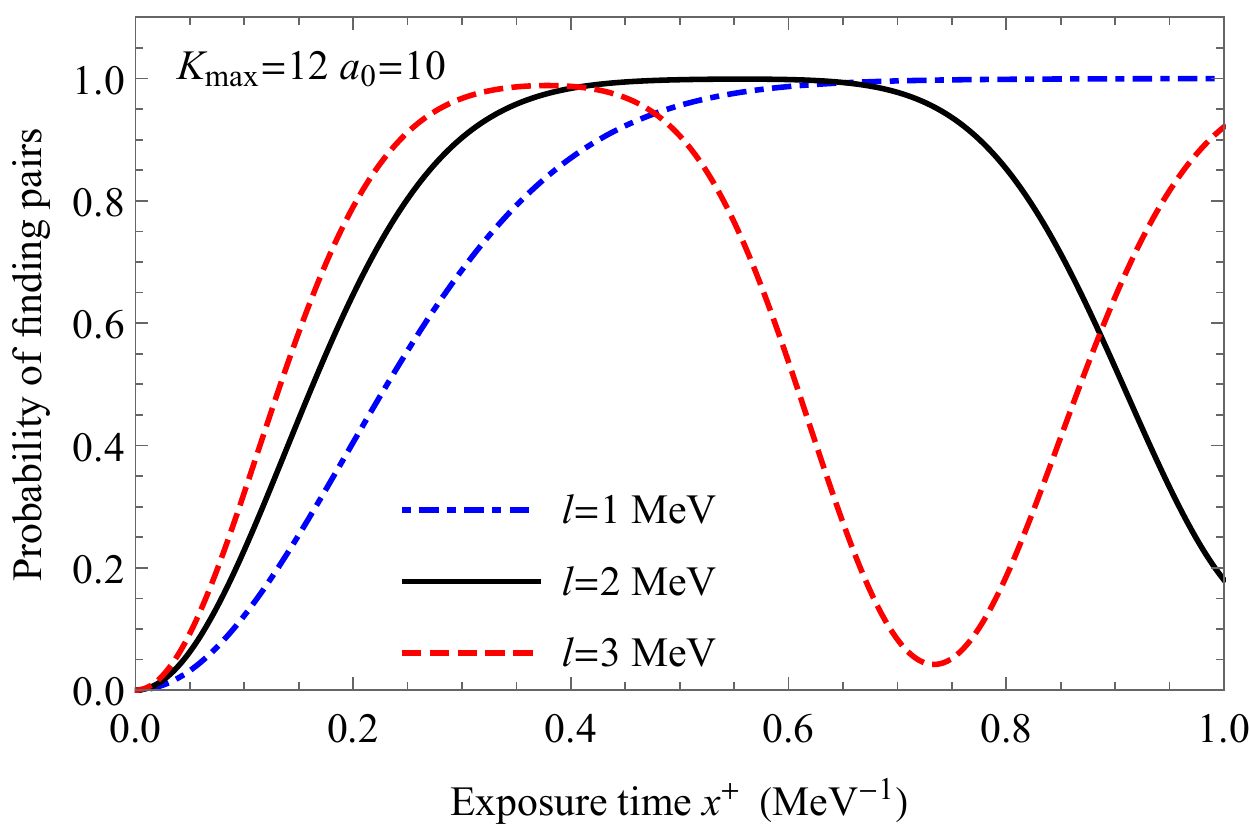}
			\includegraphics[width=.48\textwidth]{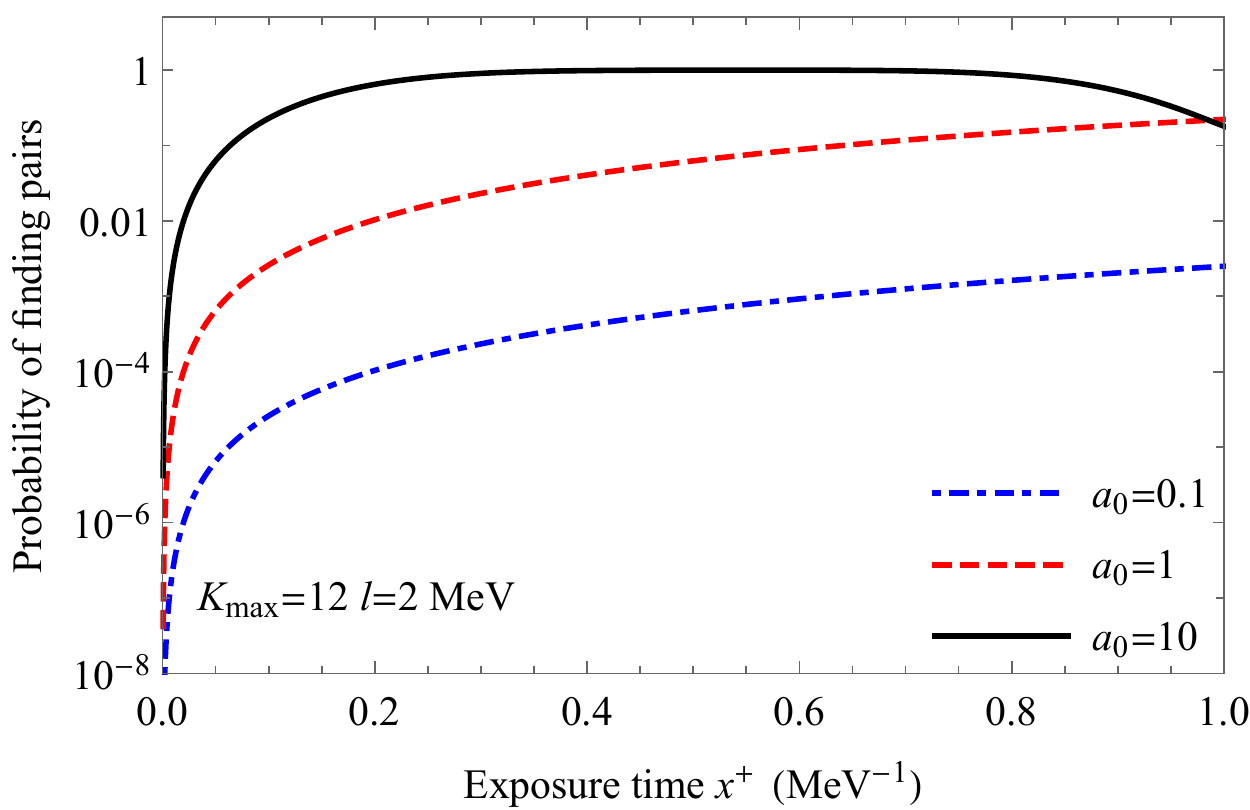}
		\end{tabular}
                \caption{Time evolution of the total probability of finding electron-positron pairs in the background field~\eqref{BG} with $f\left( x^{\LCp} \right)=1$. The left panel compares the results obtained in background fields with $3$ different longitudinal momenta $l=1\rm \,MeV$, $2\rm \,MeV$ and $3\rm \,MeV$, at the same intensity $a_{0}=10$. The right panel compares the results obtained in background fields with $3$ different intensities $a_{0}=0.1$, $1$ and $10$, at the same momentum $l=2\rm \,MeV$. Other parameters: $K_{\max}=12$, $b=m_{e}$.} 
		\label{fig:const_total}
	\end{center}
\end{figure*}

We study the dependences of the total probability of finding electron-positron pairs on the field strength $a_{0}$ and momentum $l$ in Fig.~\ref{fig:const_total}. In the left panel, we show the total probabilities of finding electron-positron pairs obtained in background fields with $3$ different longitudinal momenta $l=1\rm \,MeV$, $2\rm \,MeV$ and $3\rm \,MeV$. We find that at the early stage of the time evolution, the probabilities increase faster in fields with larger momenta; after the probabilities are saturated, we observe dramatic decreases in the $l=2\rm \,MeV$ and $l=3\rm \,MeV$ cases, which we attribute to the truncation artifacts, as already seen in Fig.~\ref{fig:const_kmax_dependence}.
The pairs are produced at larger rates in the fields with larger $l$, which are associated with stronger electric fields as seen in Fig.~\ref{fig:BG}.
The different increasing speeds stem from the different probabilities of finding multi-particle pairs, as seen in Fig.~\ref{fig:const_sector_l}, and can also be explained by the fact that the fields with larger $l$ come with stronger electric fields.
We compare the total probabilities of finding electron-positron pairs obtained in background fields with $3$ different intensities $a_{0}=0.1$, $1$ and $10$ in right panel of Fig.~\ref{fig:const_total}. Note that at the early stage of the time evolution, the probability is almost proportional to the square of the intensity parameter $a_{0}$. This behavior contrasts with Schwinger's famous result for homogeneous background fields~\cite{Schwinger:1951nm}, which exhibits a sharp exponential increase around the Schwinger limit. 

\begin{figure*}[t!]
	\centering
	\begin{center}
		\begin{tabular}{@{}cccc@{}}
                        \includegraphics[width=.48\textwidth]{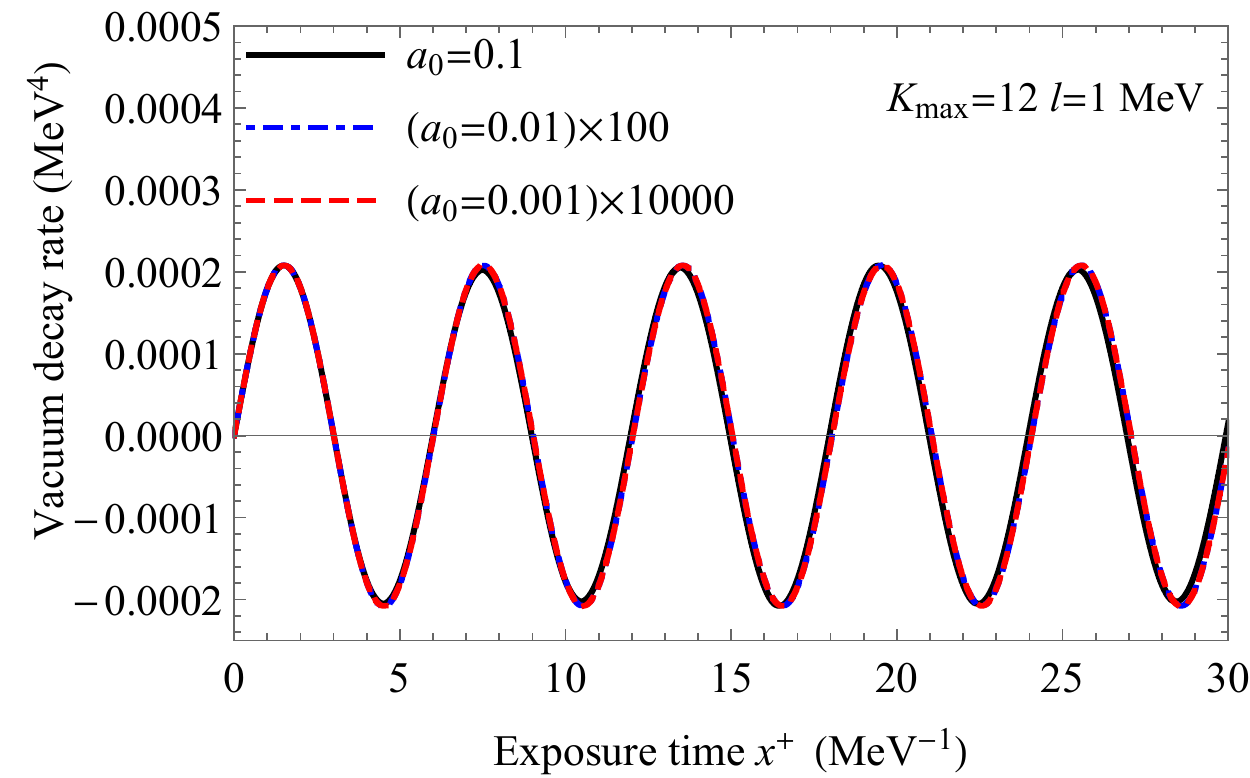} 
			\includegraphics[width=.465\textwidth]{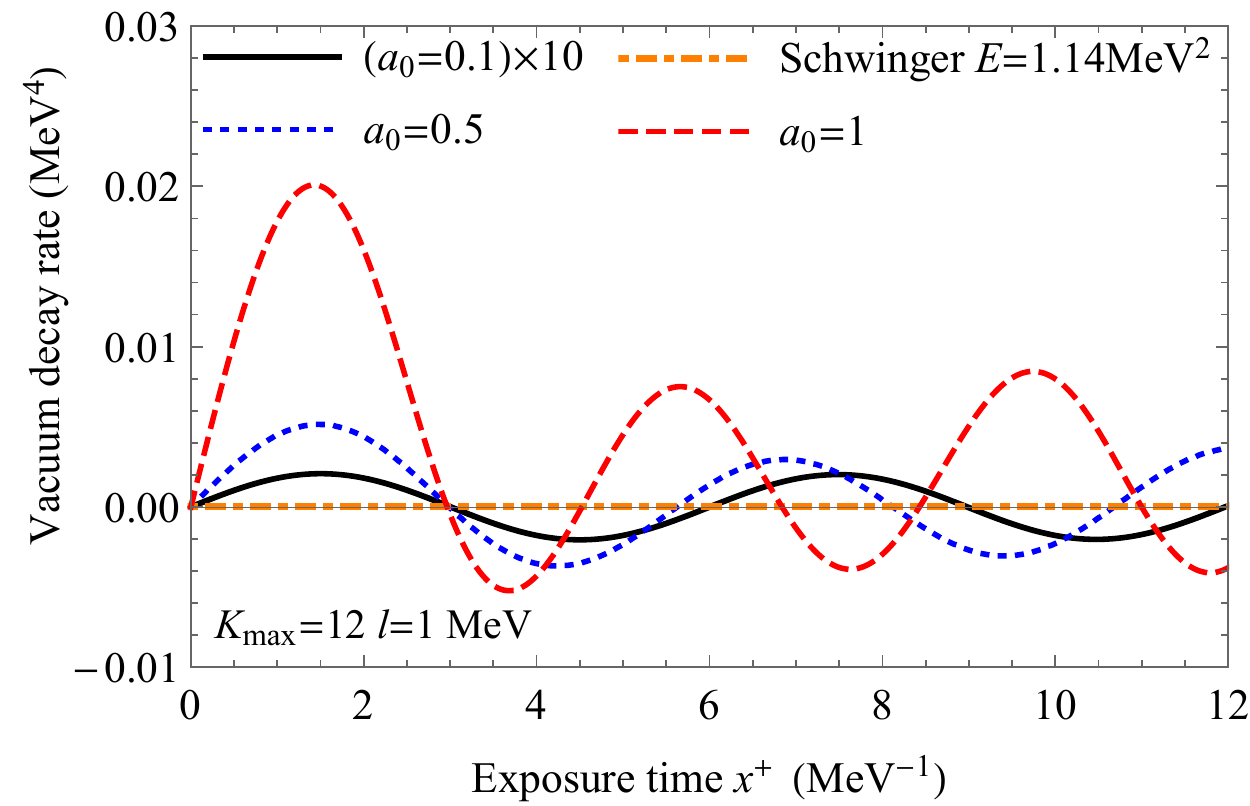} 
		\end{tabular}
                \caption{The vacuum decay rate in the background fields~\eqref{BG} with $f\left( x^{\LCp} \right)=1$ and $l=1\rm \,MeV$. The left panel shows the rates in background fields with intensities $a_{0}=0.001$, $0.01$ and $0.1$. The small results in the first and the second case are scaled by factors of $10000$ and $100$, respectively. The right panel shows the rates in background fields with intensities $a_{0}=0.1$, $0.5$ and $1$, where the result in the first case is scaled by a factor of $10$. Schwinger's result in an electric field $E=E_{\rm RMS}$ is also presented, for the comparison with the $a_{0}=1$ case. Other parameters: $K_{\max}=12$,  $b=m_{e}$.}
		\label{fig:Schwinger}
	\end{center}
\end{figure*}

Schwinger's result for the vacuum decay rate (per unit volume and time) is given by\footnote{Strictly speaking, the vacuum decay rate $\Gamma_{\rm Sch}$ is different from the pair production rate. The latter is the first term of the infinite sum in Eq.~\eqref{eqn:Schwinger}, and in extreme cases the former could be $1.64$ times as large; for a recent review see~\cite{Cohen:2008wz}. }
\begin{equation}
   \Gamma_{\rm Sch}=\frac{(eE)^2}{4\pi^3}\sum_{n=1}^\infty\frac{1}{n^2}\mathrm e^{-\frac{\pi m_{e}^2n}{eE}}\;,
   \label{eqn:Schwinger}
\end{equation}
which is obtained from the one-loop effective action, and has been verified by various theoretical studies~\cite{Tomaras:2001vs,Fried:2001ur}. The electric field $E$ in the denominator of the exponential in Eq.~\eqref{eqn:Schwinger} suggests a sharp increase of the vacuum decay rate $\Gamma_{\rm Sch}$ around the Schwinger limit $E_{c}=m_{e}^{2}/e\approx0.87\rm \,MeV^{2}\approx1.32\times10^{18}\rm\, V/m$, above which the probability of finding electron-positron pairs becomes significant. 
To make a comparison with Schwinger's result, we calculate the vacuum decay rate in our approach, which is obtained by 
\begin{equation}
   \Gamma_{\rm tBLFQ}=-\frac{2\mathrm{d} \ln \left[P_{\mathrm{vac}}\left(x^{\LCp}\right)\right]}{L\left(  L^{\LCperp}\right)^{2}\mathrm{d} x^{\LCp}}\;,
   \label{}
\end{equation}
where $P_{\mathrm{vac}}$ is the probability of the system staying in the vacuum, and the factor $2$ is the Jacobian that compensates for the difference of the volume elements in the instant form and the front form; for conventions, see Appendix~\ref{fields}. We estimate the transverse size $L^{\LCperp}$ of our problem to be $1/b$, the width of the Gaussian profile in coordinate space. We show the vacuum decay rate in our calculation obtained in background fields with different intensities in Fig.~\ref{fig:Schwinger}. The left panel shows the vacuum decay rates in the fields with intensities $a_{0}=0.001$, $0.01$ and $0.1$. The decay rates manifest periodic oscillations with a period $T\approx6\rm \,MeV^{-1}$, and the magnitude of the oscillations is roughly proportional to the square of the intensity. The periodic oscillations can be understood from time-dependent perturbation theory, which simply follows from the energy difference between the vacuum and the states containing an electron-positron pair with $p_{1}^{\LCp}=p_{3}^{\LCp}=1/2\rm\, MeV$. The detailed comparison with time-dependent perturbation theory for small $a_{0}$ can be found in Appendix~\ref{perturbation_theory}. In the cases of small $a_{0}$, the alternating positive and negative parts cancel each other and thus in the long term result in vanishing average decay rates. As $a_{0}$ increases, however, as shown in the right panel of Fig.~\ref{fig:Schwinger}, the decay rate begins to lose the periodicity at a value around $a_{0}=0.1$, above which the average decay rate accumulates with time. We also present Schwinger's result $\Gamma_{\rm Sch}=3.475\times10^{-4}\rm \,MeV$ at $E=E_{\rm RMS}=E_{\max}/\sqrt{2}$, in comparison with our $a_{0}=1$ case. 
The curve corresponding to Schwinger's result is orders of magnitude below our result at $a_{0}=1$. Although we do not aim for a quantitative comparison with Schwinger's result, Fig.~\ref{fig:Schwinger} suggests that at certain instants the vacuum decay rate could be much larger than Schwinger's results even when the field strength is smaller than the Schwinger limit. This is as expected since our background field (in the longitudinal direction) is oscillating with a frequency comparable to the mass of the electron.

\begin{figure*}[t!]
	\centering
	\begin{center}
		\begin{tabular}{@{}cccc@{}}
                        \includegraphics[width=.48\textwidth]{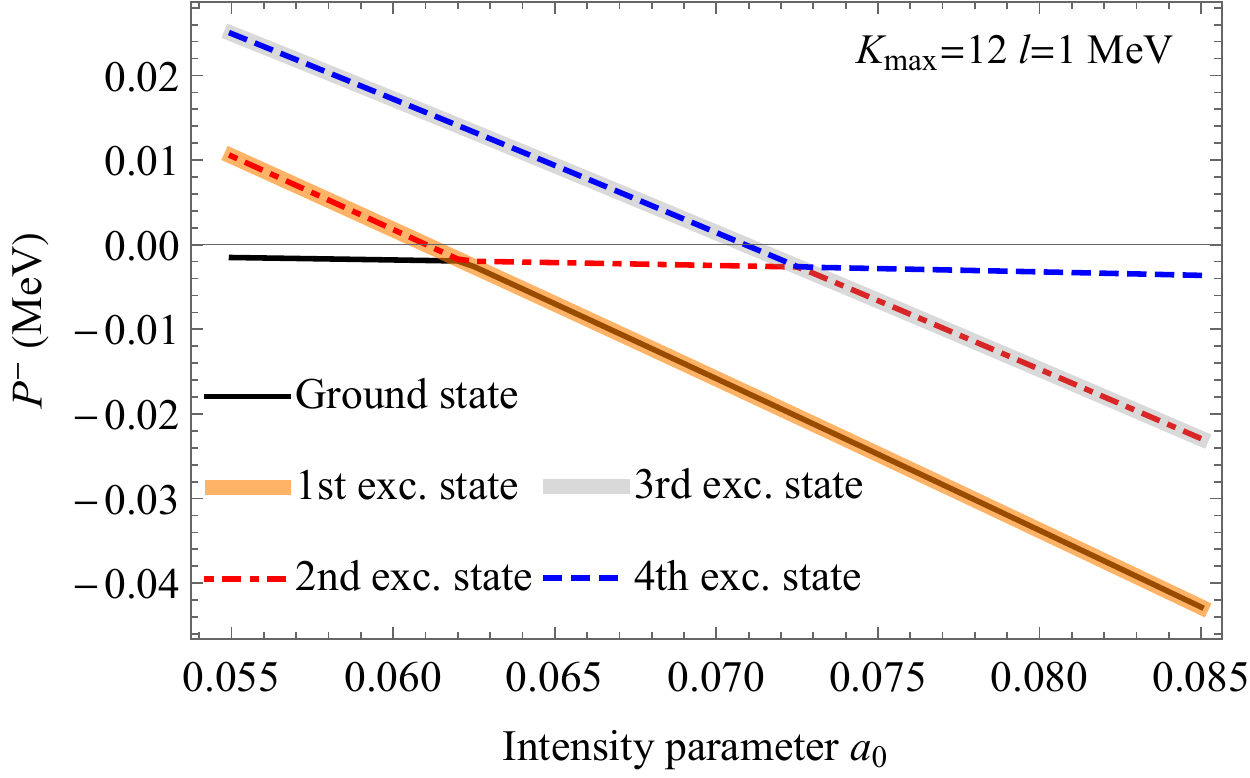} 
			\includegraphics[width=.46\textwidth]{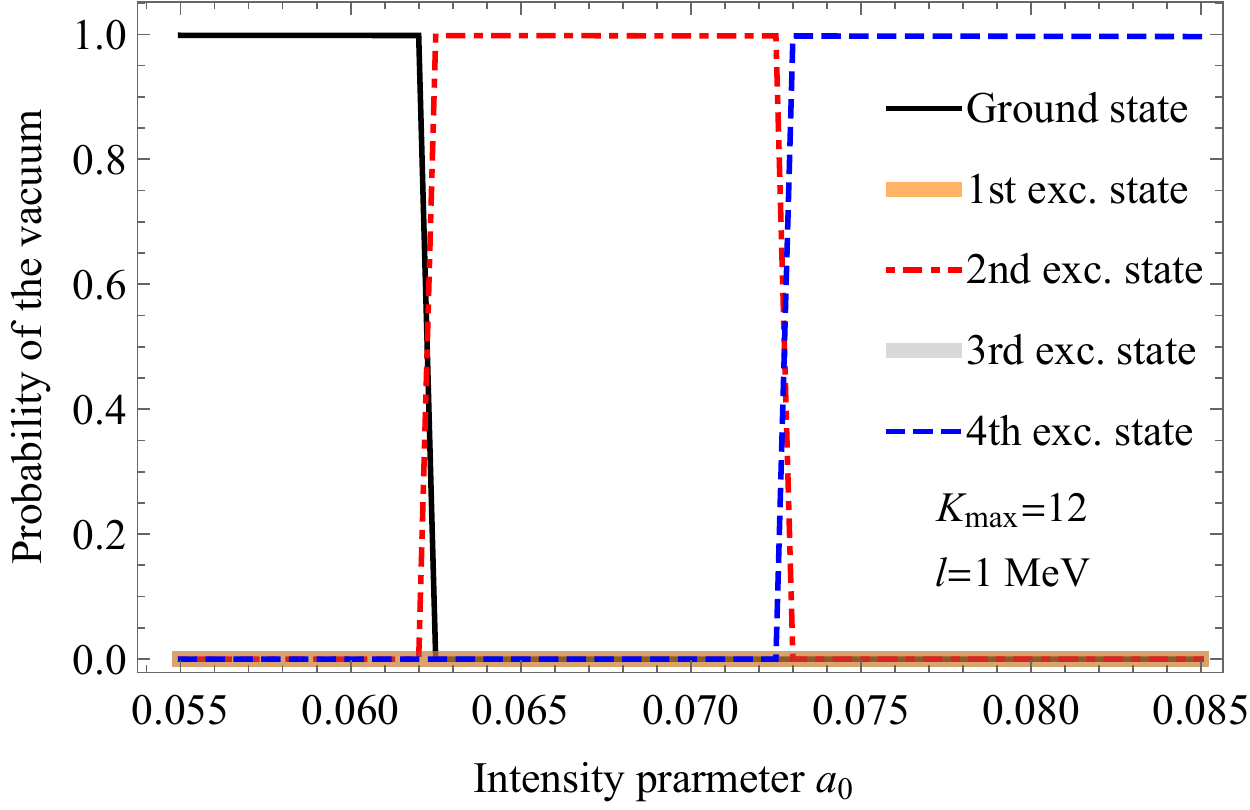} 
		\end{tabular}
                \caption{Left: eigenvalues of several low-lying states as functions of $a_{0}$. Right: the probabilities of finding the vacuum $|\,0\,\rangle$ in several low-lying states as functions $a_{0}$. Other parameters: $K_{\max}=12$, $l=1\rm \,MeV$, $b=m_{e}$.}
		\label{fig:Hg}
	\end{center}
\end{figure*}

In order to have a better understanding of the transition around the intensity parameter $a_{0}=0.1$, we study the eigenvalue problem of the full Hamiltonian $P^{\LCm}=P^{\LCm}_{0}+V$ with different $a_{0}$. We show the resulting light-front energy, as well as the probabilities of finding the vacuum state $|\,0\,\rangle$, of several low-lying states as functions of $a_{0}$ in Fig.~\ref{fig:Hg}. We find that when $a_{0}$ is below the critical value of $0.063$, the ground state energy is close to zero, as shown in the left panel. At the same time the ground state is dominated by the vacuum, as shown in the right panel. As $a_{0}$ increases past the critical value, we observe a ``level crossing'': the background state energy starts to drop with $a_{0}$, and from the right panel of Fig.~\ref{fig:Hg} we observe the ground state is dominated by the pairs. The original ground state seems to be replaced by the second excited states in both $P^{\LCm}$ and the probability of the vacuum. At $a_{0}$ around $0.73$ there is a second ``level crossing'': the light-front energy of the second excited state starts to drop, with the vacuum replaced by the fourth excited state in both $P^{\LCm}$ and the probability of the vacuum. The series of ``level crossings'' indicates the vacuum decay in the presence of background fields. However, we find that these low-lying states are made of particles with large longitudinal momentum of almost $K_{\max}/2$, and therefore are not directly connected to the vacuum by the background field. In the language of perturbation theory, the transitions to these states are higher order effects, so we expect the transition to be efficient only when the transition matrix elements are of the same magnitude of the energy difference. For the states directly connected to the vacuum we estimate the energy difference $\Delta P^{\LCm}_{0}\sim m_{e}^{2}L$ and the transition matrix element $\sim m_{e}a_{0}$, which indicates the critical value of the electric field 
\begin{equation}
   E_{z}=\partial_{\LCm}\mathcal{A}_{\LCp}\sim\frac{m_{e}a_{0}l}{e}\sim\frac{m_{e}^{2}Ll}{e}\sim\frac{m_{e}^{2}}{e}\;.
   \label{eqn:critical}
\end{equation}
Note that this result is independent of the period of the longitudinal direction $L$. When $L$ approaches infinity our field will approach a homogeneous field in space, and our critical value is consistent with that of Schwinger.

\begin{figure*}[t!]
	\centering
	\begin{center}
		\begin{tabular}{@{}cccc@{}}
			\includegraphics[width=.455\textwidth]{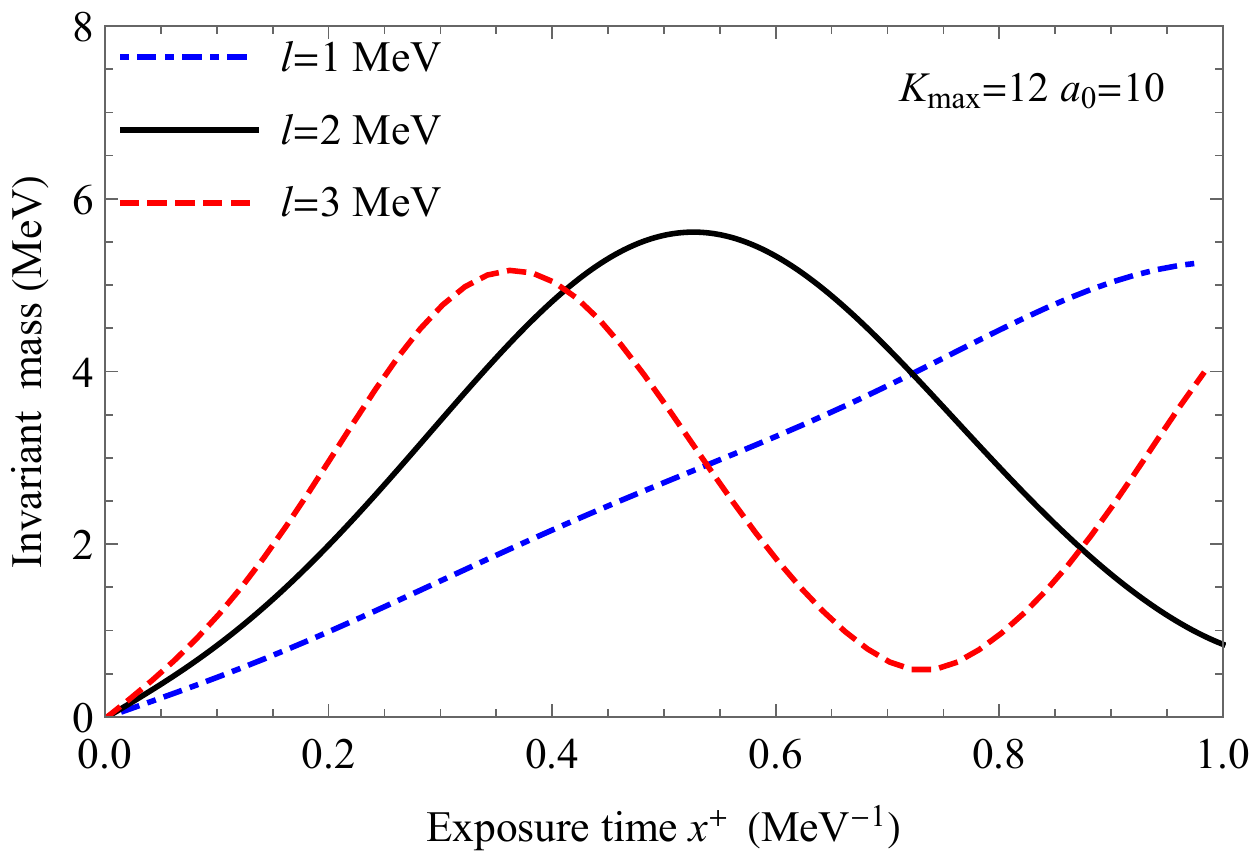}
			\includegraphics[width=.482\textwidth]{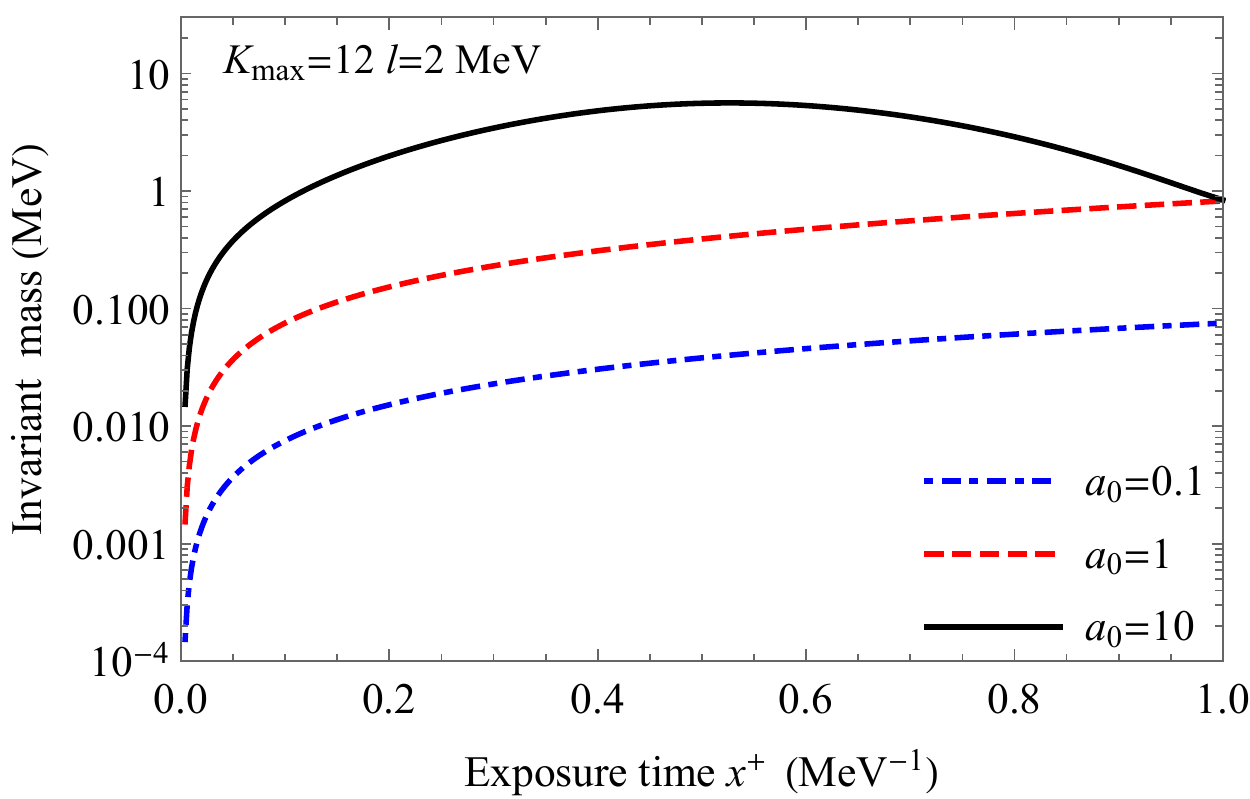} 
		\end{tabular}
                \caption{Time evolution of the invariant mass of all the particles in the background fields~\eqref{BG} with $f\left( x^{\LCp} \right)=1$. The left panel shows results obtained in fields with different longitudinal momenta $l=1\rm \,MeV$, $2\rm \,MeV$ and $3\rm \,MeV$, at the intensity $a_{0}=10$. The right panel shows results obtained in fields with different intensities $a_{0}=0.1$, $a_{0}=1$ and $a_{0}=10$, at the momentum $l=2\rm \,MeV$. Other parameters: $K_{\max}=12$, $b=m_{e}$.} 
		\label{fig:const_invmass}
	\end{center}
\end{figure*}

We present the dependences of the invariant mass~\eqref{eqn:invmass} on the momentum $l$ and the intensity parameter $a_{0}$ of the background field in Fig.~\ref{fig:const_invmass}. The left panel compares the results obtained in fields with $3$ different longitudinal momenta $l=1\rm \,MeV$, $2\rm \,MeV$ and $3\rm \,MeV$. The invariant mass increases faster in fields with larger longitudinal momentum, implying that background fields with larger momentum (corresponding to stronger electric field) inject energy into the system at faster rates. The decreases of the invariant masses for $l=1\rm MeV$ and $2\rm MeV$ are possibly due to the truncation of the finite basis, as seen in Fig.~\ref{fig:const_kmax_dependence}.
The right panel compares the invariant mass obtained in the background fields~\eqref{BG} with different intensities $a_{0}=0.1$, $1$ and $10$. We observe that the invariant mass is roughly proportional to the square of the intensity of the background field. 

\begin{figure*}[t!]
	\centering
	\begin{center}
		\begin{tabular}{@{}cccc@{}}
			\includegraphics[width=.46\textwidth]{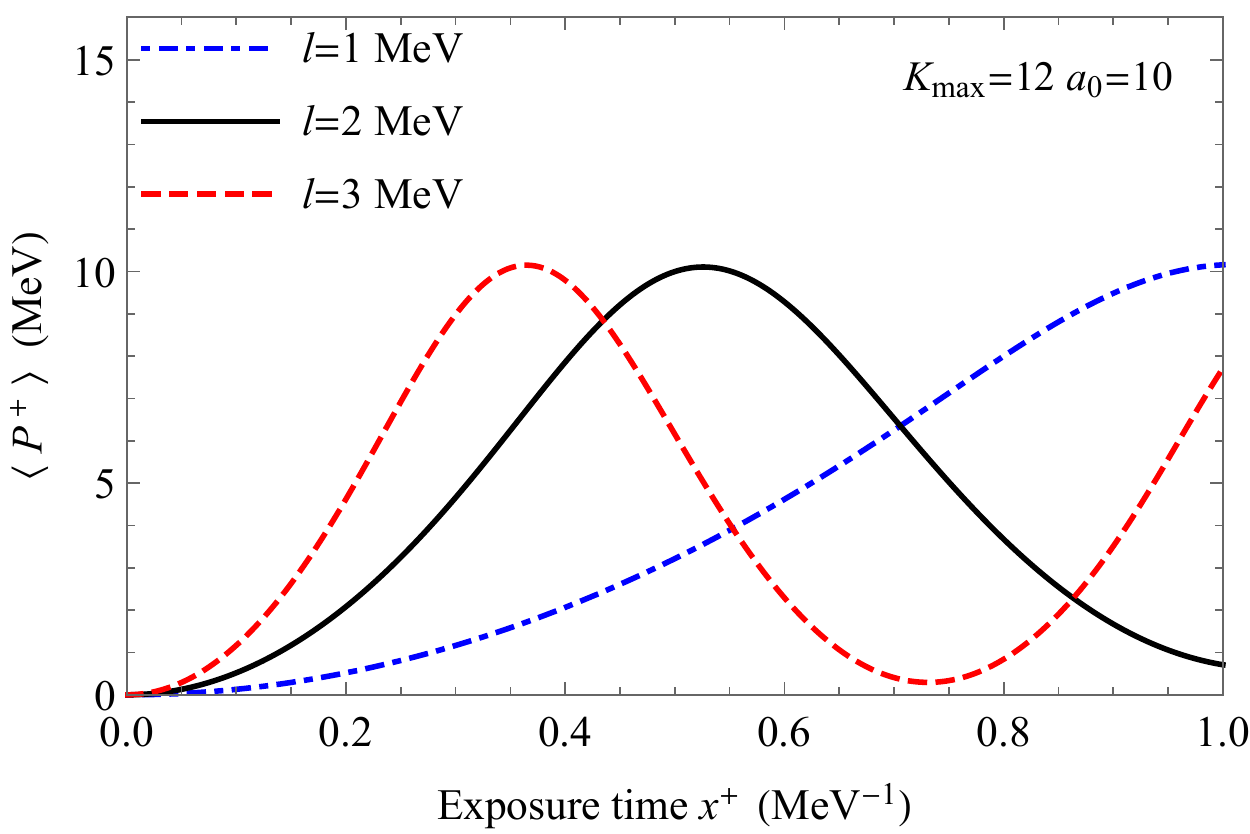}
			\includegraphics[width=.48\textwidth]{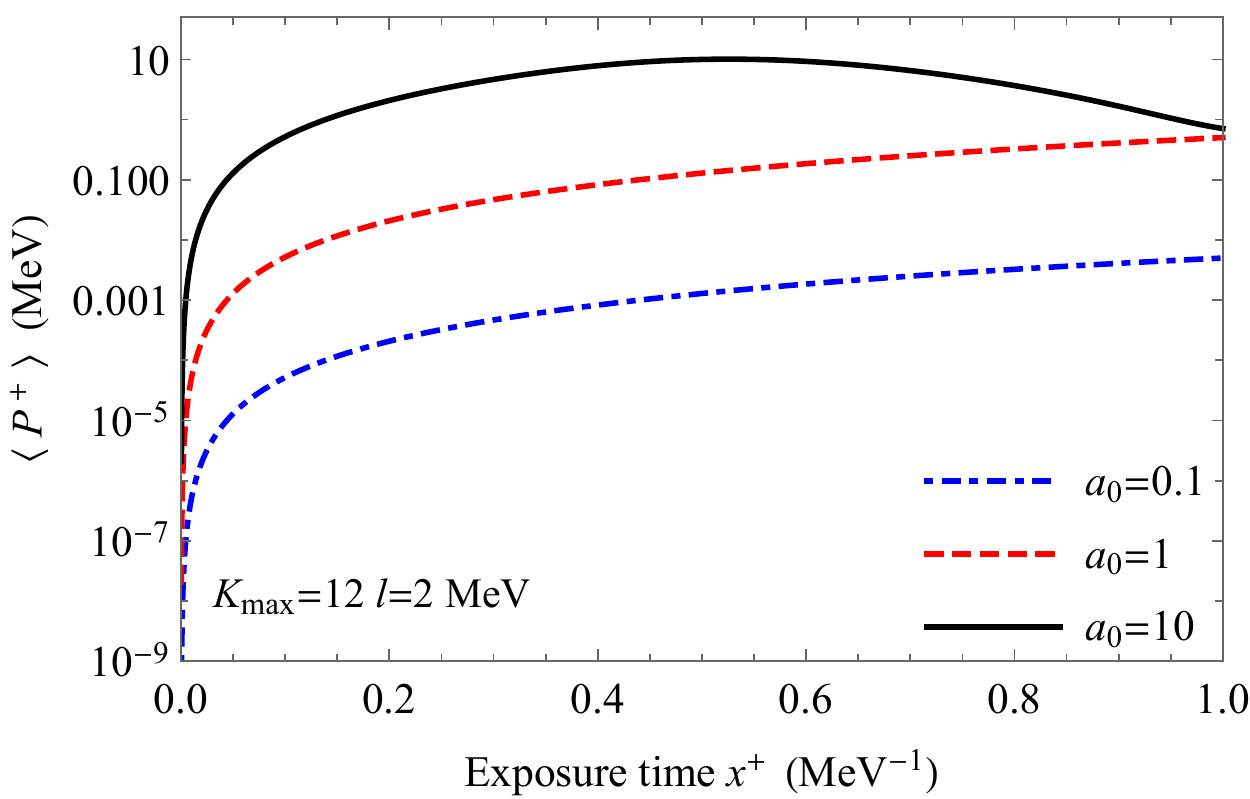}
		\end{tabular}
                \caption{Time evolution of the expectation value of the total momentum of all the particles in the background field~\eqref{BG} with $f\left( x^{\LCp} \right)=1$. The left panel shows results obtained in fields with different longitudinal momenta $l=1\rm \,MeV$, $2\rm \,MeV$ and $3\rm \,MeV$, at the intensity $a_{0}=10$. The right panel shows results obtained in fields with different intensities $a_{0}=0.1$, $a_{0}=1$ and $a_{0}=10$, at the same momentum $l=2\rm \,MeV$. Other parameters: $K_{\max}=12$, $b=m_{e}$.} 
		\label{fig:const_total_momentum}
	\end{center}
\end{figure*}
\begin{figure*}[t!]
	\centering
	\begin{center}
		\begin{tabular}{@{}cccc@{}}
			\includegraphics[width=.47\textwidth]{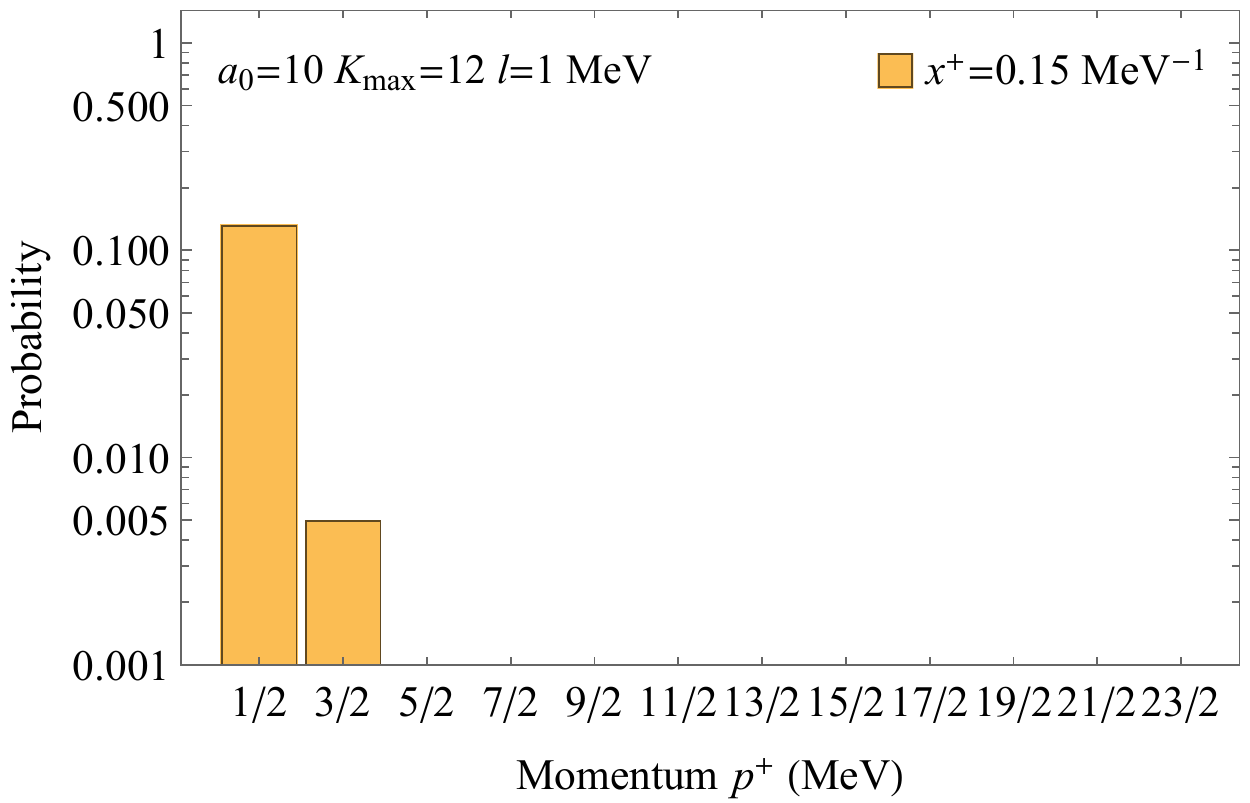} 
			\includegraphics[width=.47\textwidth]{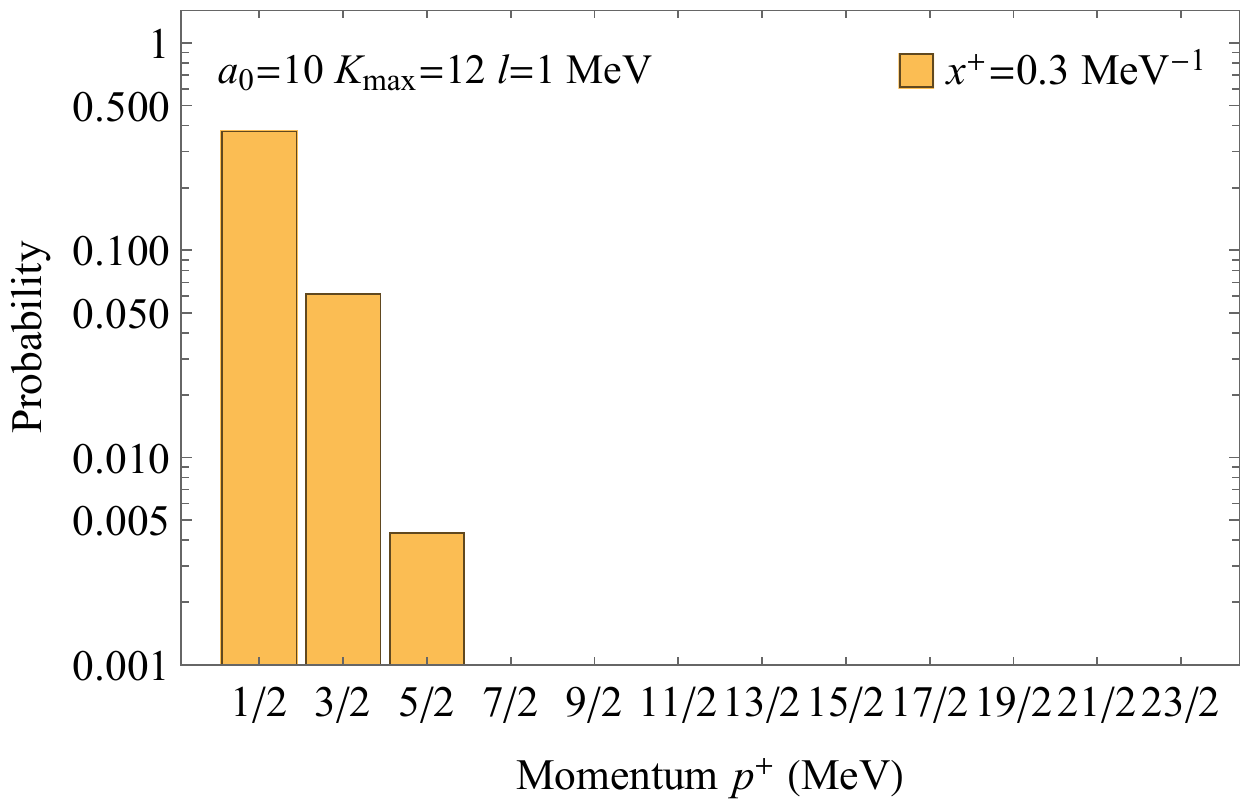}\\
			\includegraphics[width=.47\textwidth]{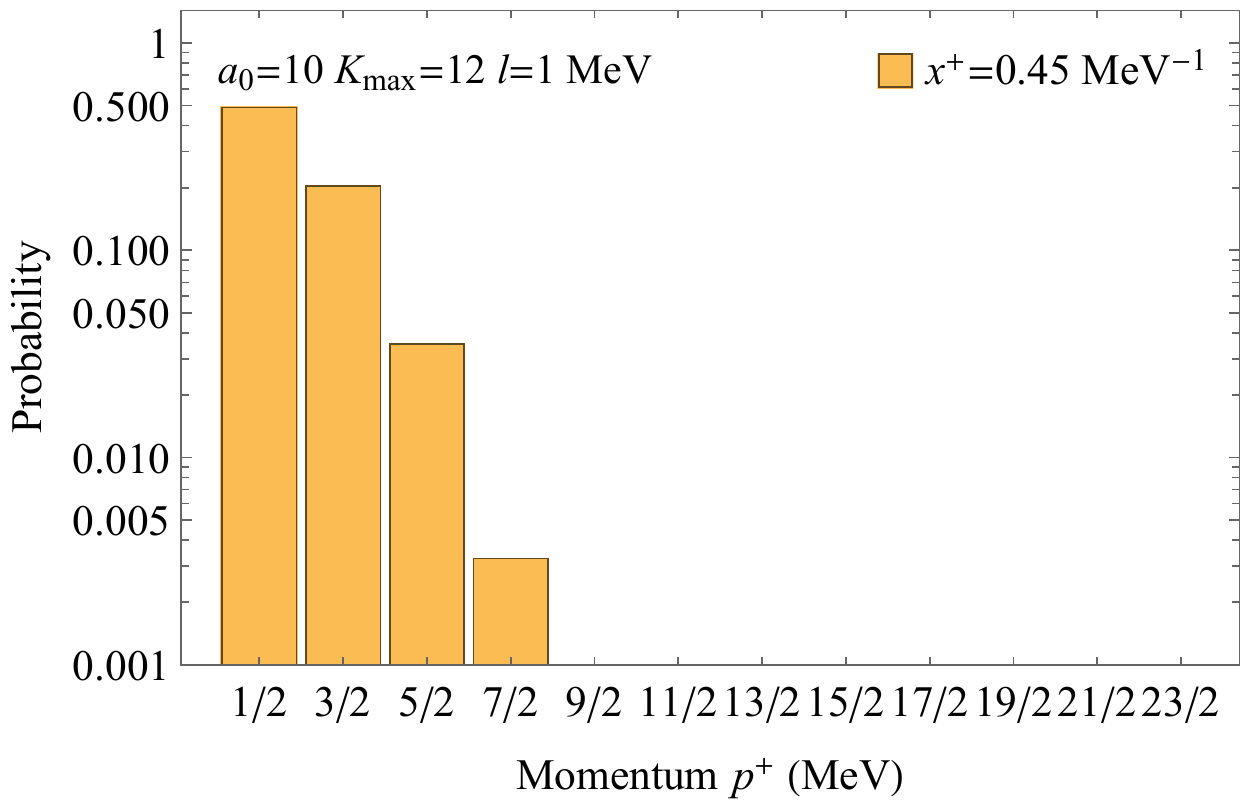} 
			\includegraphics[width=.47\textwidth]{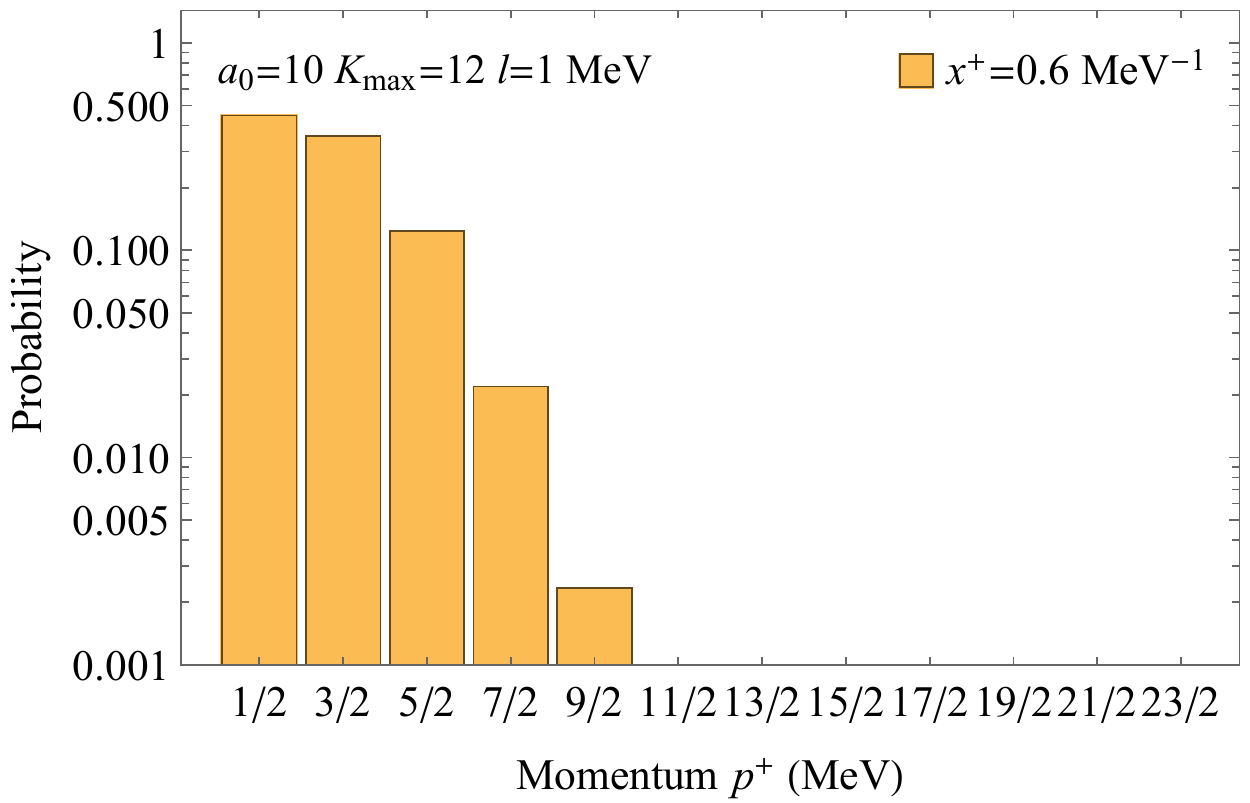} 
		\end{tabular}
                \caption{Time evolution of the longitudinal momentum distribution of the electron in the background field~\eqref{BG} with $f\left( x^{\LCp} \right)=1$, $a_{0}=10$, and $l=1\rm \,MeV$. Other parameters: $K_{\max}=12$, $b=m_{e}$.}
		\label{fig:const_md_l1}
	\end{center}
\end{figure*}
\begin{figure*}[t!]
	\centering
	\begin{center}
		\begin{tabular}{@{}cccc@{}}
			\includegraphics[width=.47\textwidth]{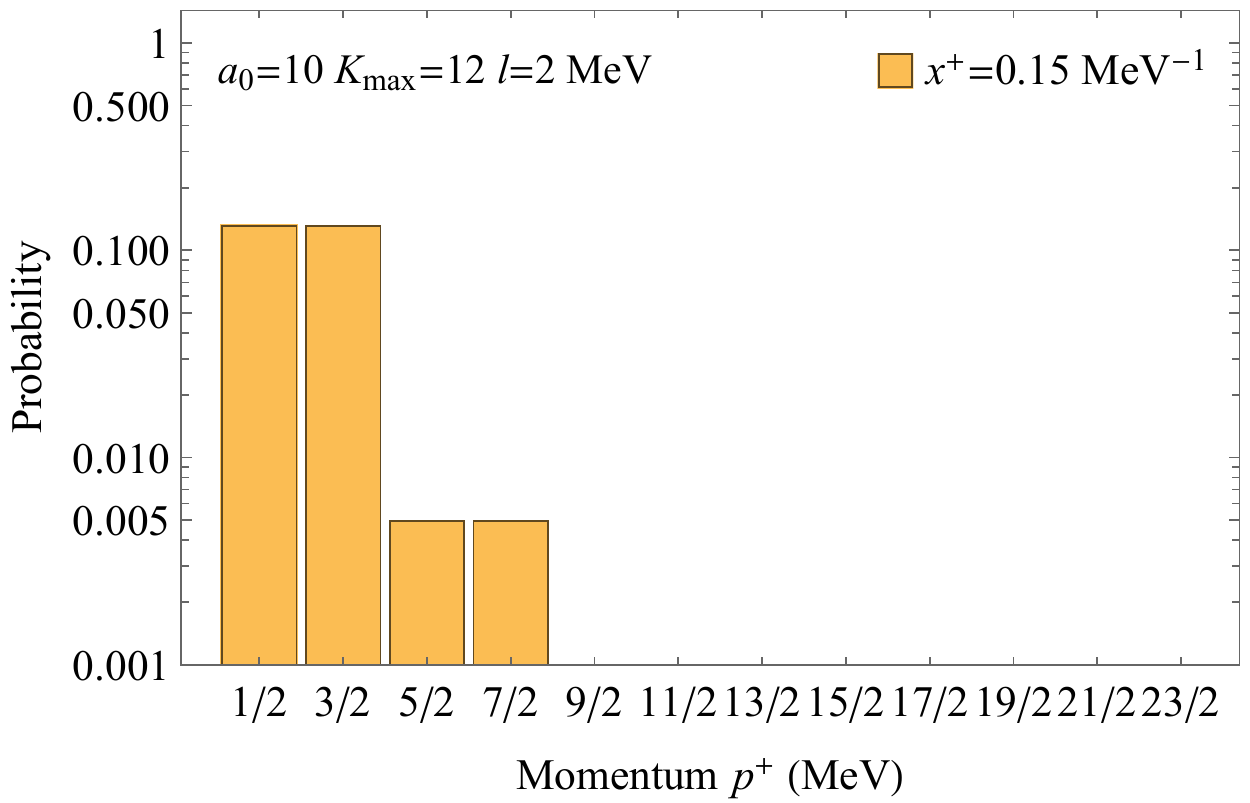} 
			\includegraphics[width=.47\textwidth]{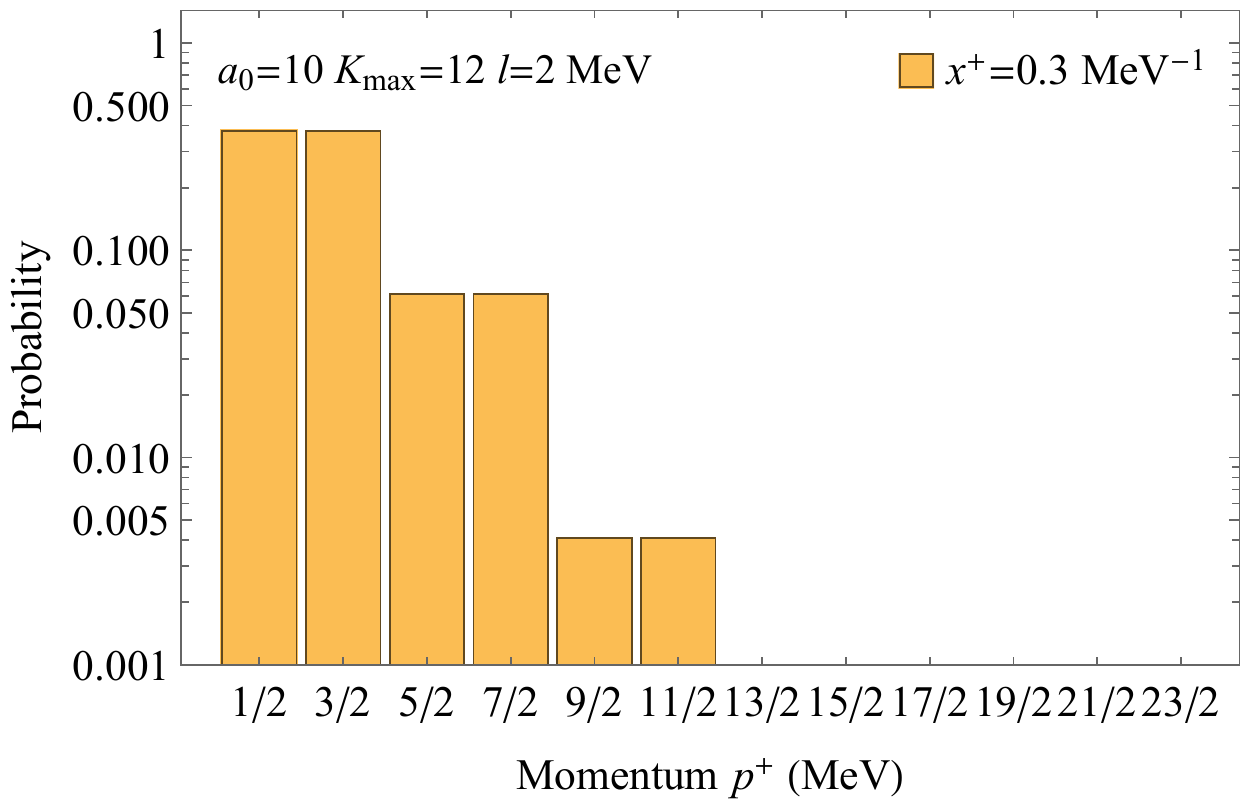}\\
			\includegraphics[width=.47\textwidth]{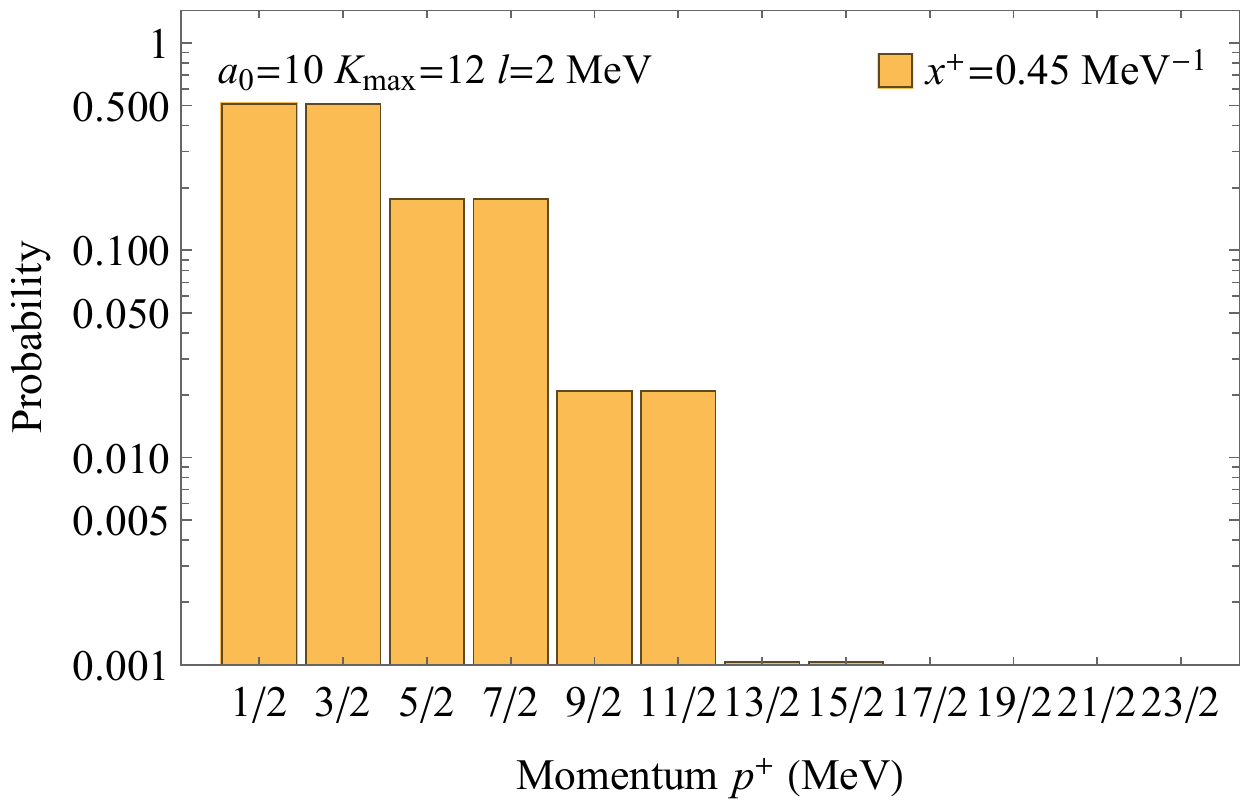} 
			\includegraphics[width=.47\textwidth]{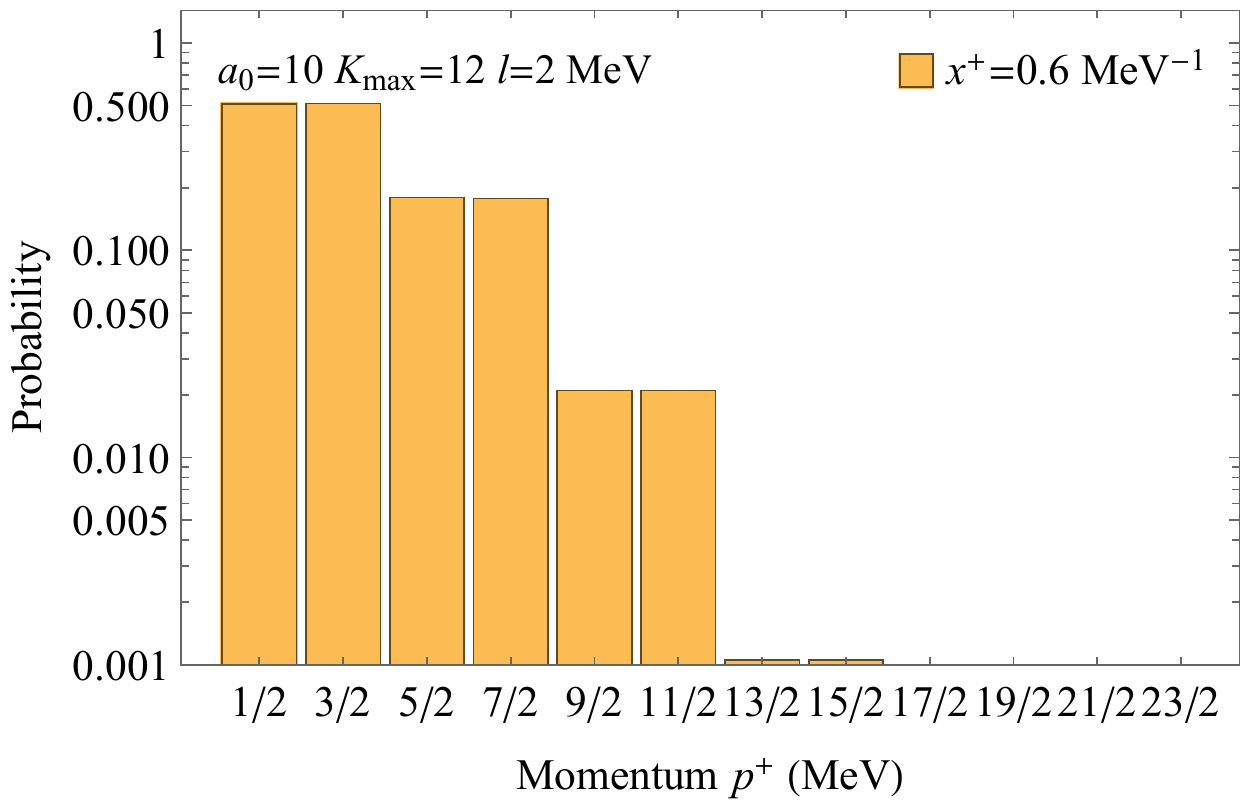} 
		\end{tabular}
                \caption{Time evolution of the longitudinal momentum distribution of the electron in the background field~\eqref{BG} with $f\left( x^{\LCp} \right)=1$, $a_{0}=10$, and $l=2\rm \,MeV$. Other parameters: $K_{\max}=12$, $b=m_{e}$.}
		\label{fig:const_md_l2}
	\end{center}
\end{figure*}
\begin{figure*}[t!]
	\centering
	\begin{center}
		\begin{tabular}{@{}cccc@{}}
			\includegraphics[width=.47\textwidth]{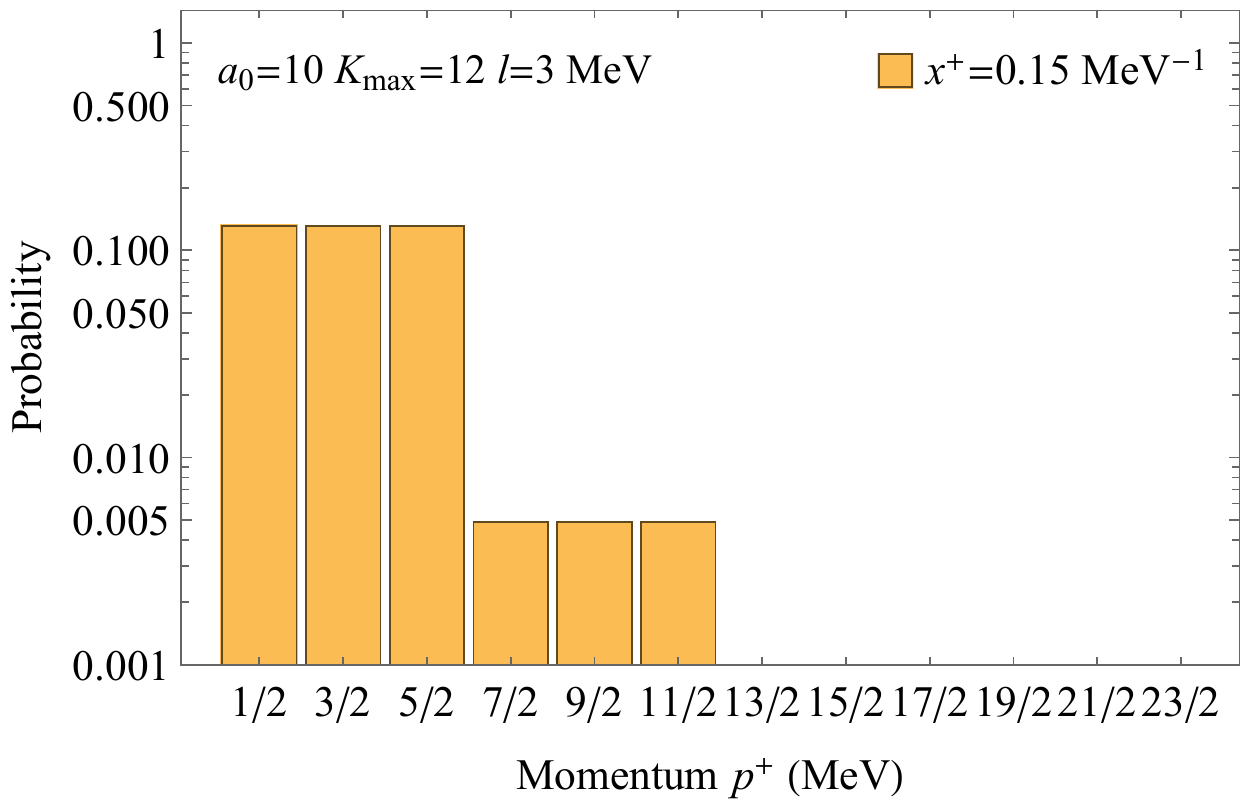} 
			\includegraphics[width=.47\textwidth]{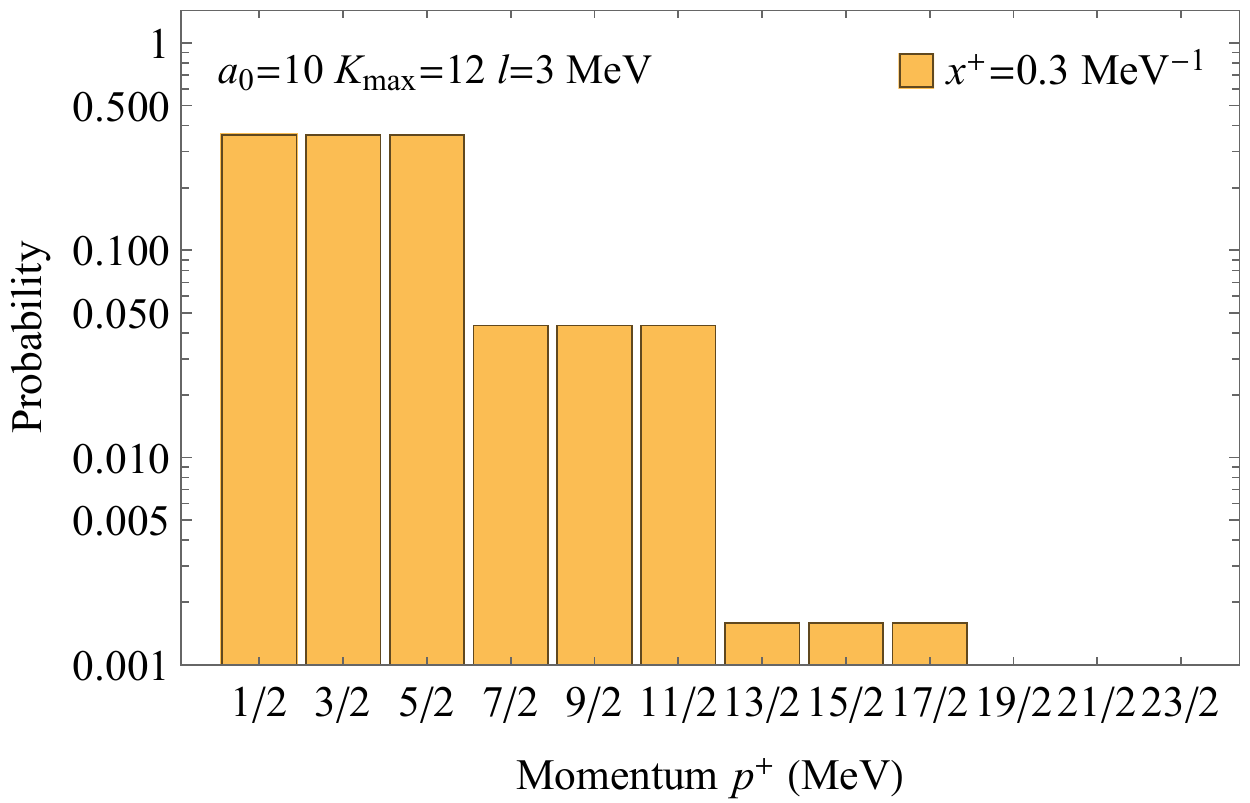} \\
			\includegraphics[width=.47\textwidth]{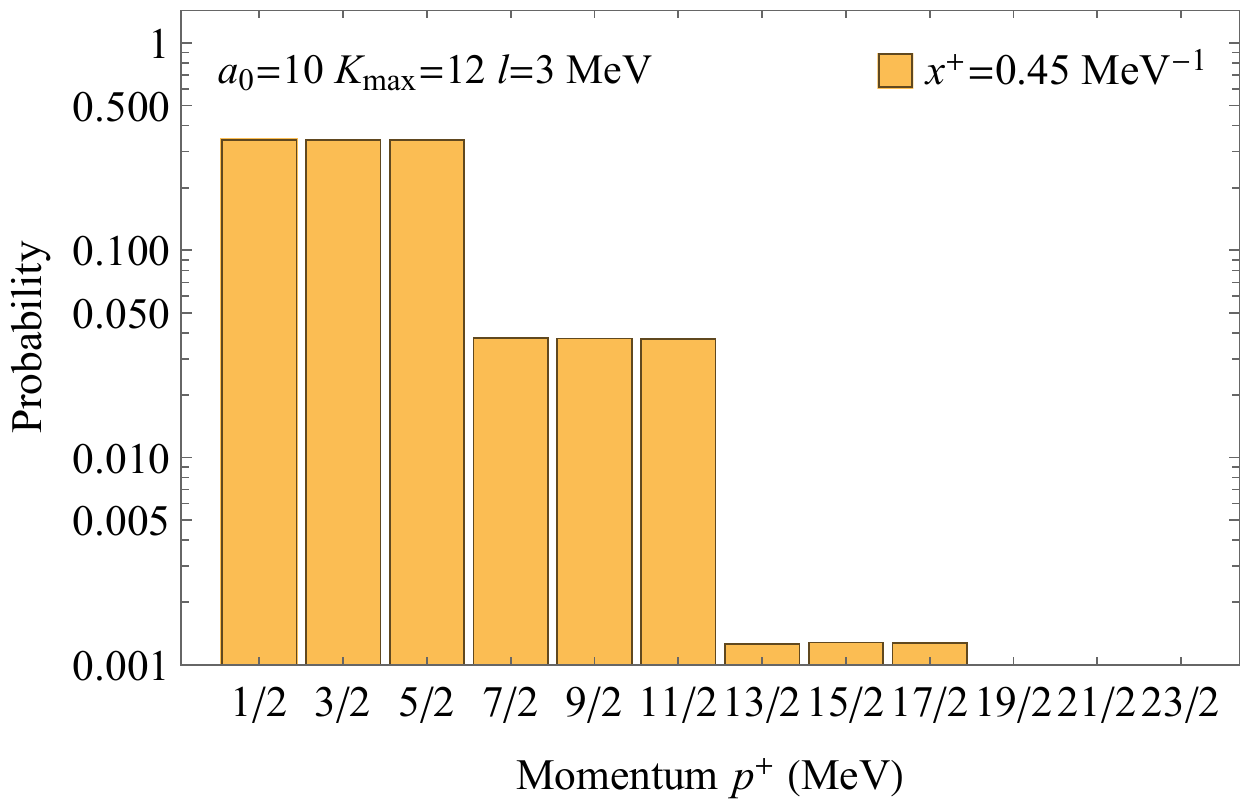} 
			\includegraphics[width=.47\textwidth]{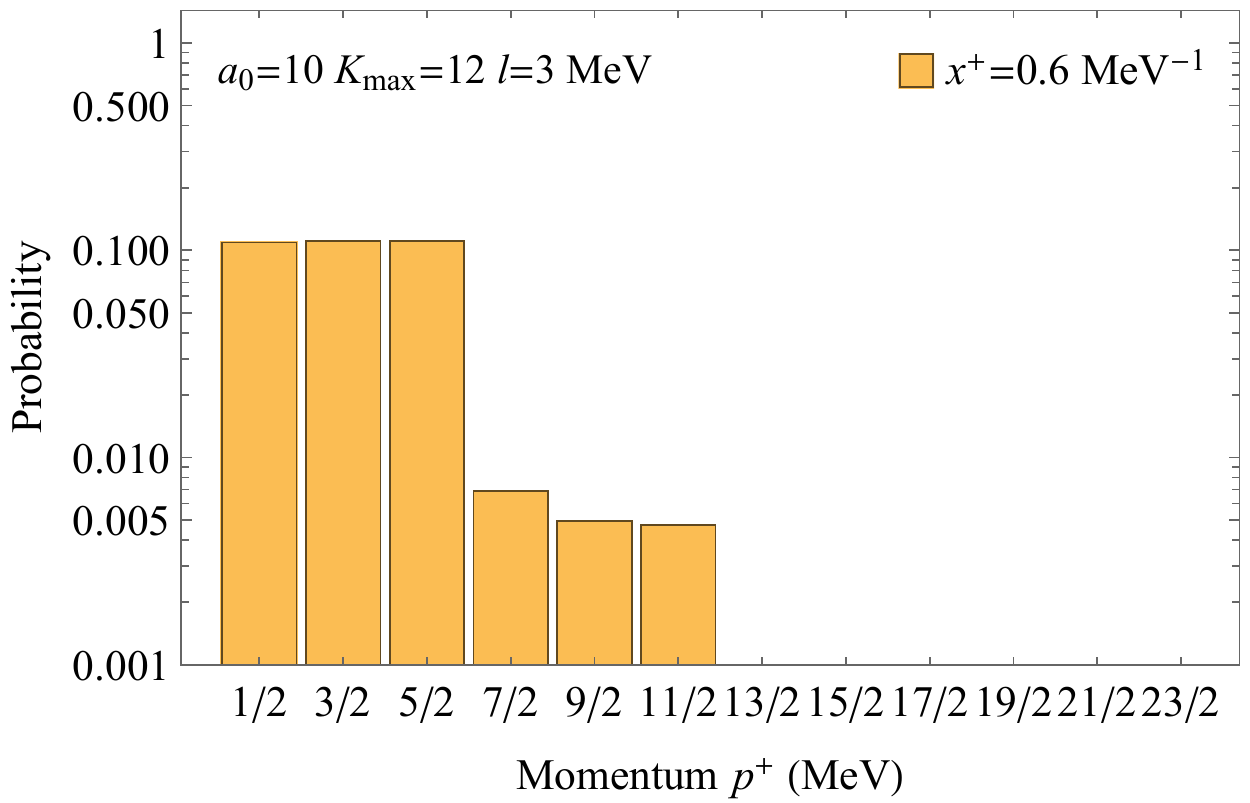} 
		\end{tabular}
                \caption{Time evolution of the longitudinal momentum distribution of the electron in the background field~\eqref{BG} with $f\left( x^{\LCp} \right)=1$, $a_{0}=10$, and $l=3\rm \,MeV$. Other parameters: $K_{\max}=12$, $b=m_{e}$.}
		\label{fig:const_md_l3}
	\end{center}
\end{figure*}

We present the dependences of the expectation value of the total longitudinal momentum on the momentum $l$ and the intensity parameter $a_{0}$ of the background field in Fig.~\ref{fig:const_total_momentum}. The left panel shows results obtained in background fields with $3$ different longitudinal momenta $l=1\rm \,MeV$, $2\rm \,MeV$ and $3\rm \,MeV$; the right panel shows results obtained in fields with $3$ different intensities $a_{0}=0.1$, $a_{0}=1$ and $a_{0}=10$. In both panels the curves show similar trends to those in Fig.~\ref{fig:const_invmass}.

We show the time evolution of the longitudinal momentum distribution~\eqref{eqn:md} in background fields with momenta $l=1\rm \,MeV$, $2\rm \,MeV$ and $3\rm \,MeV$ in Figs.~\ref{fig:const_md_l1}, \ref{fig:const_md_l2} and \ref{fig:const_md_l3}, respectively. The momentum distribution describes the probability of finding a particle with momentum $k_{i}$ and helicity $\lambda$ in the system, and provides a more differential description of the dynamics of the system interacting with the background field. In our background fields, there are exact symmetries between the spin-up and the spin-down particles and between electrons and positrons, so we only need to study the spin-up electrons in the momentum distribution. The figures show that, as time passes, the probabilities build up initially at small longitudinal momenta and then gradually spread to larger ones, implying the electrons are both created and accelerated by the background field. With larger $l$, the momentum distributions are increasing and broadening more quickly, implying that the pairs have larger probabilities of being created and accelerated by fields with larger longitudinal momentum. Another main feature in the momentum distributions is that for $l=2\rm \,MeV$ and $l=3\rm \,MeV$, the probabilities are almost the same for adjacent longitudinal momenta. For example, in Fig.~\ref{fig:const_md_l2} at $p^{\LCp}=1/2\rm \,MeV$ and $p^{\LCp}=3/2\rm \,MeV$ the probability of finding an electron is almost the same, which is not the case in Fig.~\ref{fig:const_md_l1}. This difference can be explained by the interaction part of the Hamiltonian, which shows an exact symmetry in creating the $p^{\LCp}=1/2\rm \,MeV$ and $p^{\LCp}=3/2\rm \,MeV$ electrons in the $l=2\rm \,MeV$ case, and this symmetry is barely broken by the kinetic energy part and the truncation of the basis.  Note that the reduction of the longitudinal momentum distribution in the $l=3\rm \, MeV$ case at $x^{\LCp}=0.6\rm \, MeV^{-1}$ is likely due to the truncation artifacts.

\begin{figure*}[t!]
	\centering
	\begin{center}
		\begin{tabular}{@{}cccc@{}}
                   \includegraphics[width=.485\textwidth]{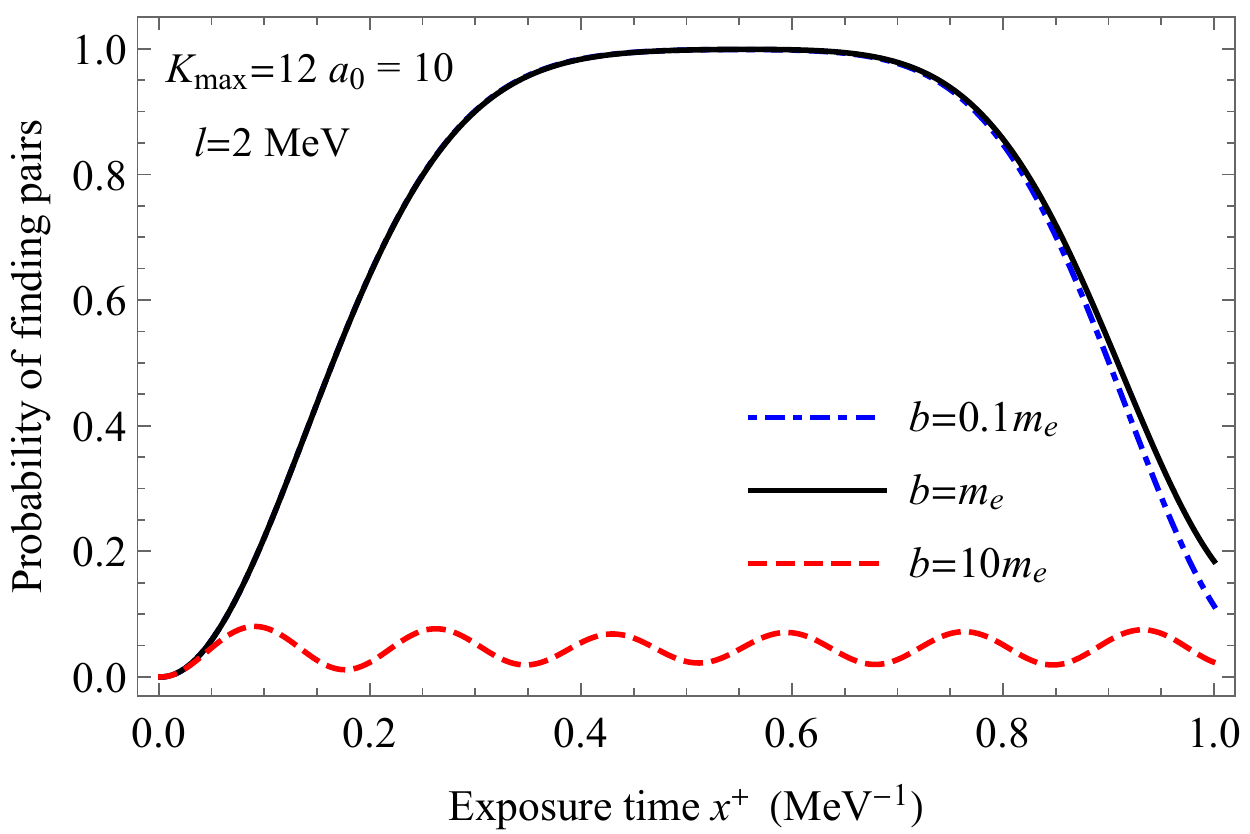} 
                        \includegraphics[width=.47\textwidth]{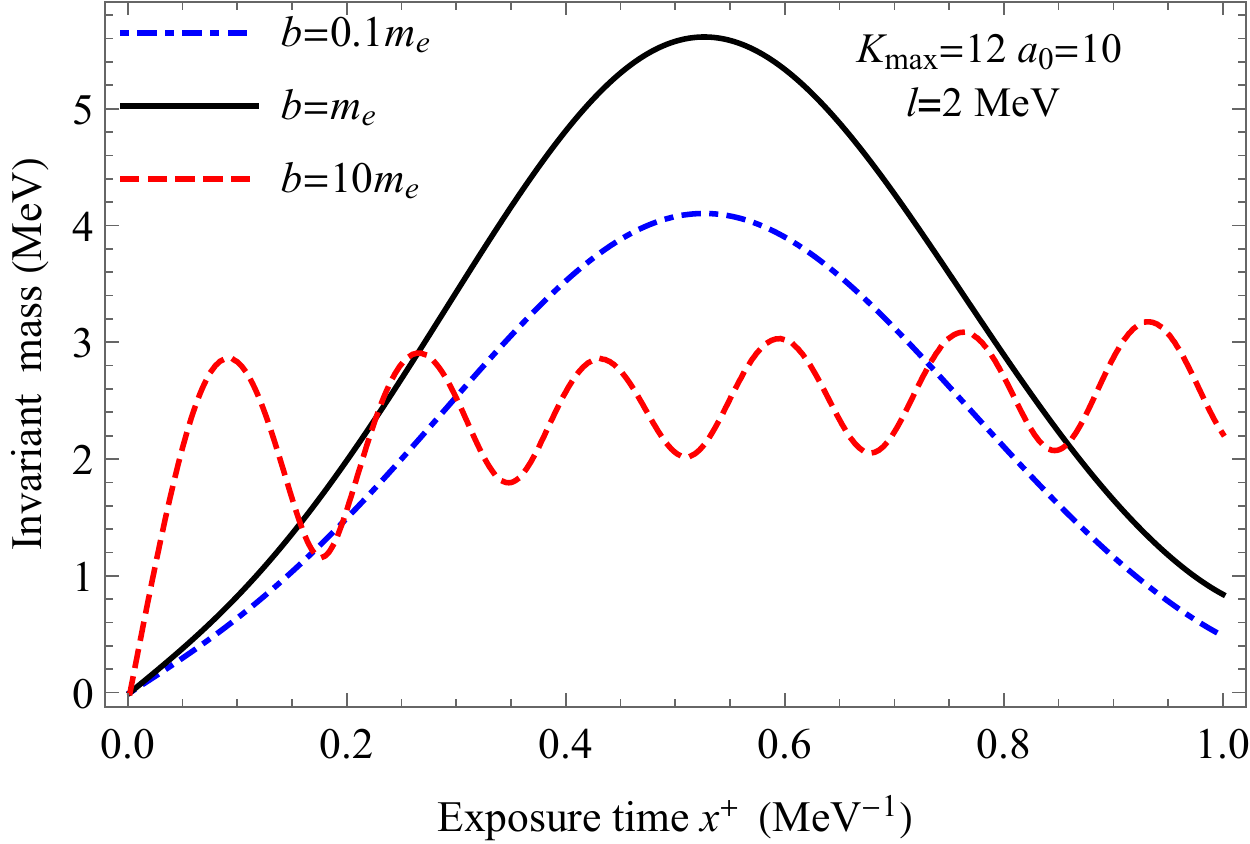} 
		\end{tabular}
                \caption{Left: time evolution of the total probability of finding electron-positron pairs in bases with $3$ different widths $b=0.1m_{e}$, $m_{e}$ and $10m_{e}$. Right: time evolution of the invariant mass in bases with $3$ different widths $b=0.1m_{e}$, $m_{e}$ and $10m_{e}$. Other parameters: $K_{\max}=12$, $a_{0}=10$, $l=2\rm \,MeV$.}
		\label{fig:b_dependence}
	\end{center}
\end{figure*}

Finally we study the dependences of the total probability and the invariant mass on the transverse size $b$. In the left panel of Fig.~\ref{fig:b_dependence}, we show the total probability of finding electron-positron pairs in bases with $3$ different transverse widths $b=0.1m_{e}$, $m_{e}$ and $10m_{e}$. We observe that the curves for $b=0.1m_{e}$ and $m_{e}$ coincide with each other from $x^{\LCp}=0\rm\, MeV^{-1}$ to $x^{\LCp}\approx0.6\rm\, MeV^{-1}$. However, for $b=10m_{e}$ the probability show periodic oscillations as in Fig.~\ref{fig:Schwinger}. This is because $b=10m_{e}$ will dramatically increase the kinetic energy of the basis states (see Appendix~\ref{melements}), so the transition matrix elements and the kinetic energy differences will be at different magnitudes, which will lead to the suppression of the excitation of the vacuum to higher states. Again we attribute the decreases of the probabilities for $b=0.1m_{e}$ and $m_{e}$ around $x^{\LCp}=0.7\rm\,MeV^{-1}$ to truncation artifacts as seen in Fig.~\ref{fig:const_kmax_dependence}. In the right panel of Fig.~\ref{fig:b_dependence}, we show the time evolution of the invariant mass in bases with $3$ different transverse widths $b=0.1 m_{e}$, $m_{e}$ and $10m_{e}$. The invariant masses and the total probabilities show similar trends in all $3$ widths: the curve for $b=10m_{e}$ is oscillating and does not accumulate with time; the curves for $b=0.1m_{e}$ and $m_{e}$ increase and then decrease with time, with the slight difference coming from the kinetic energy which also contributes to the invariant mass. The main difference is that the curves for the invariant mass do not have plateaus as seen in the curves for the probability. This is because even when the total probability is saturated, the background field is still able to accelerate the particles and generate new pairs in the higher sectors, which will change the invariant mass.

\subsection{Background fields with time profile $f\left( x^{\LCp} \right)=\mathrm{sin}\left({\omega x^{\LCp}}\right), \omega=\pi\rm\,MeV$}\label{pi}
In this subsection, we take the frequency of time profile $ \omega=\pi\rm\,MeV$ in the background field~\eqref{BG}. We start and stop the evolution at $x^{\LCp}=0 \rm \,MeV^{-1}$ and $x^{\LCp}=1\rm \,MeV^{-1}$ respectively. This time profile mimics the pulse structure in modern laser facilities. It also gives the unambiguous definition of the particle number in the final state after the time evolution~\cite{Dabrowski:2016tsx}. We study the time evolution of observables and their dependences on the truncation parameter $K_{\max}$, as well as the field intensity parameter $a_{0}$ and momentum $l$. The observables include the probability of finding electron-positron pairs, the invariant mass, and the momentum distributions, as we did in Sec.~\ref{1}.
The results obtained in the fields with this sinusoidal time profile show similar overall patterns to their counterparts in the previous subsection. The main differences between the results with the $2$ time profiles are that most of the results with this sinusoidal time profile are smaller in magnitude, which can be explained by the fact that fields with the latter time profile translate to lower intensities averaged over the entire cycle. For conciseness, we will only mention the differences between the previous results.

\begin{figure*}[t!]
   \centering
   \begin{center}
      \begin{tabular}{@{}cccc@{}}
         \includegraphics[width=.50\textwidth]{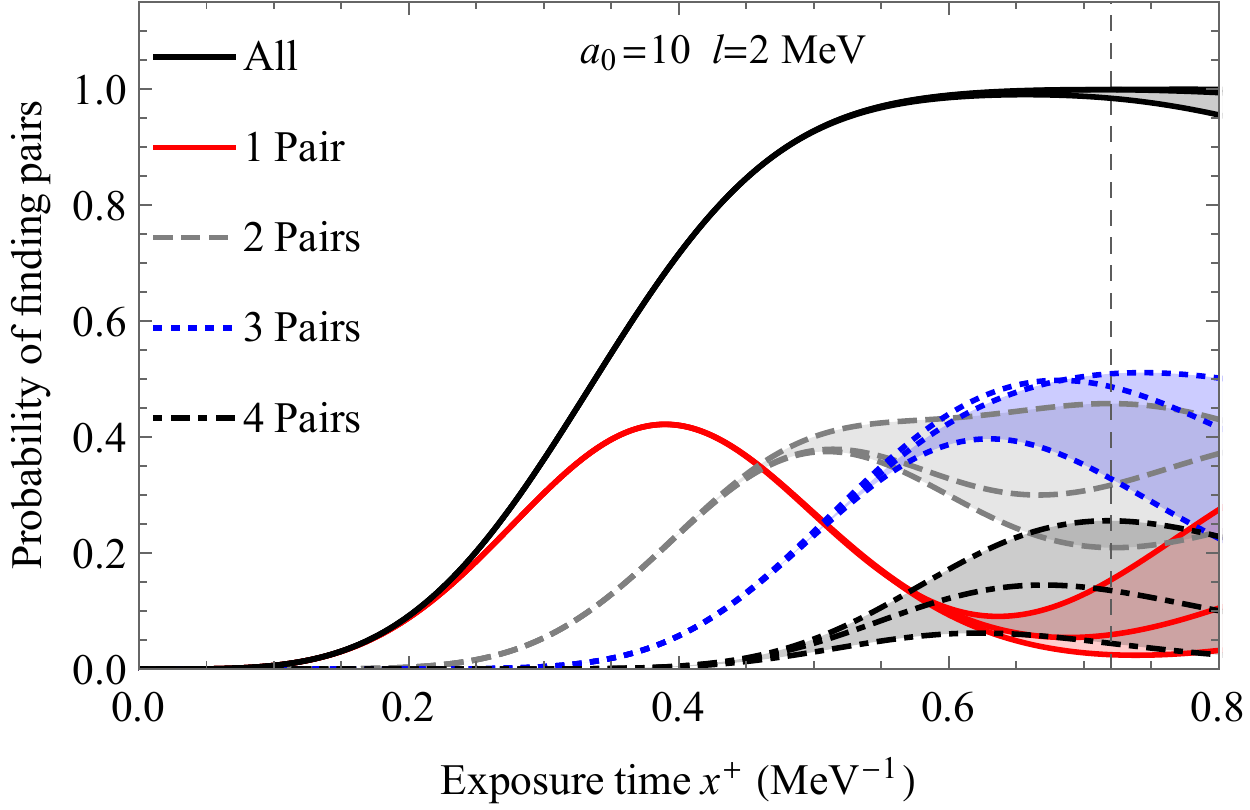} 
      \end{tabular}
      \caption{Time evolution of the probabilities of finding $n$ pairs of electrons and positrons (and their total) in the background field~\eqref{BG} with $f\left( x^{\LCp} \right)=\sin\left( \omega x^{\LCp} \right), \omega=\pi\mathrm{\,MeV}$, $a_{0}=10$ and $l=2\rm \,MeV$. Results in bases with $K_{\max}=8$, $10$ and $12$ are presented in different curves with the same plot style, roughly corresponding to the top, middle and bottom curves in each shaded area. One exception is the curve corresponding to the probability of finding $3$ pairs for $K_{\max}=10$, which is at the top before $x^{\LCp}\approx0.7 {\rm \,MeV}$.
      Other parameter:  $b=m_{e}$.}
      \label{fig:1pi_kmax_dependence}
   \end{center}
\end{figure*}

\begin{figure*}[t!]
	\centering
	\begin{center}
		\begin{tabular}{@{}cccc@{}}
			\includegraphics[width=.47\textwidth]{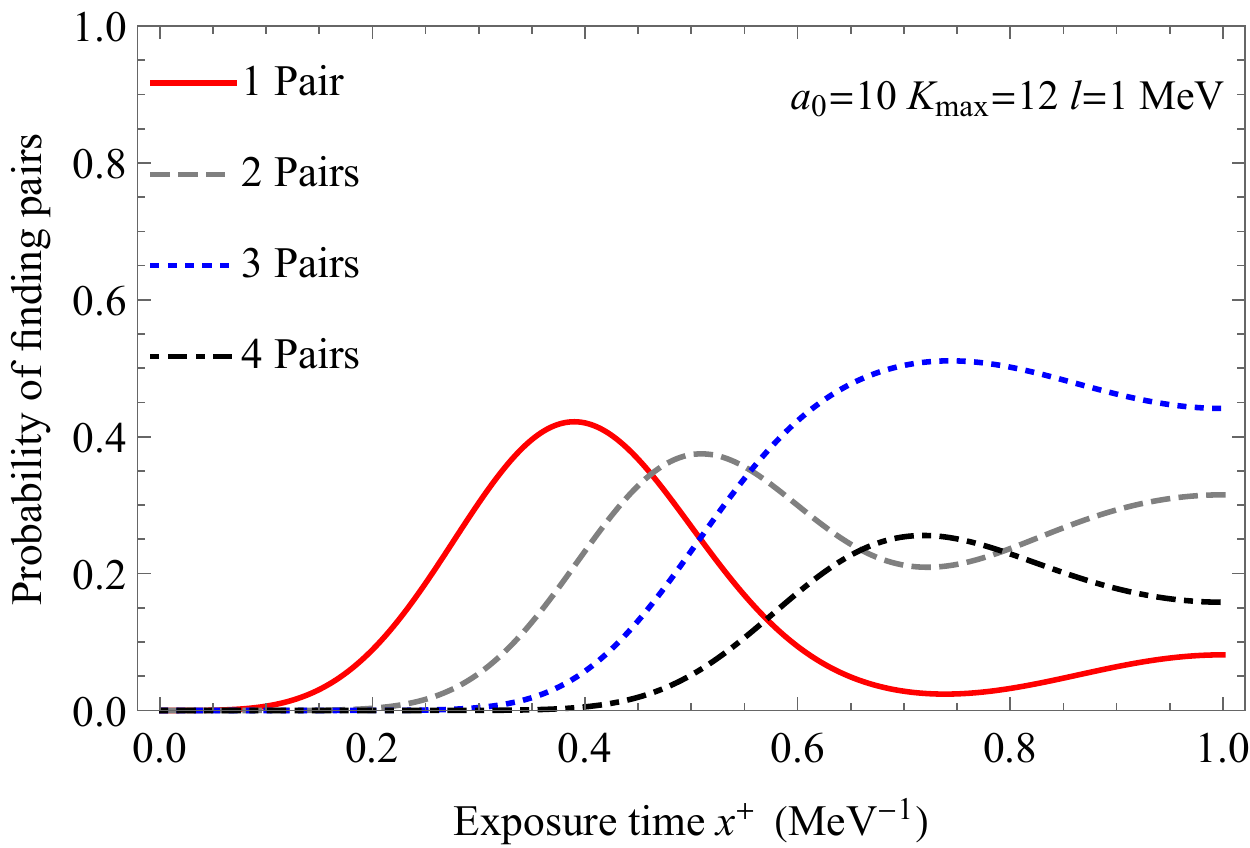} 
			\includegraphics[width=.47\textwidth]{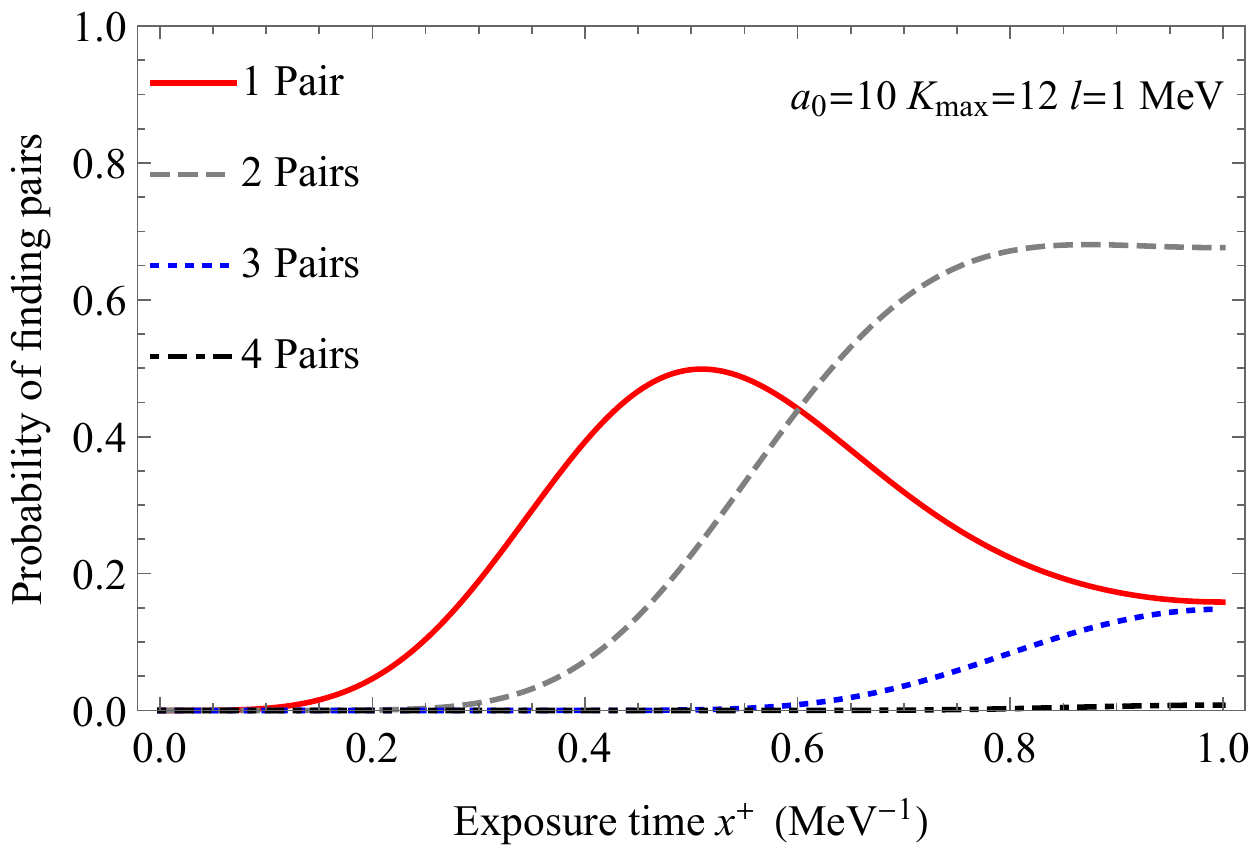} 
		\end{tabular}
                \caption{Time evolution of the probabilities of finding $n$ pairs of electrons and positrons in the background field~\eqref{BG} with $f\left( x^{\LCp} \right)=\sin\left( \omega x^{\LCp} \right), \omega=\pi\mathrm{\,MeV}$ and $a_{0}=10$. In the left panel the longitudinal momentum of the background field $l=2\rm \,MeV$ and in the right panel $l=1\rm \,MeV$. Other parameters: $K_{\max}=12$, $b=m_{e}$.}
		\label{fig:1pi_sector_l}
	\end{center}
\end{figure*}

The convergence of the probabilities of finding $n$ pairs of electrons and positrons with respect to $K_{\max}$ is presented in Fig.~\ref{fig:1pi_kmax_dependence}. Because the background field sets in more slowly, the discrepancies in the probabilities among different $K_{\max}$'s show up at a later time ($x^{\LCp}\approx0.5\rm \,MeV^{-1}$) compared with $x^{\LCp}\approx0.3\rm \,MeV^{-1}$ in Fig.~\ref{fig:const_kmax_dependence}. At around $x^{\LCp}=0.7\rm \,MeV^{-1}$ the probability of finding $4$ pairs of electrons and positrons starts to decrease, after which the results are subject to considerable truncation artifacts. In Fig.~\ref{fig:1pi_sector_l}, we show the time evolution of the probabilities of finding $n$ pairs of electrons and positrons in fields with $l=2\rm \,MeV$ (left) and $l=1\rm \,MeV$ (right). We find that the probabilities start to stabilize after the background fields reach their peaks at $x^{\LCp}=0.5\rm \,MeV^{-1}$ (see Fig.~\ref{fig:BG}). Compared with Fig.~\ref{fig:const_sector_l}, all the curves in both panels of Fig.~\ref{fig:1pi_sector_l} start off at slower rates but still reach peak values comparable with those of the constant time profile. 

\begin{figure*}[t!]
	\centering
	\begin{center}
		\begin{tabular}{@{}cccc@{}}
			\includegraphics[width=.47\textwidth]{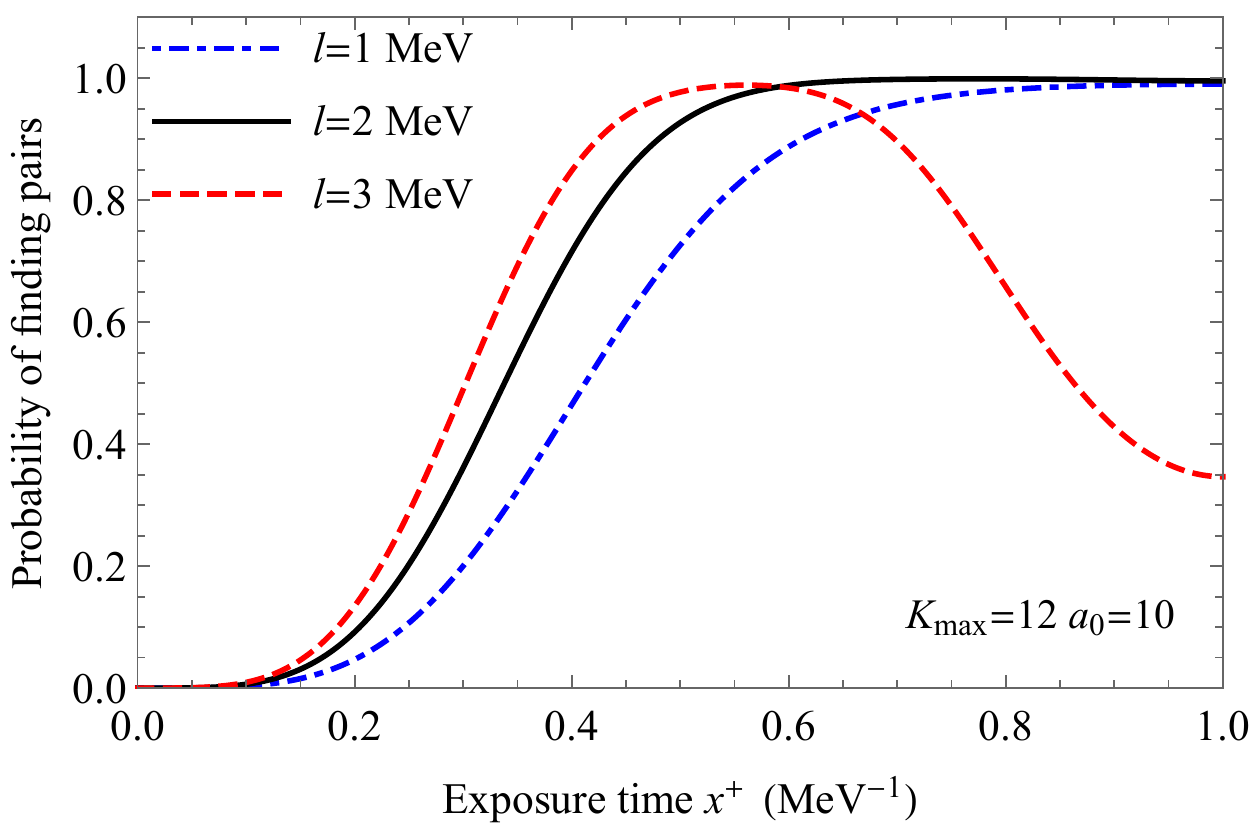}
			\includegraphics[width=.48\textwidth]{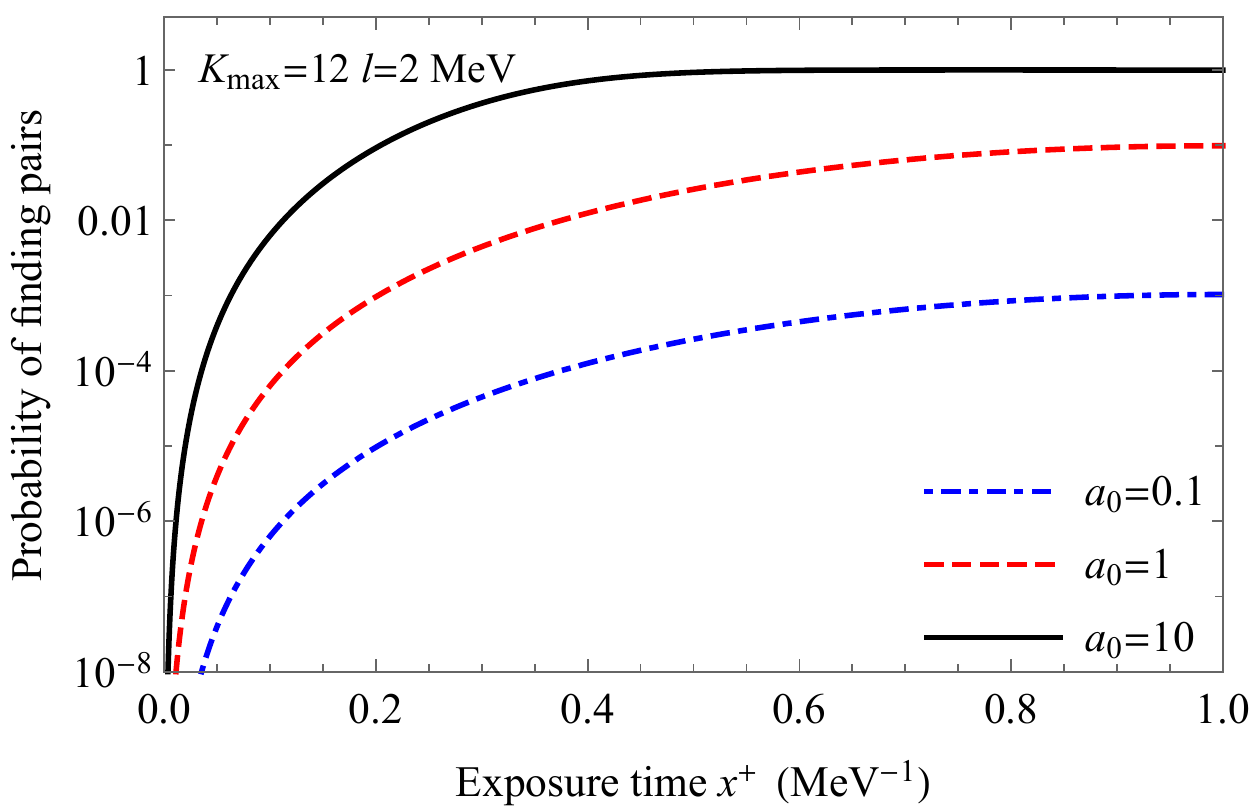}
		\end{tabular}
                \caption{Time evolution of the total probability of finding electron-positron pairs in the background field~\eqref{BG} with $f\left( x^{\LCp} \right)=\sin\left( \omega x^{\LCp} \right), \omega=\pi\mathrm{\,MeV}$. The left panel shows results obtained in fields with different longitudinal momenta $l=1\rm \,MeV$, $2\rm \,MeV$ and $3\rm \,MeV$, at the intensity $a_{0}=10$. The right panel shows results obtained in fields with different intensities $a_{0}=0.1$, $1$ and $10$, at the momentum $l=2\rm \,MeV$. Other parameters: $K_{\max}=12$, $b=m_{e}$.} 
		\label{fig:1pi_total}
	\end{center}
\end{figure*}

\begin{figure*}[t!]
	\centering
	\begin{center}
		\begin{tabular}{@{}cccc@{}}
			\includegraphics[width=.455\textwidth]{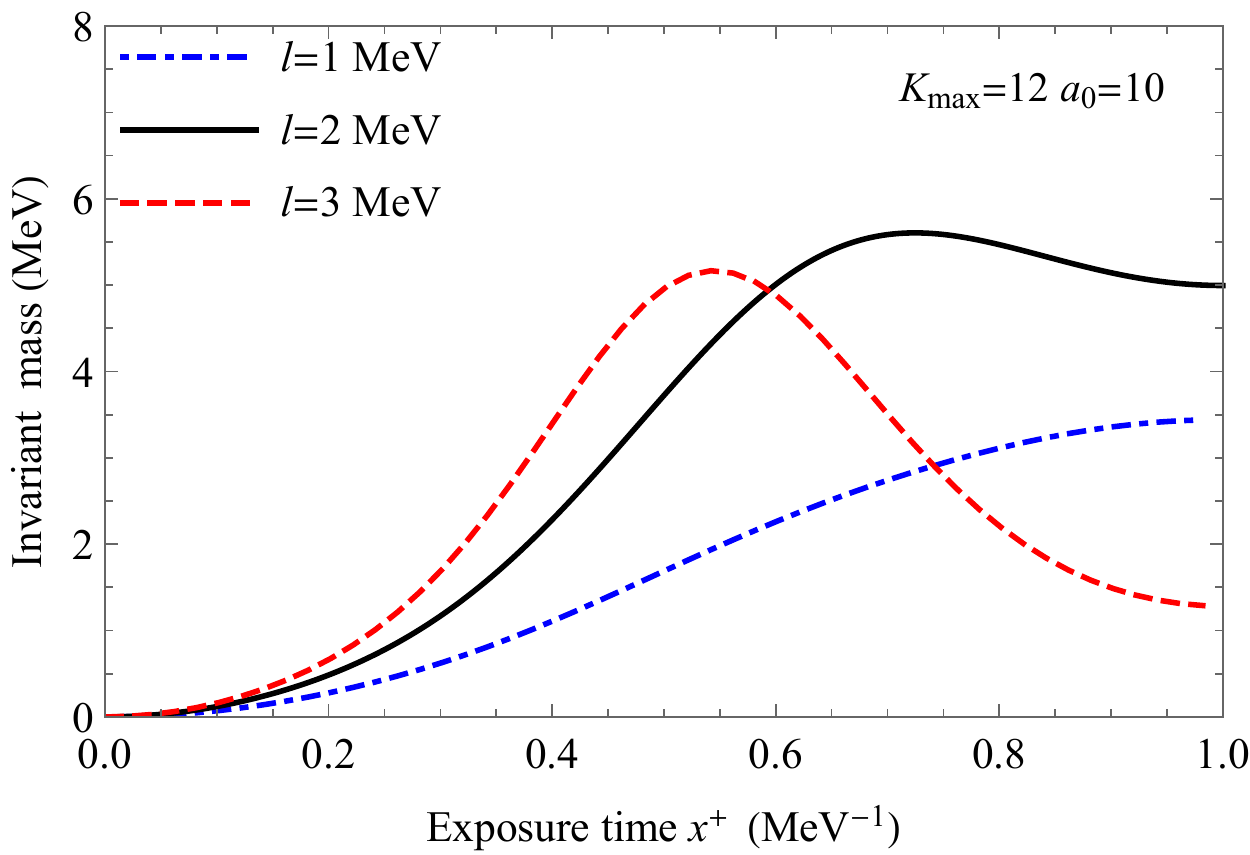}
			\includegraphics[width=.482\textwidth]{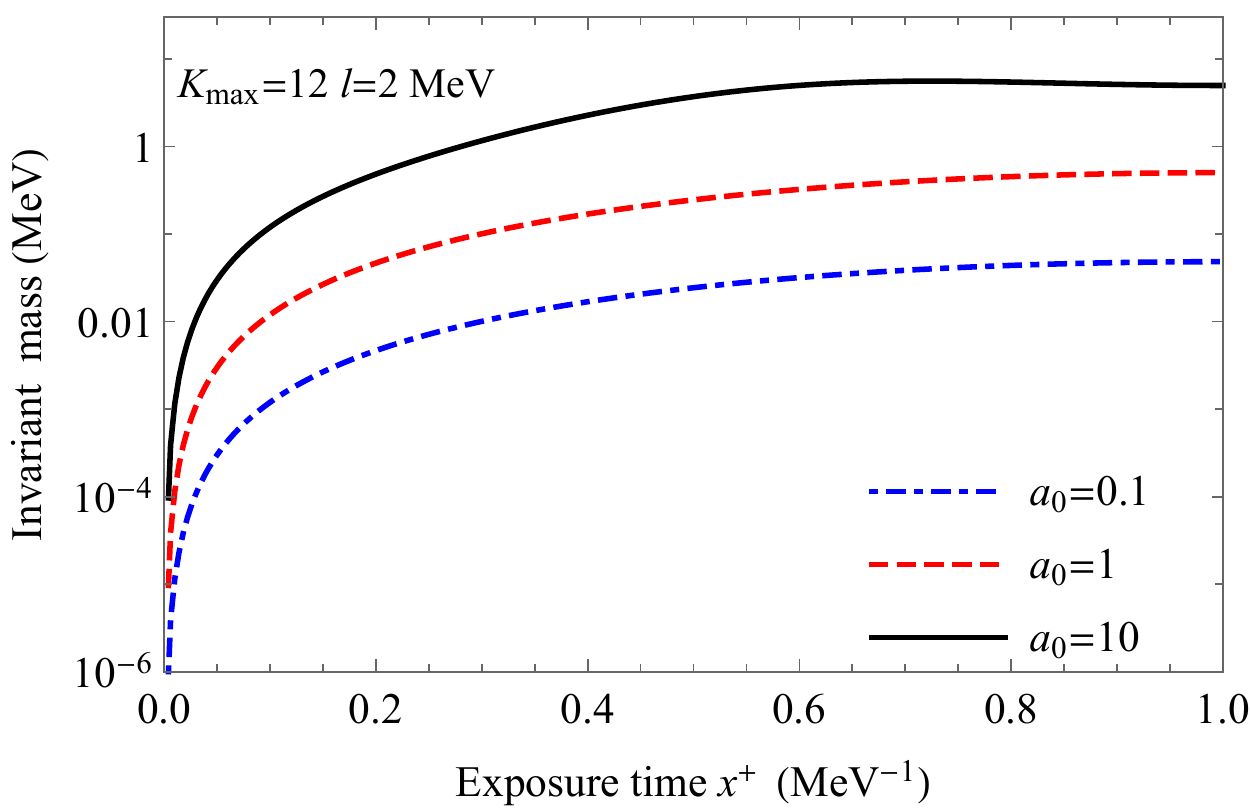}
		\end{tabular}
                \caption{Time evolution of the invariant mass of all the particles in the background fields~\eqref{BG} with $f\left( x^{\LCp} \right)=\sin\left( \omega x^{\LCp} \right), \omega=\pi\mathrm{\,MeV}$. The left panel shows results obtained in fields with different longitudinal momenta $l=1\rm \,MeV$, $2\rm \,MeV$ and $3\rm \,MeV$, at the intensity $a_{0}=10$. The right panel shows results obtained in fields with different intensities $a_{0}=0.1$, $1$ and $10$, at the momentum $l=2\rm \,MeV$. Other parameters: $K_{\max}=12$, $b=m_{e}$.} 
		\label{fig:1pi_invmass}
	\end{center}
\end{figure*}

\begin{figure*}[t!]
	\centering
	\begin{center}
		\begin{tabular}{@{}cccc@{}}
			\includegraphics[width=.46\textwidth]{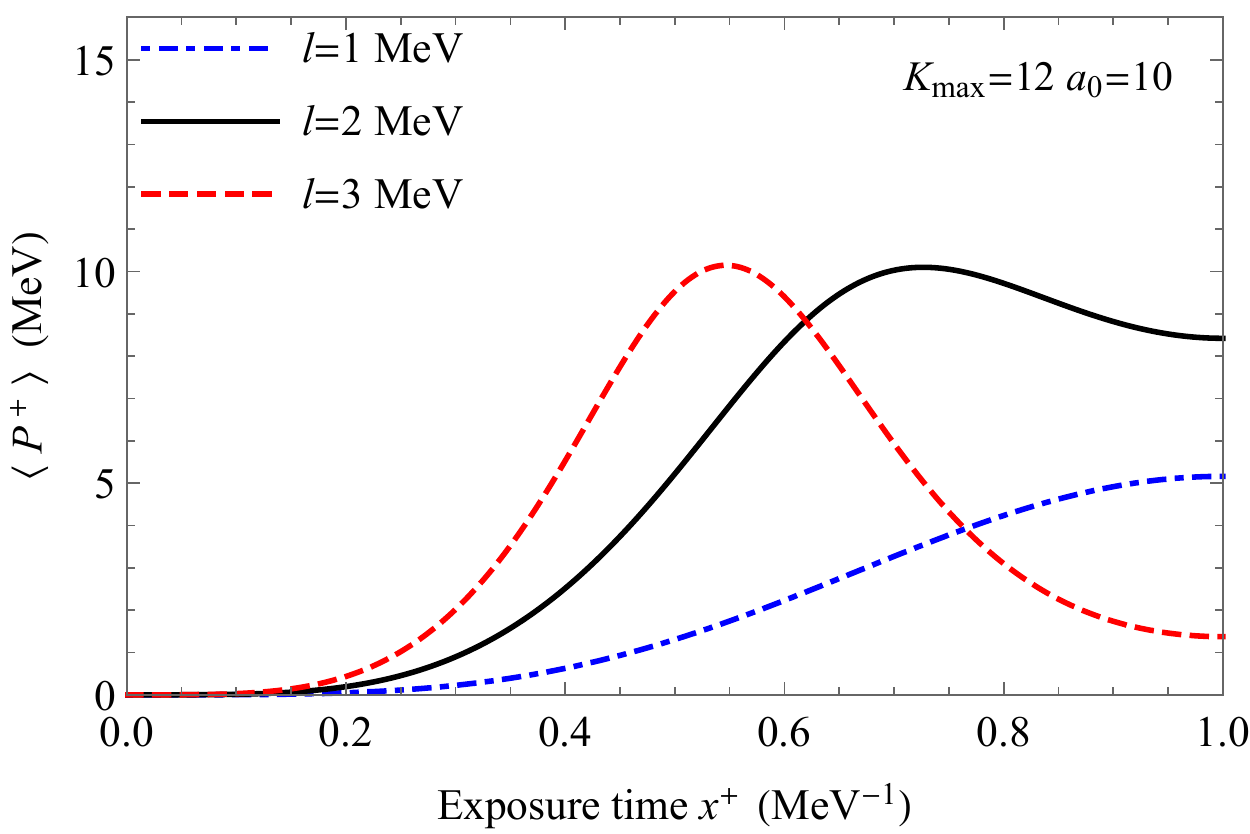}
			\includegraphics[width=.48\textwidth]{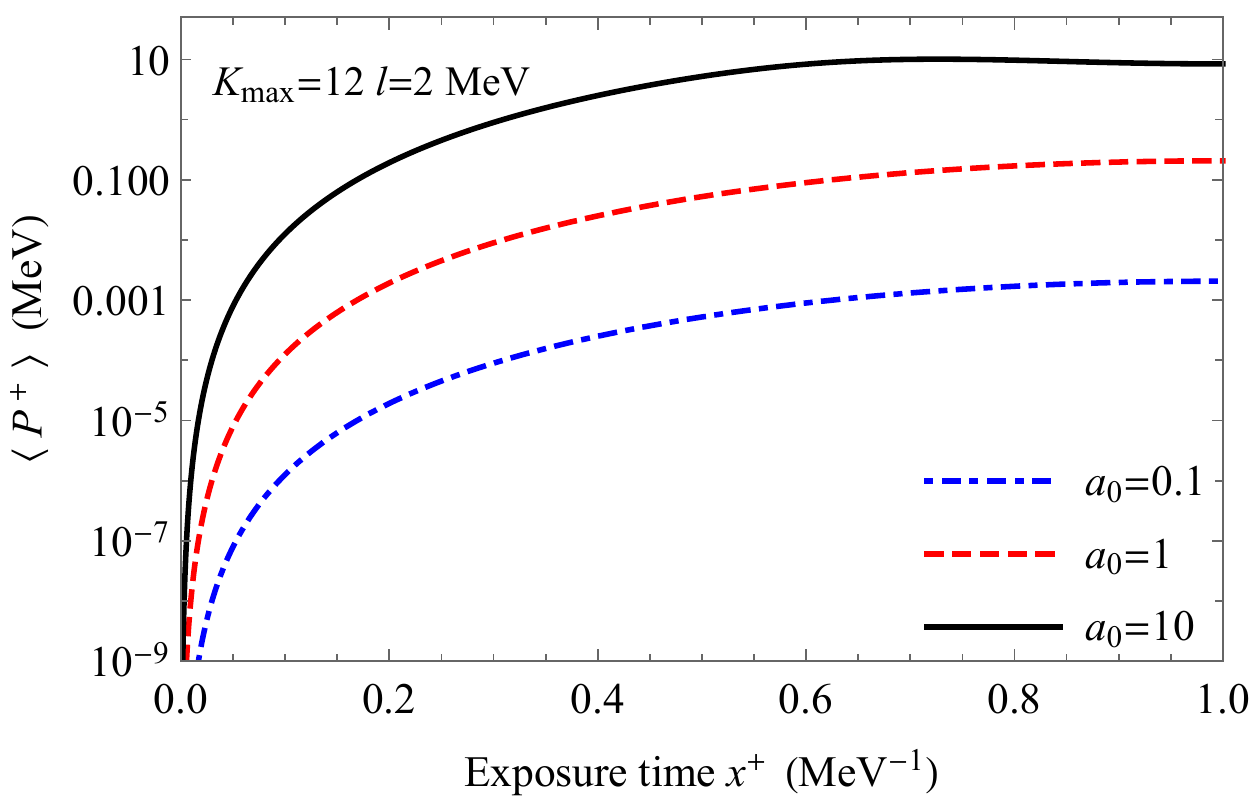}
		\end{tabular}
                \caption{Time evolution of the expectation value of the total momentum of all the particles in the background field~\eqref{BG} with $f\left( x^{\LCp} \right)=\sin\left( \omega x^{\LCp} \right), \omega=\pi\mathrm{\,MeV}$. The left panel shows results obtained in fields with different longitudinal momenta $l=1\rm \,MeV$, $2\rm \,MeV$ and $3\rm \,MeV$, at the intensity $a_{0}=10$. The right panel shows results obtained in fields with different intensities $a_{0}=0.1$, $1$ and $10$, at the momentum $l=2\rm \,MeV$. Other parameters: $K_{\max}=12$, $b=m_{e}$.}
		\label{fig:1pi_total_momentum}
	\end{center}
\end{figure*}

\begin{figure*}[t!]
	\centering
	\begin{center}
		\begin{tabular}{@{}cccc@{}}
			\includegraphics[width=.47\textwidth]{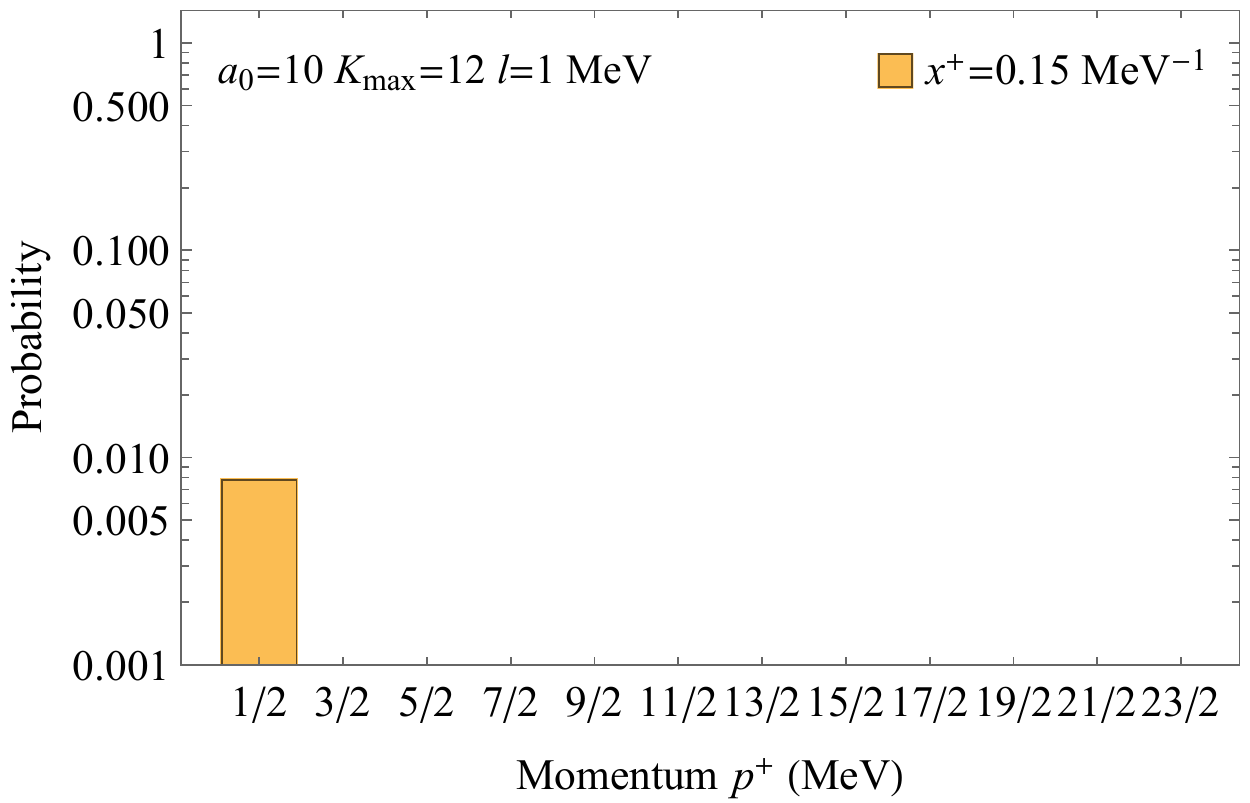} 
			\includegraphics[width=.47\textwidth]{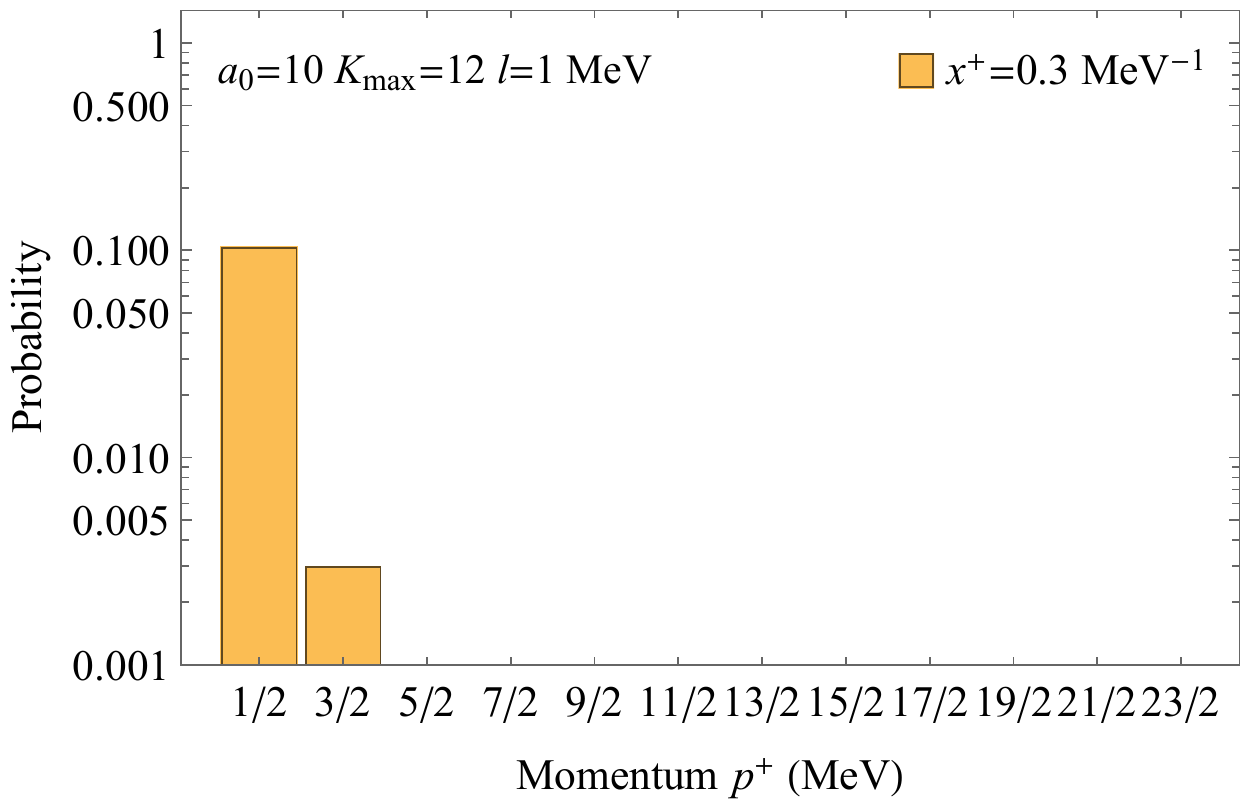}\\
			\includegraphics[width=.47\textwidth]{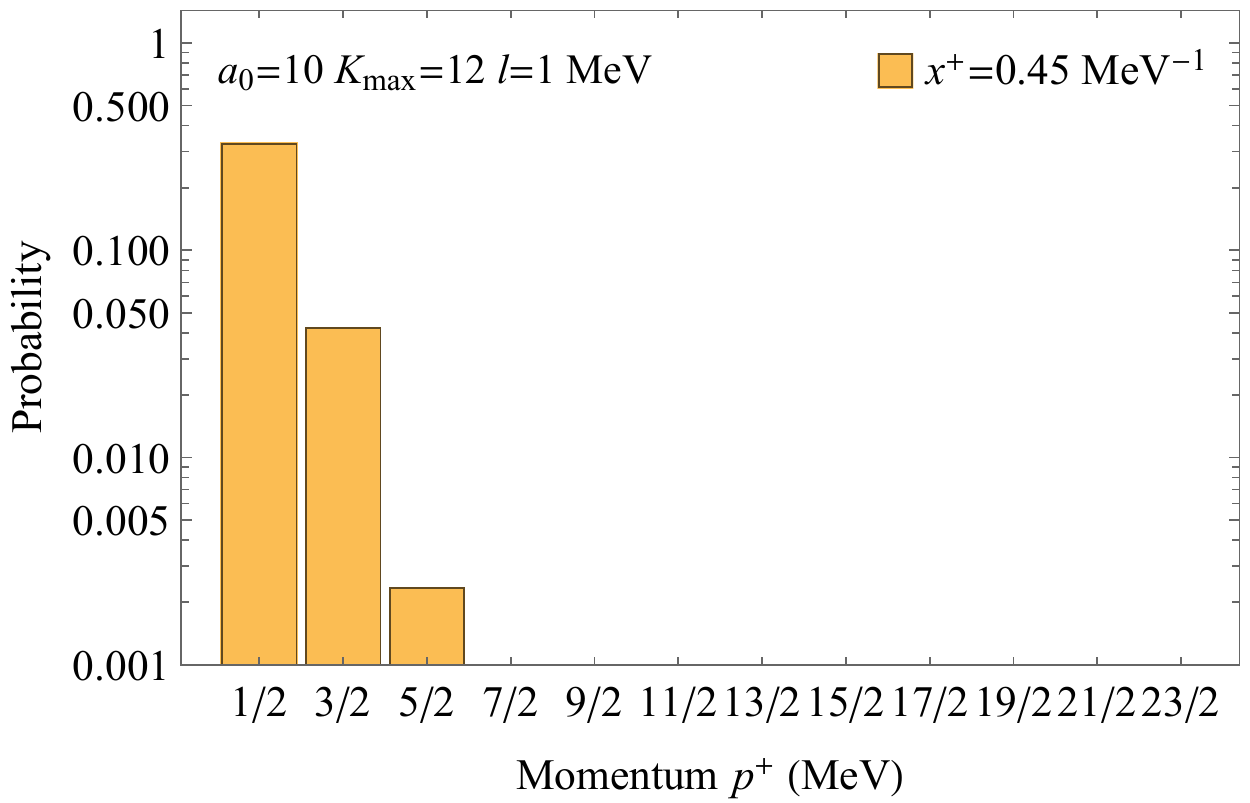} 
			\includegraphics[width=.47\textwidth]{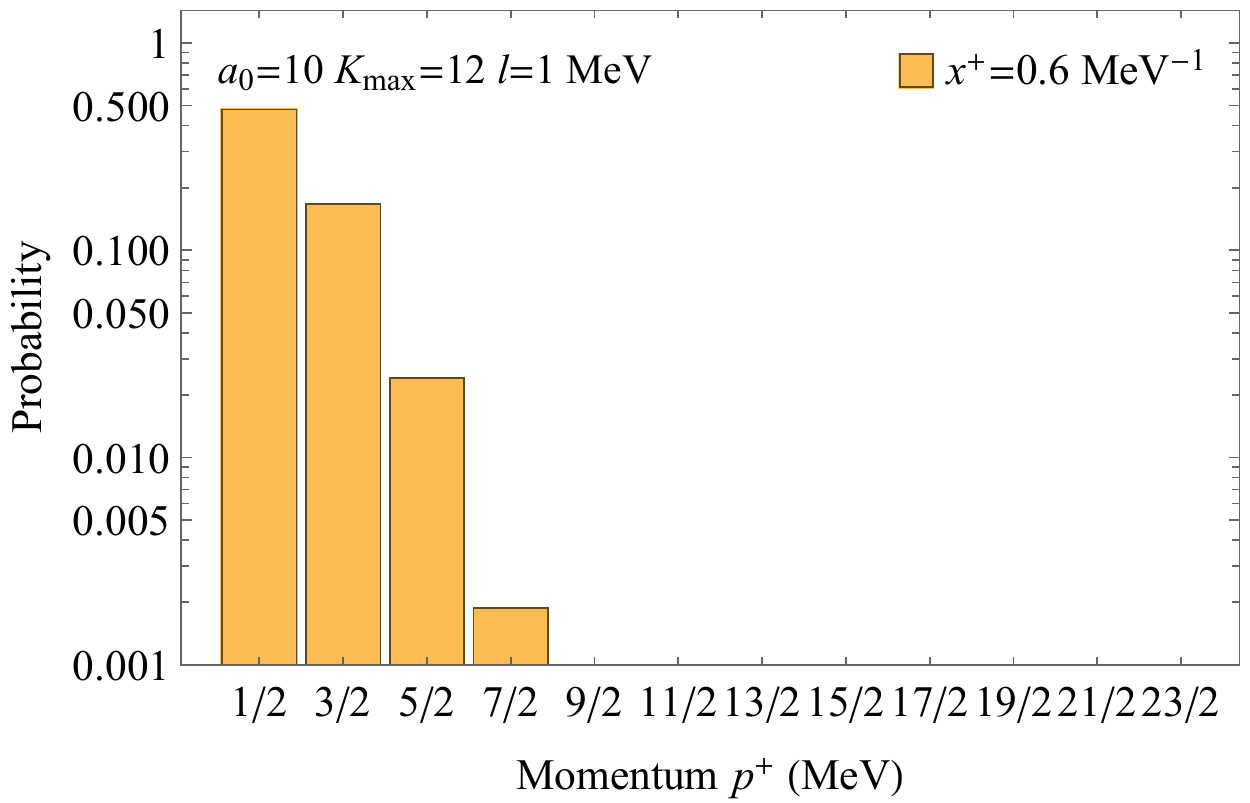} 
		\end{tabular}
                \caption{Time evolution of the longitudinal momentum distribution of the electron in the background field~\eqref{BG} with $f\left( x^{\LCp} \right)=\sin\left( \omega x^{\LCp} \right), \omega=\pi\mathrm{\,MeV}$, $a_{0}=10$ and $l=1\rm \,MeV$. Other parameters: $K_{\max}=12$, $b=m_{e}$.}
		\label{fig:1pi_md_l1}
	\end{center}
\end{figure*}

\begin{figure*}[t!]
	\centering
	\begin{center}
		\begin{tabular}{@{}cccc@{}}
			\includegraphics[width=.47\textwidth]{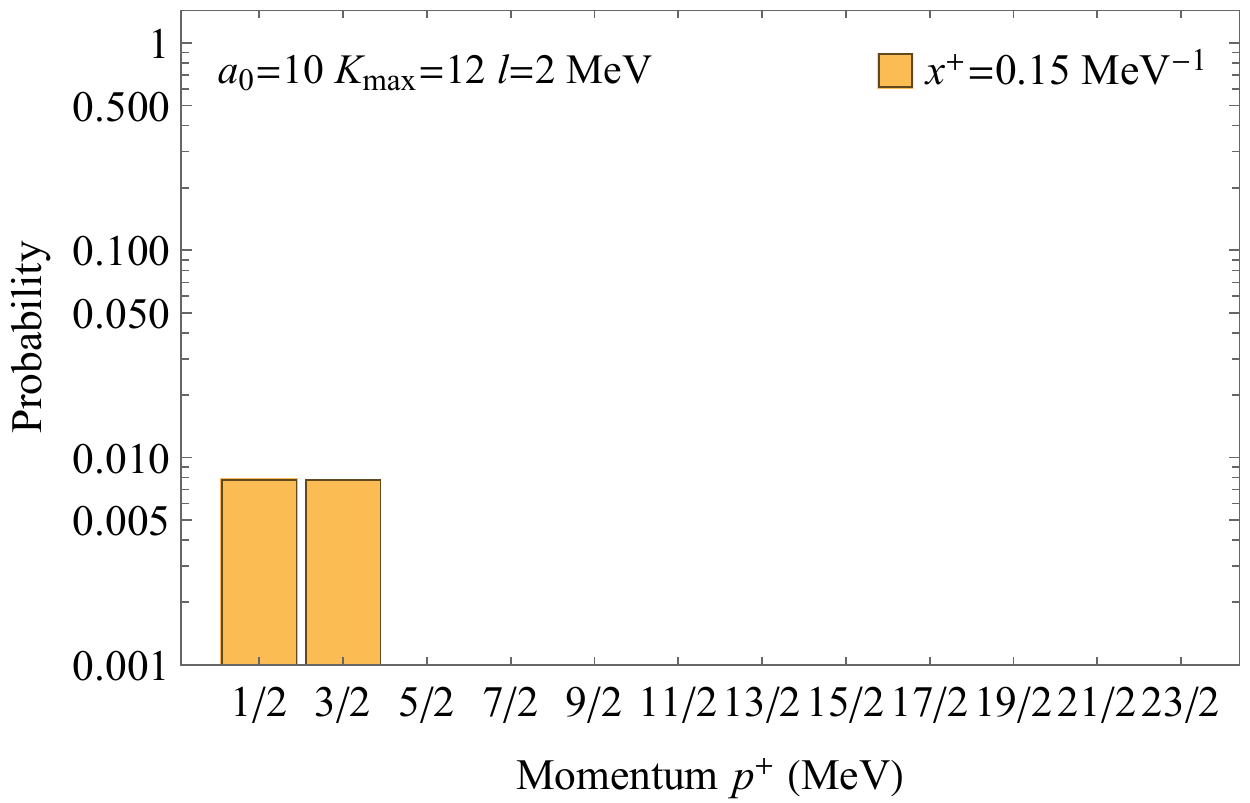} 
			\includegraphics[width=.47\textwidth]{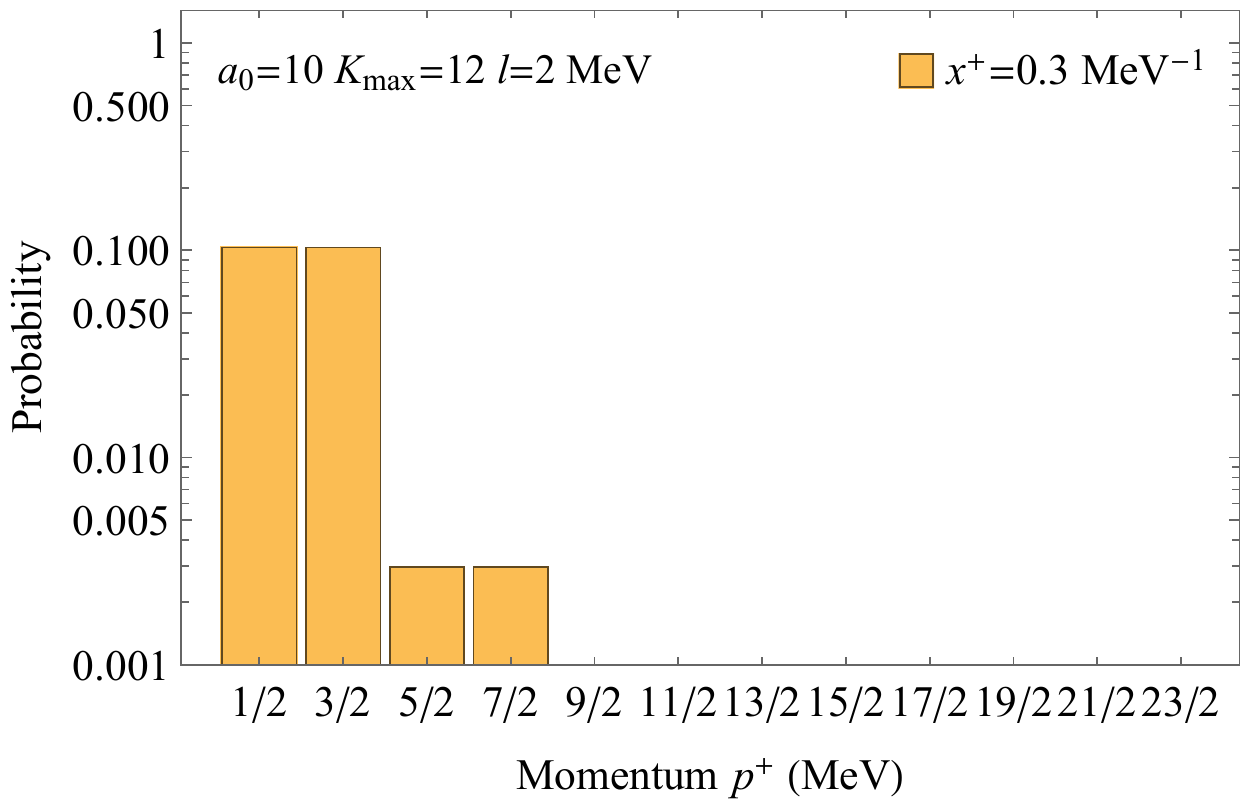}\\
			\includegraphics[width=.47\textwidth]{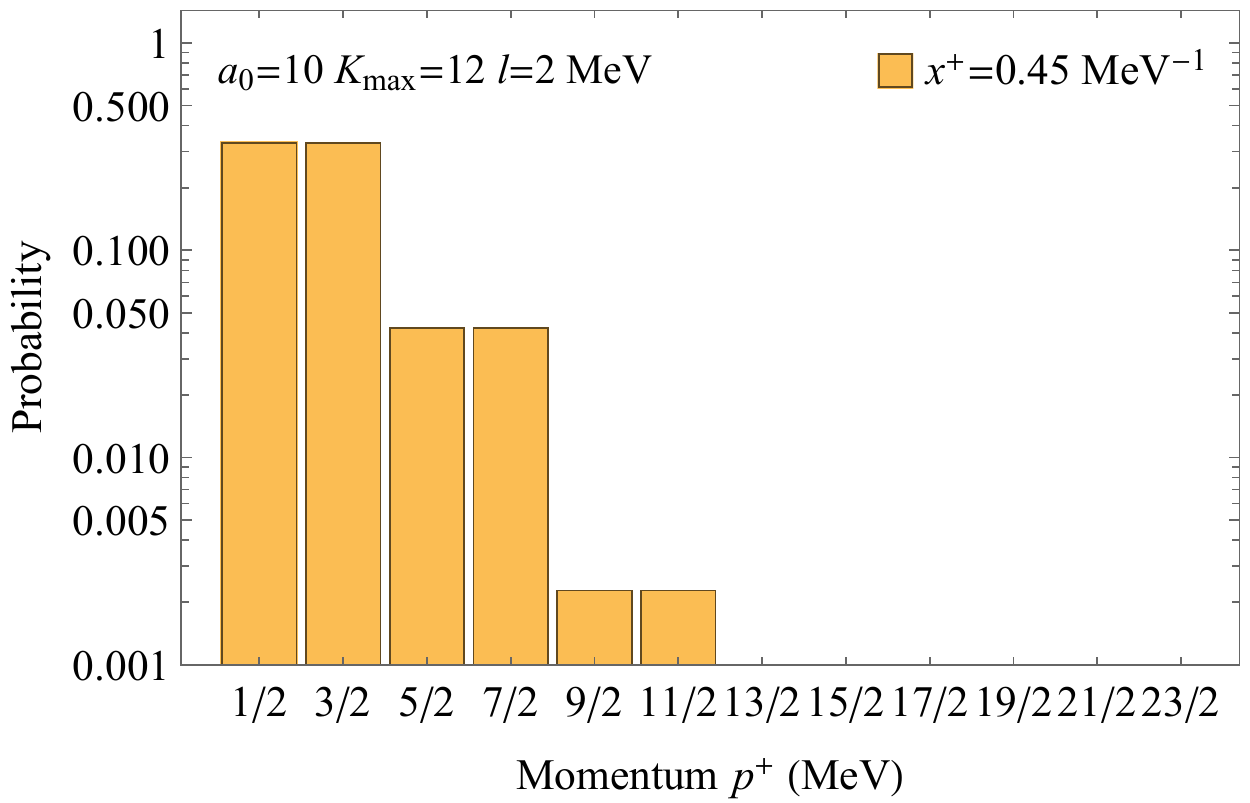} 
			\includegraphics[width=.47\textwidth]{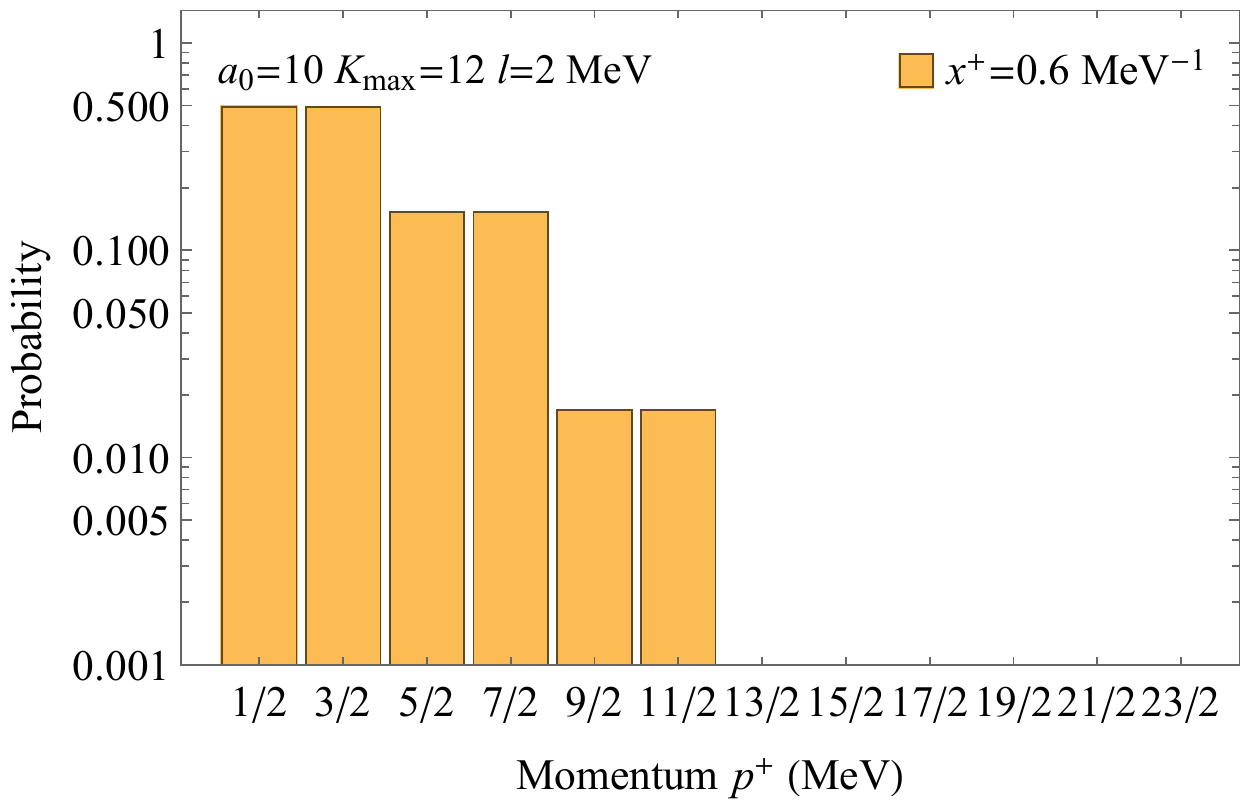} 
		\end{tabular}
                \caption{Time evolution of the longitudinal momentum distribution of the electron in the background field~\eqref{BG} with $f\left( x^{\LCp} \right)=\sin\left( \omega x^{\LCp} \right), \omega=\pi\mathrm{\,MeV}$, $a_{0}=10$ and $l=2\rm \,MeV$. Other parameters: $K_{\max}=12$, $b=m_{e}$.}
		\label{fig:1pi_md_l2}
	\end{center}
\end{figure*}

\begin{figure*}[t!]
	\centering
	\begin{center}
		\begin{tabular}{@{}cccc@{}}
			\includegraphics[width=.47\textwidth]{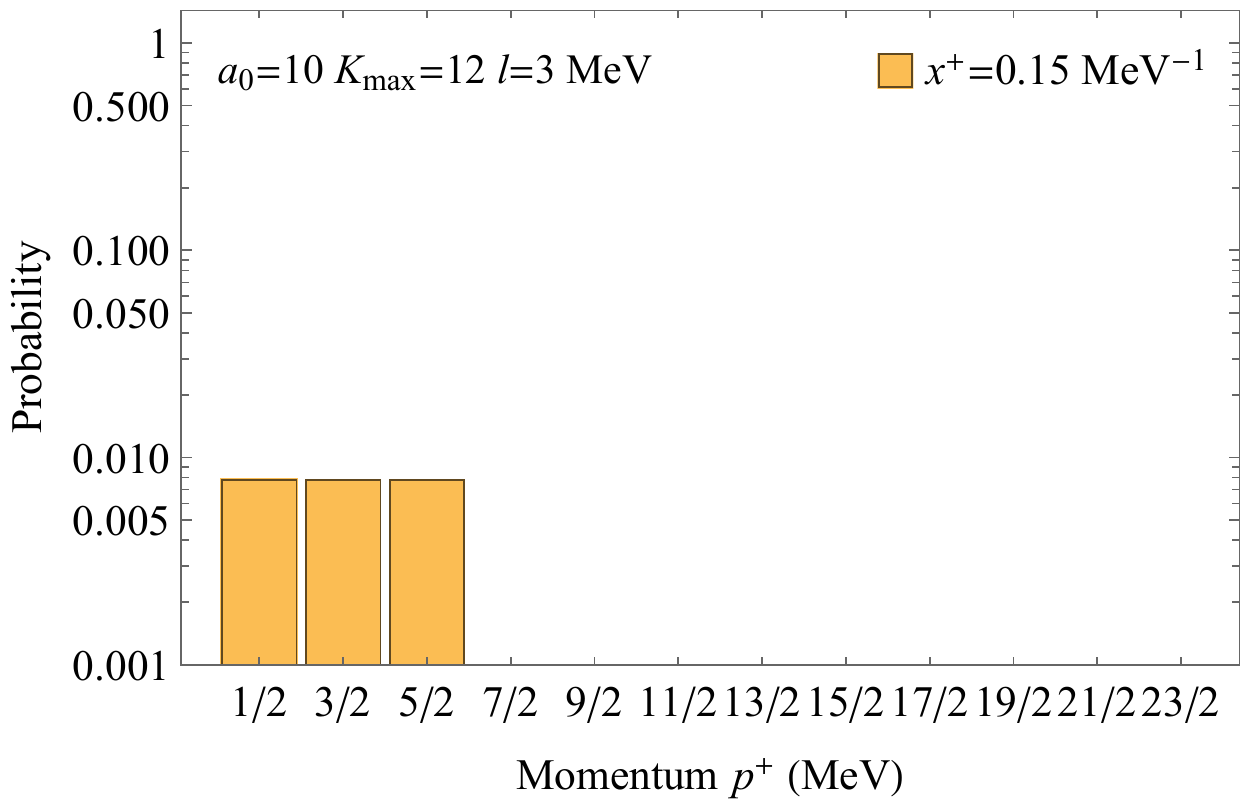} 
			\includegraphics[width=.47\textwidth]{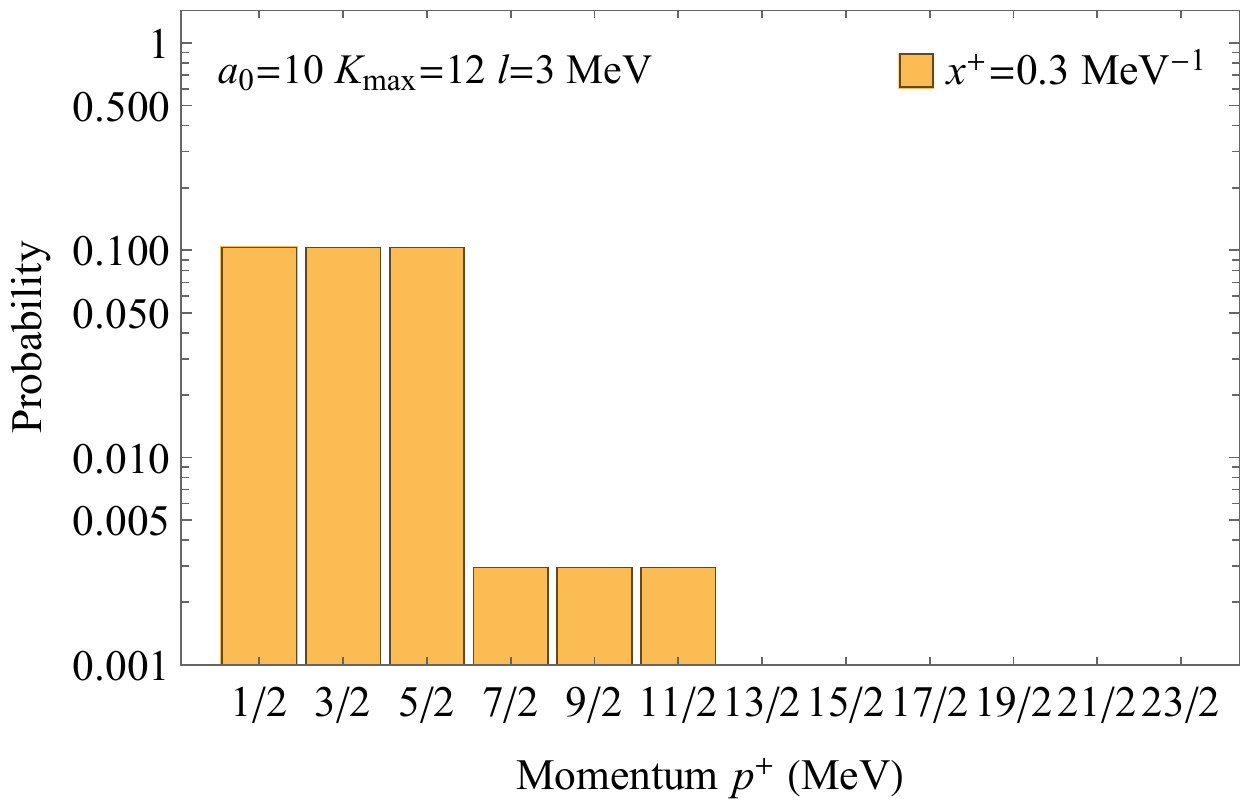} \\
			\includegraphics[width=.47\textwidth]{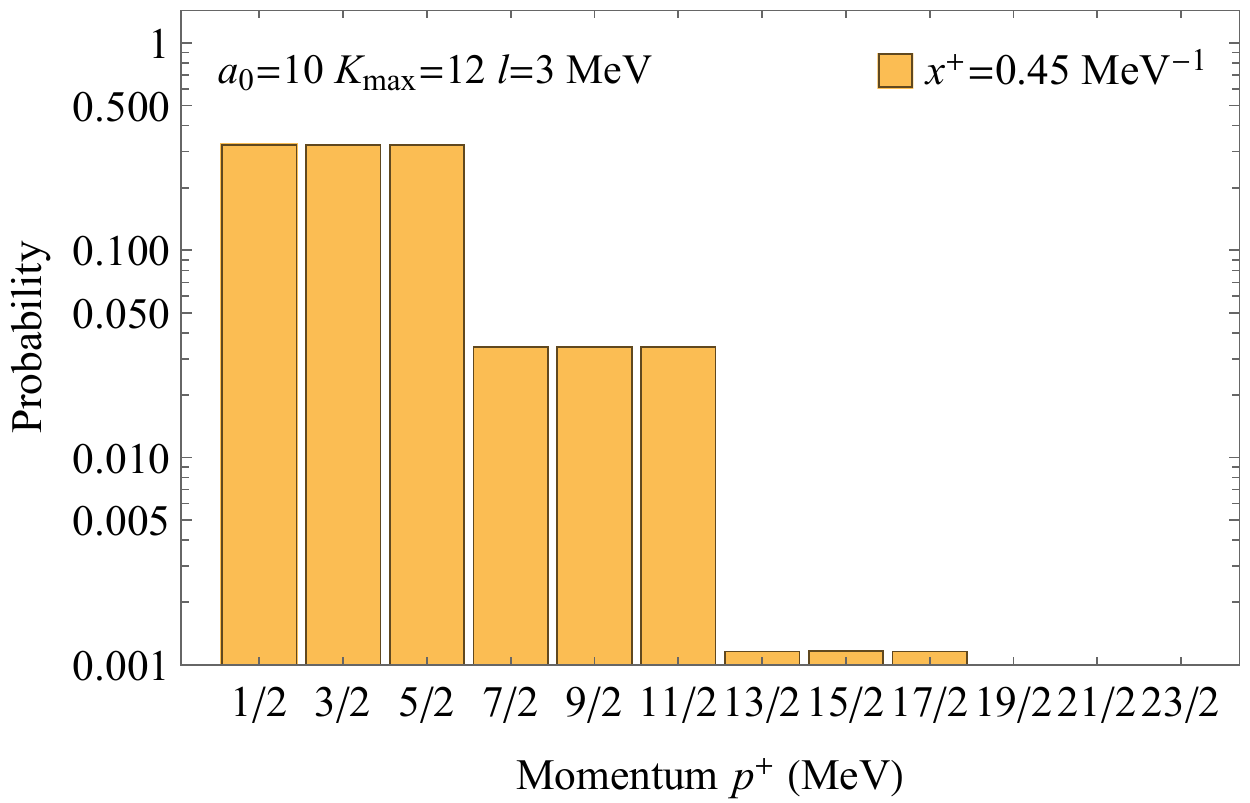} 
			\includegraphics[width=.47\textwidth]{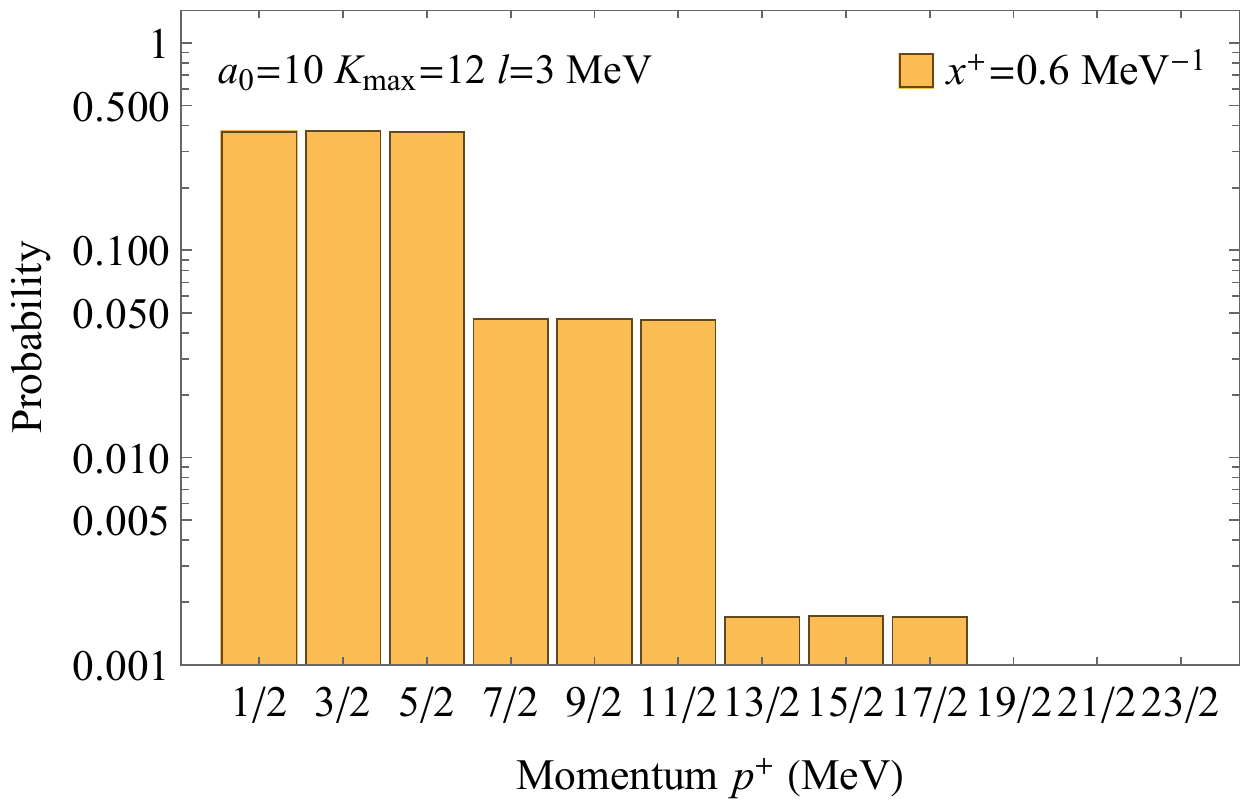} 
		\end{tabular}
                \caption{Time evolution of the longitudinal momentum distribution of the electron in the background field~\eqref{BG} with $f\left( x^{\LCp} \right)=\sin\left( \omega x^{\LCp} \right), \omega=\pi\mathrm{\,MeV}$, $a_{0}=10$ and $l=3\rm \,MeV$. Other parameters: $K_{\max}=12$, $b=m_{e}$.}
		\label{fig:1pi_md_l3}
	\end{center}
\end{figure*}

The total probability of finding pairs in Fig.~\ref{fig:1pi_total}, the invariant mass in Fig.~\ref{fig:1pi_invmass} and the total momentum in Fig.~\ref{fig:1pi_total_momentum} all show similar trends to their counterparts in Sec.~\ref{1}, with the main differences again being that the slopes of the curves are smaller in both the initial and the final stage of the evolution when the background strengths are small. We show the momentum distributions of the system in fields with $3$ different momenta $l=1\rm \,MeV$, $2\rm \,MeV$ and $3\rm \,MeV$ in Figs.~\ref{fig:1pi_md_l1},~\ref{fig:1pi_md_l2} and~\ref{fig:1pi_md_l3}, respectively. We observe that the probabilities in the panels with $x^{\LCp}=0.15\rm \,MeV^{-1}$ in these $3$ figures are all largely reduced compared to the constant time profile case, since the fields have not yet built up their strengths at this moment. Nevertheless, after the background fields reach their peak values at $x^{\LCp}=0.5\rm \,MeV^{-1}$, the longitudinal momentum distributions with both time profiles are comparable.
\subsection{Background fields with time profiles $f\left( x^{\LCp} \right)=\mathrm{sin}\left(\omega x^{\LCp}\right), \omega=2\pi\mathrm{\,MeV}, 3\pi\mathrm{\,MeV}$, and $5\pi\mathrm{\,MeV}$}\label{other}
\begin{figure*}[t!]
	\centering
	\begin{center}
		\begin{tabular}{@{}cccc@{}}
			\includegraphics[width=.47\textwidth]{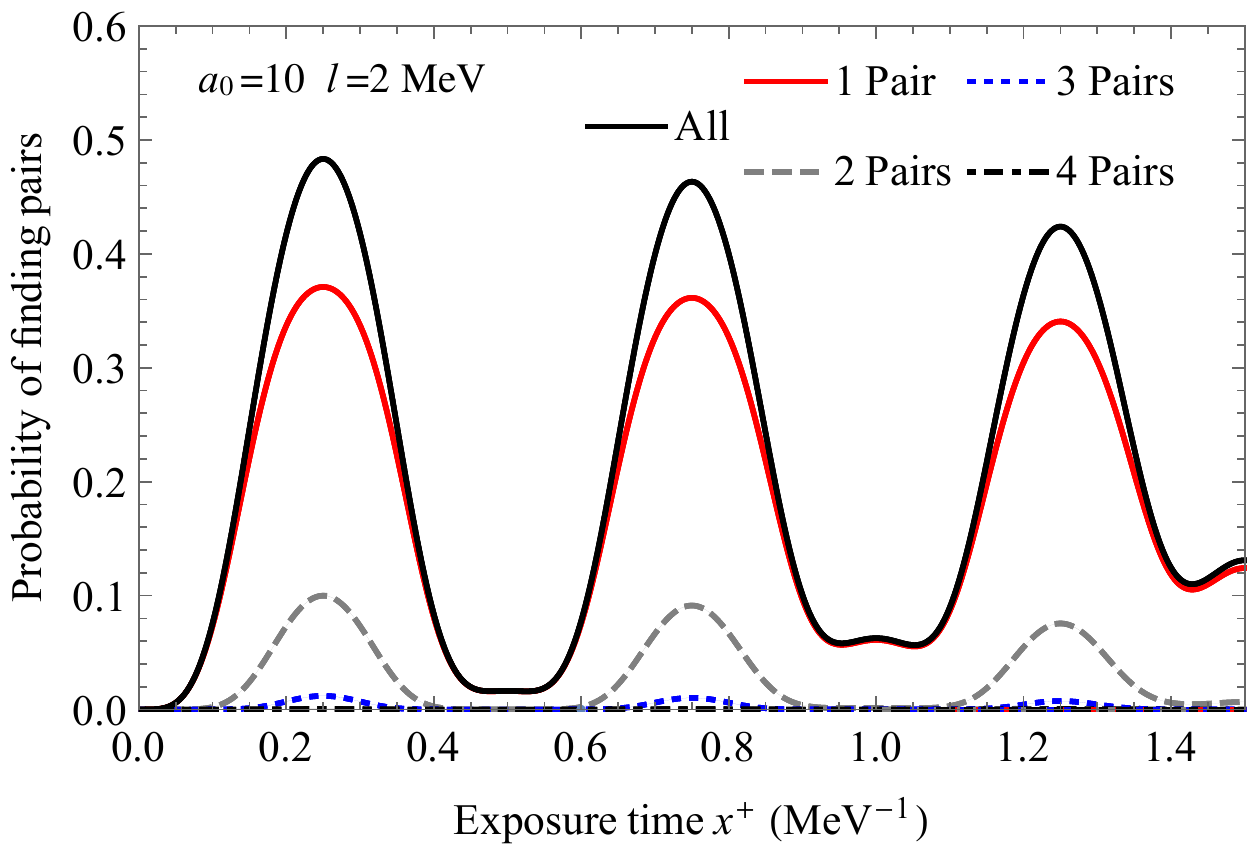}
			\includegraphics[width=.47\textwidth]{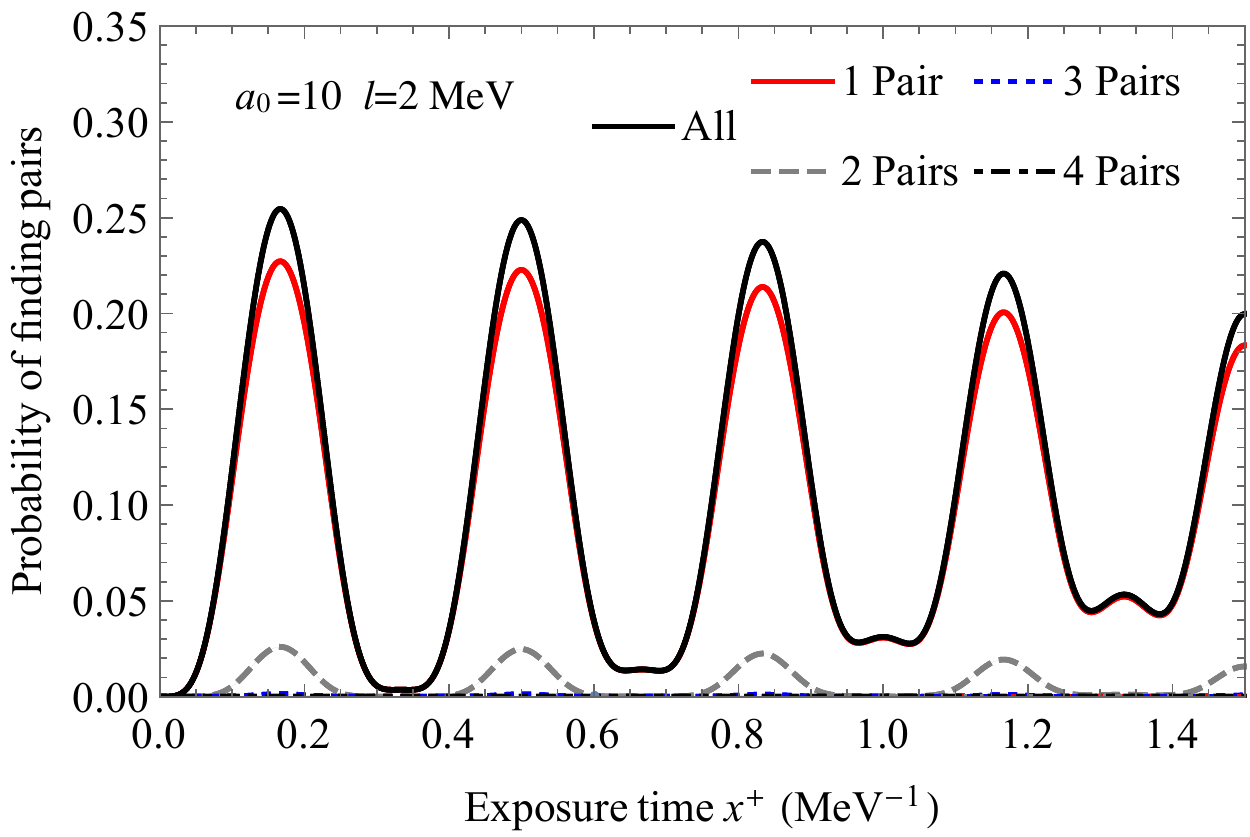}
		\end{tabular}
		\caption{Time evolution of the probabilities of finding $n$ pairs of electrons and positrons (and their total) in the background field~\eqref{BG}. In the left panel the time profile $f\left( x^{\LCp} \right)=\sin\left( \omega x^{\LCp} \right), \omega=2\pi\mathrm{\,MeV}$ and in the right panel $\omega=3\pi\mathrm{\,MeV}$. In both panels, results obtained in bases with $K_{\max}=8$, $10$ and $12$ are presented in the same plot style, which roughly coincide with each other. 
                   Other parameters: $a_{0}=10$, $l=2\rm \,MeV$, $b=m_{e}$.
			}
		\label{fig:other_kmax_dependence}
	\end{center}
\end{figure*}

\begin{figure*}[t!]
	\centering
	\begin{center}
		\begin{tabular}{@{}cccc@{}}
			\includegraphics[width=.47\textwidth]{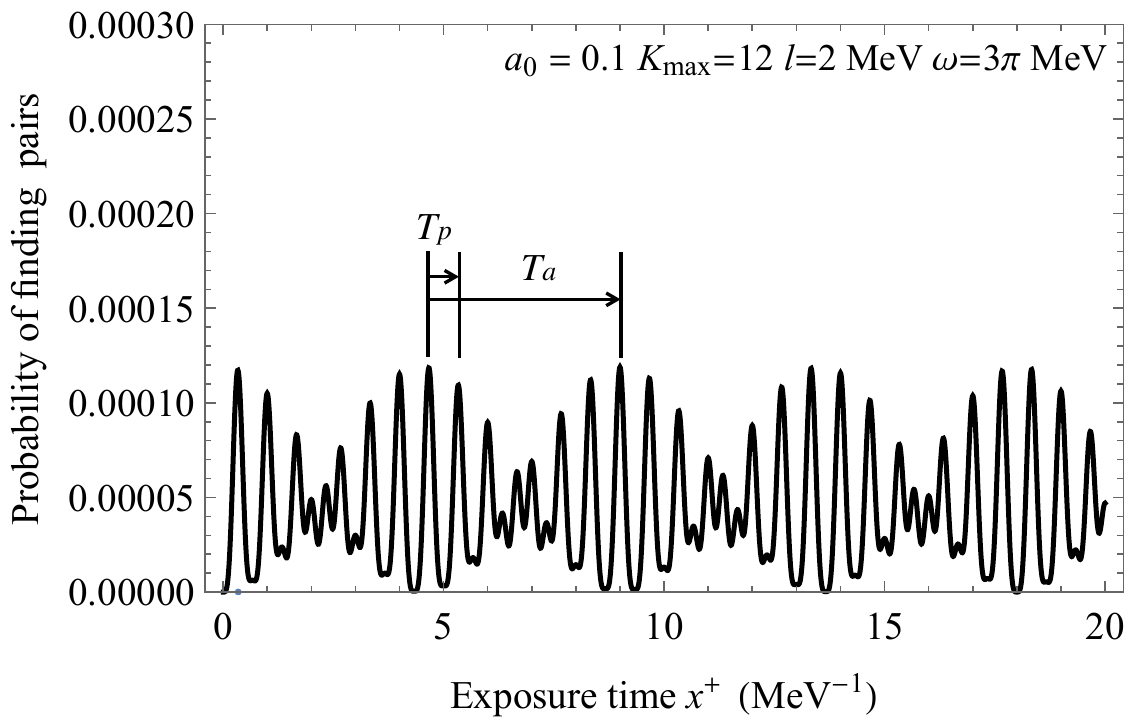}
			\includegraphics[width=.47\textwidth]{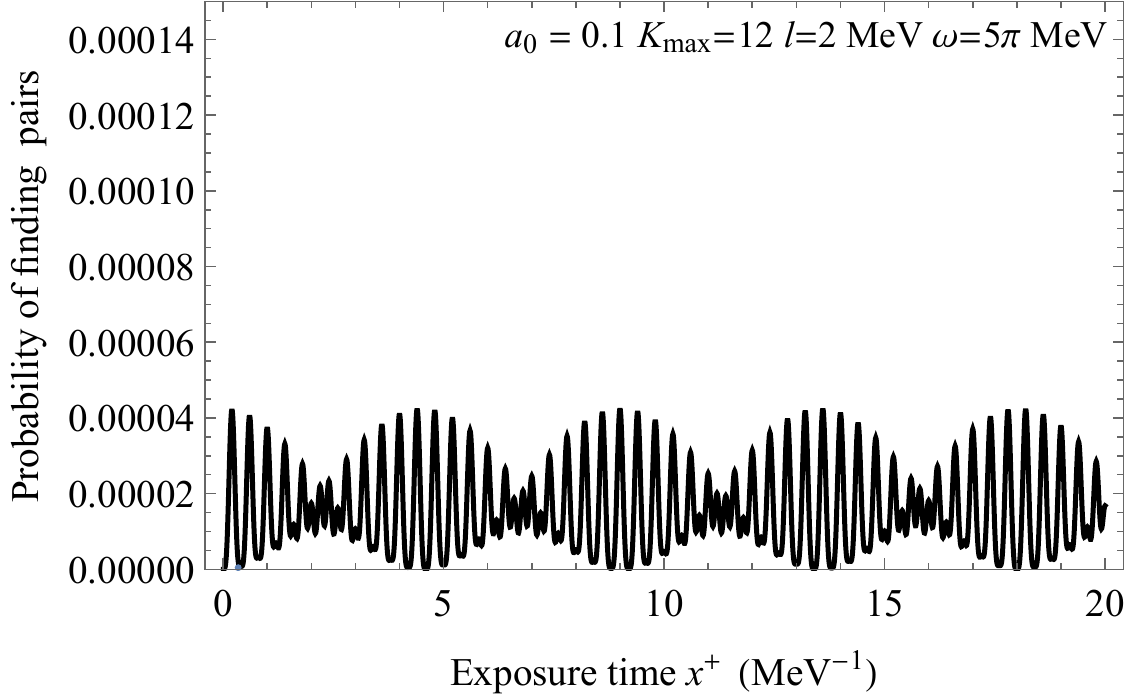}\\
			\includegraphics[width=.47\textwidth]{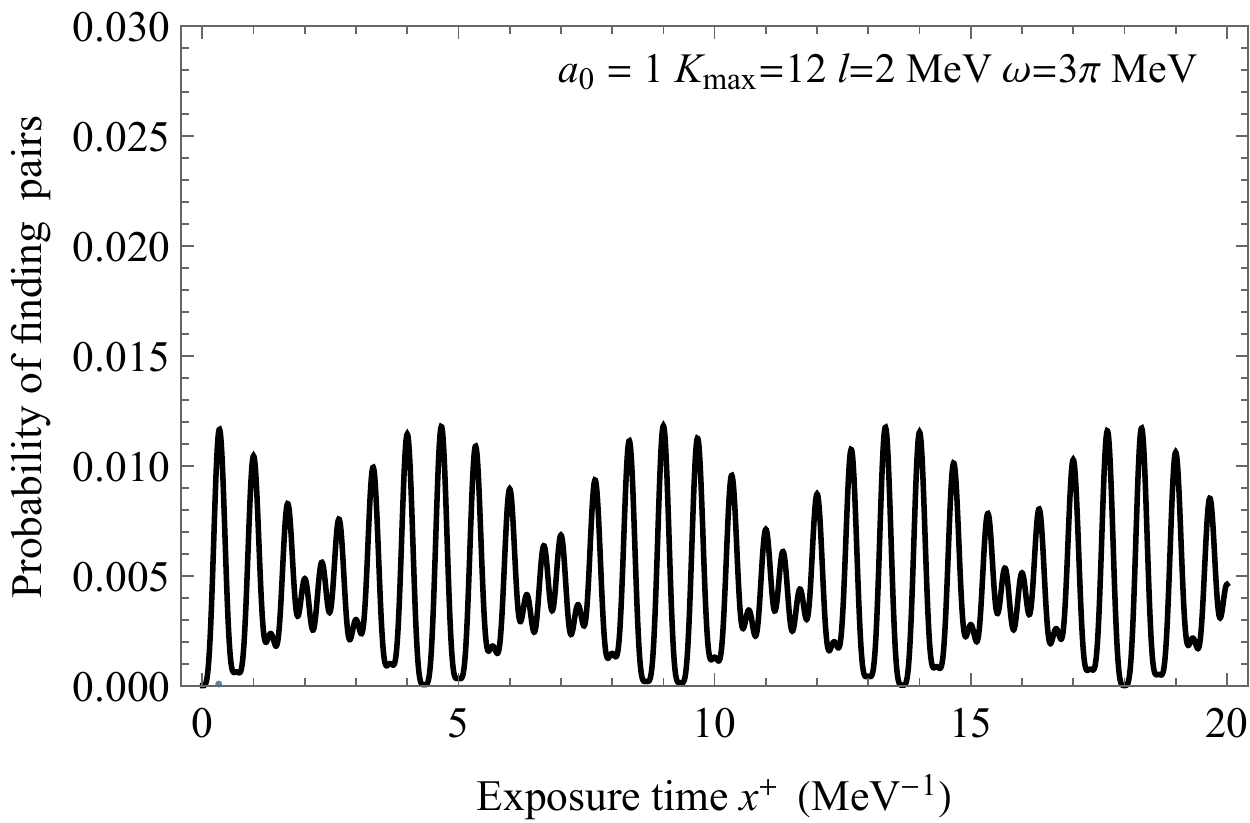}
			\includegraphics[width=.47\textwidth]{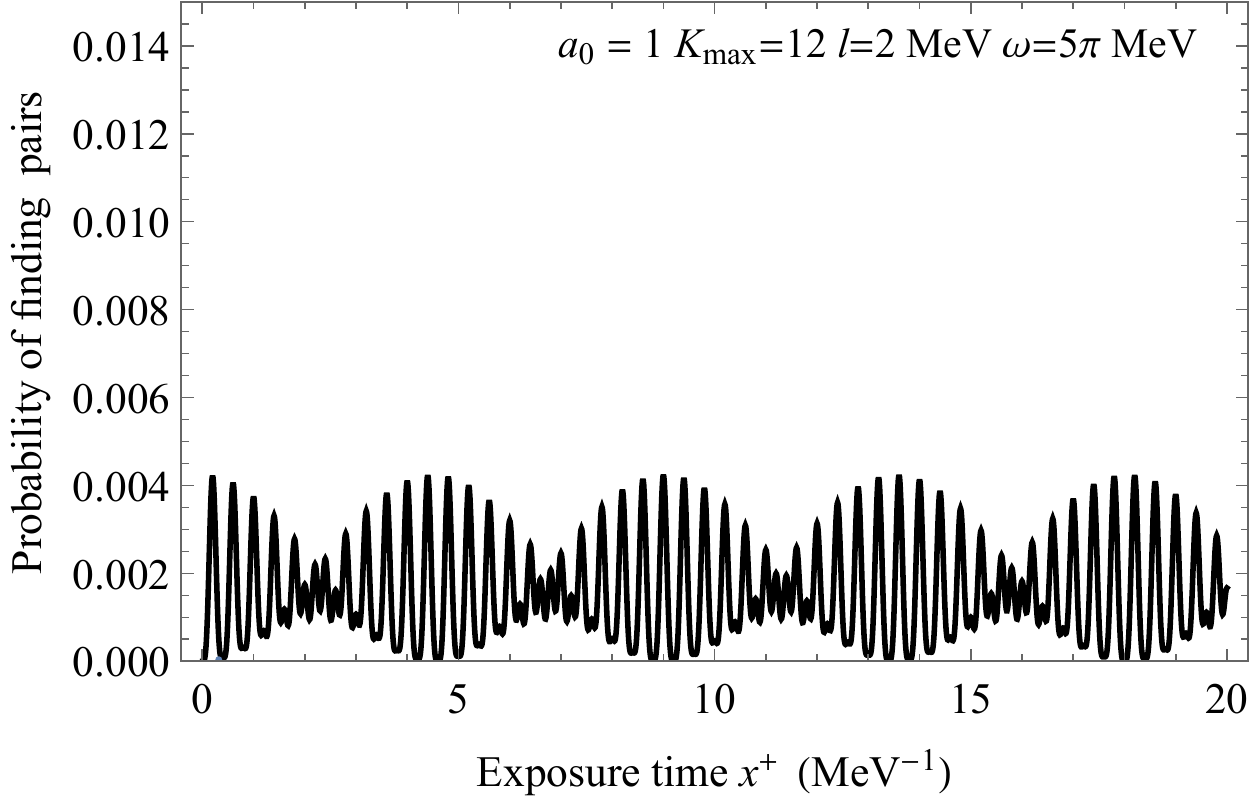}\\
			\includegraphics[width=.47\textwidth]{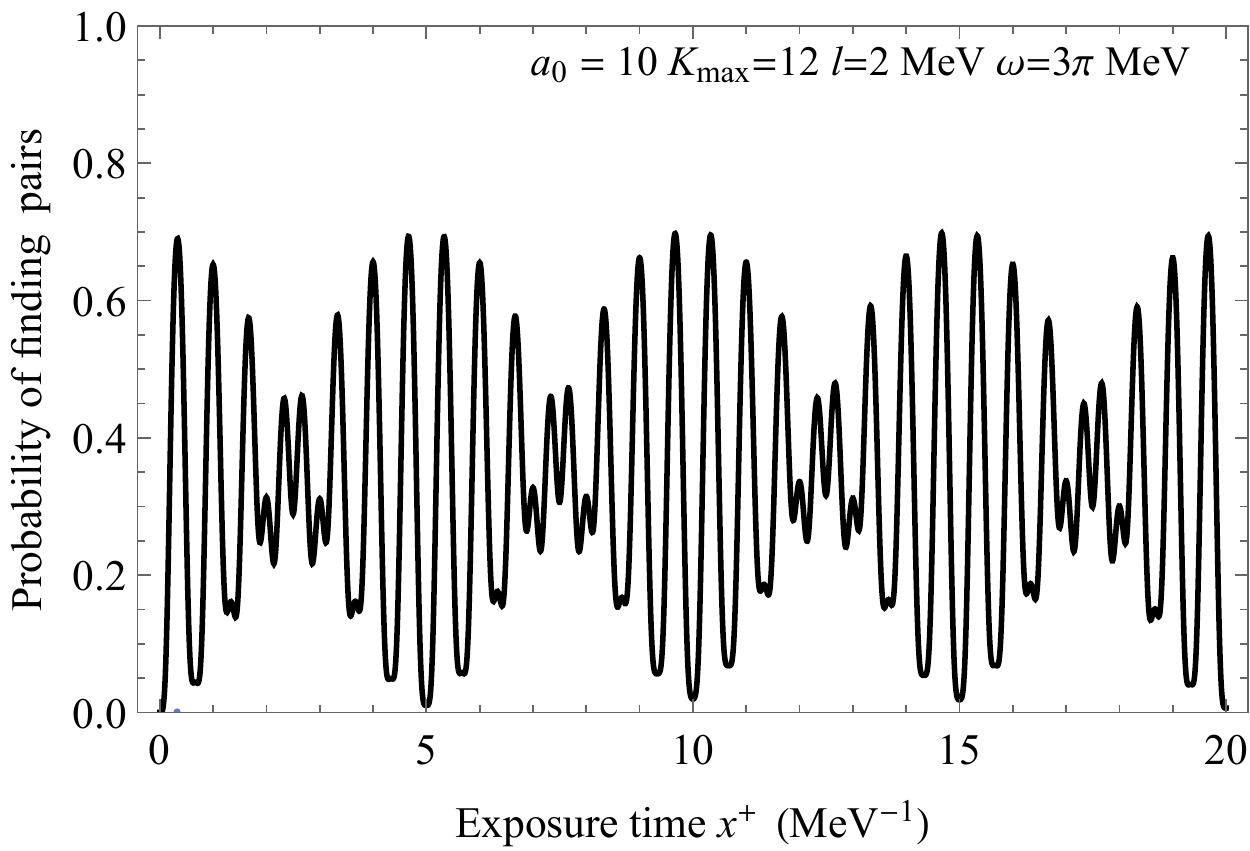}
			\includegraphics[width=.47\textwidth]{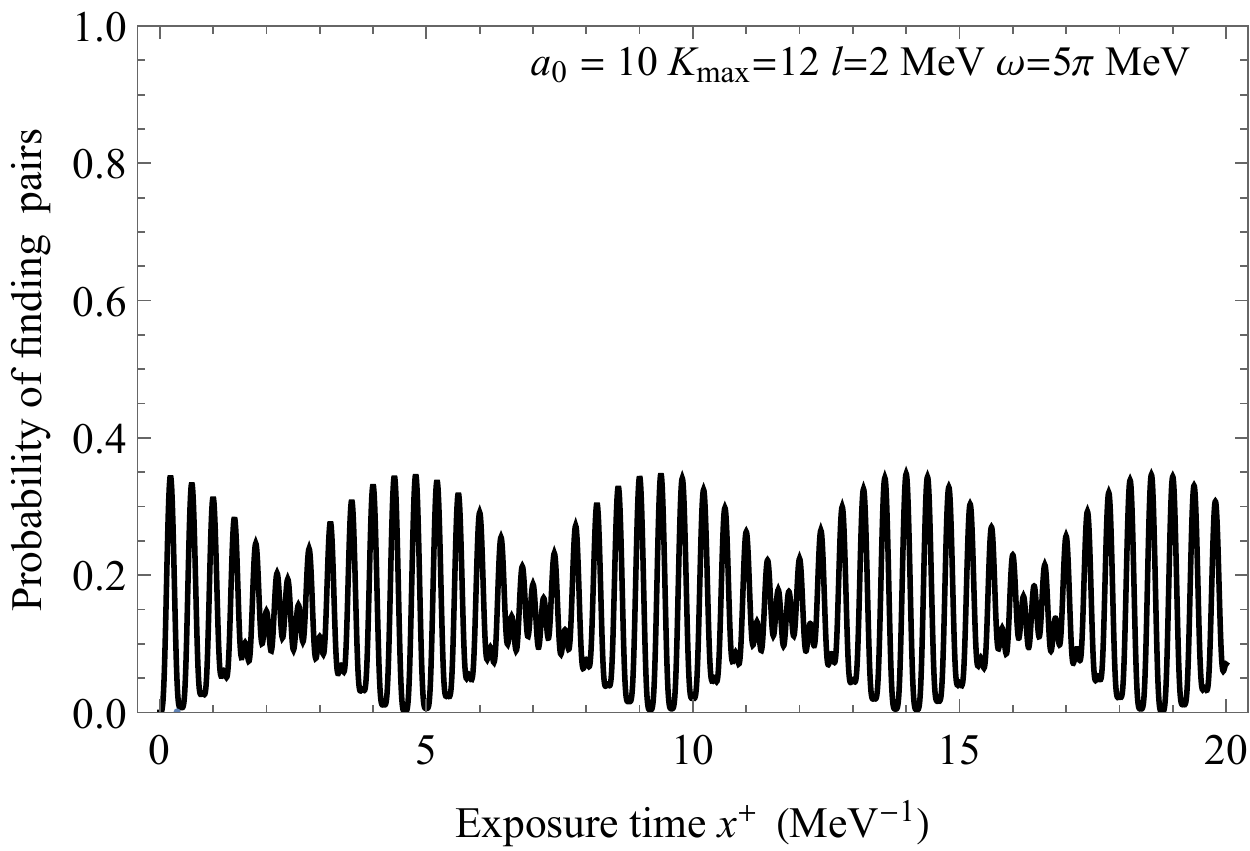}\\
			\includegraphics[width=.47\textwidth]{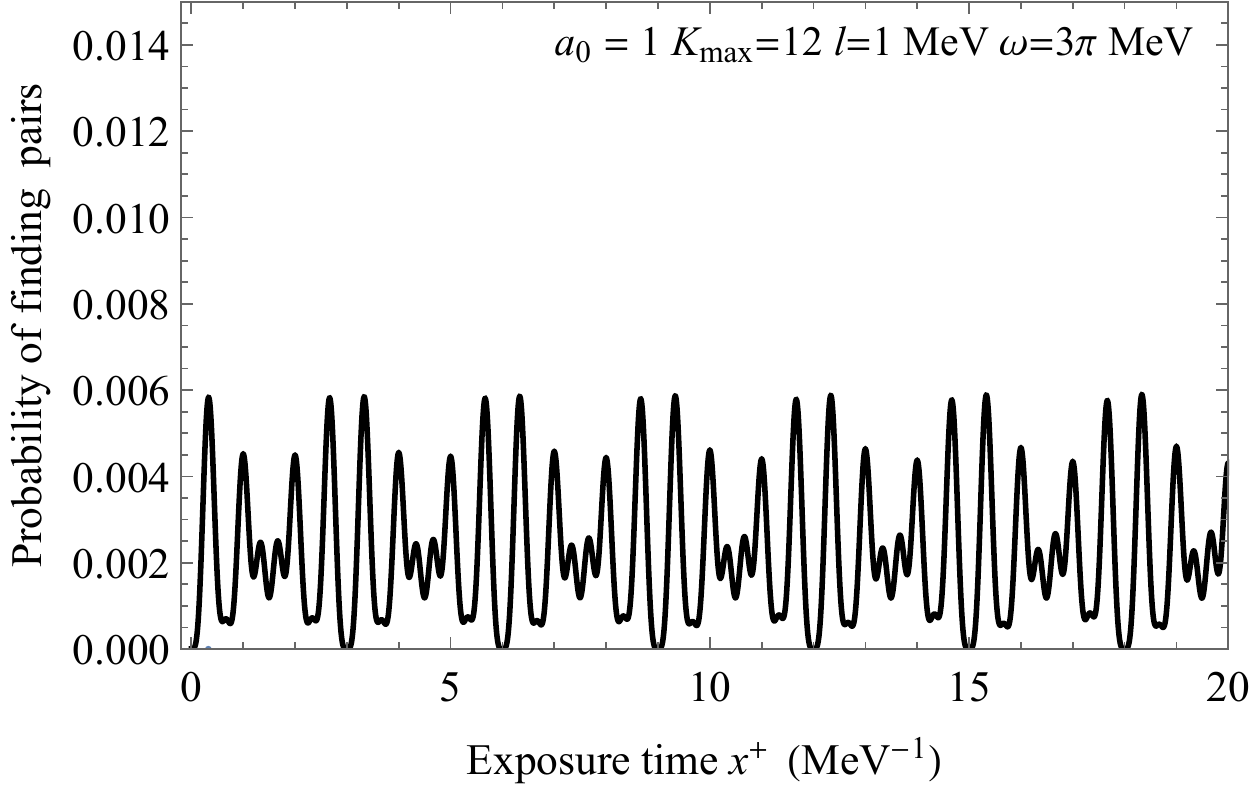}
			\includegraphics[width=.47\textwidth]{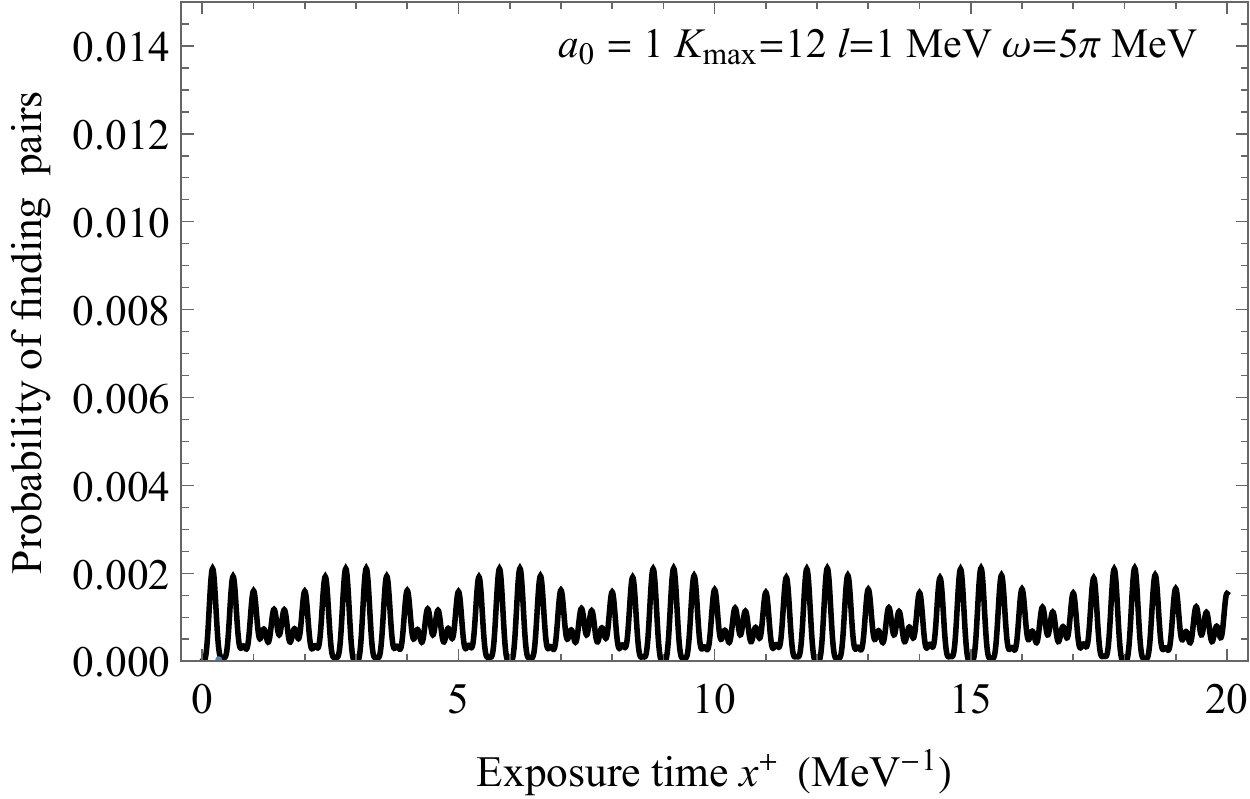}
		\end{tabular}
		\caption{Time evolution of the total probability of finding electron-positron pairs in the background field~\eqref{BG}. In the first $3$ rows of the figure, from the top to the bottom, probabilities of finding electron-positron pairs in fields with increasing intensities are presented. Panels in the left column show results with time profile $f\left( x^{\LCp} \right)=\sin\left( \omega x^{\LCp} \right), \omega=3\pi\mathrm{\,MeV}$ and panels in the right column show results with $\omega=5\pi\mathrm{\,MeV}$. The first $3$ rows show results for $l=2\rm \,MeV$ and the fourth row shows results for $l=1\rm \,MeV$. Other parameters: $K_{\max}=12$, $b=m_{e}$.}
		\label{fig:other_period}
	\end{center}
\end{figure*}

In order to study the pair production process in the background fields with richer time structure, we consider time profiles with various frequencies, specifically $\omega=2\pi\mathrm{\,MeV}, 3\pi\mathrm{\,MeV}$ and $ 5\pi\mathrm{\,MeV}$. In this subsection, we keep the background on for multiple cycles which allows for a more comprehensive study of the time structure in the pair production process. We show the time evolution of the probabilities of finding $n$ pairs of electrons and positrons, and their convergences with respect to $K_{\max}$ in Fig.~\ref{fig:other_kmax_dependence}. The left panel corresponds to the frequency of time profile $\omega=2\pi\rm\,MeV$ and the right panel corresponds to $\omega=3\pi\rm\,MeV$. In both panels, the probabilities at different $K_{\max}$'s almost coincide with each other, implying good convergences. This is as expected, because as the frequencies increase the basis states with larger number of pairs do not have time to be populated, but we still take $K_{\max}=12$ in this subsection for consistency. Note that in the left panel the probability of finding $4$ pairs of electrons and positrons starts to decrease around $x^{\LCp}=0.25\rm \,MeV^{-1}$, but since this decrease is triggered by the high-frequency time profile which changes the direction of the background, we still consider the result after this moment to be reliable. The results in both panels of Fig.~\ref{fig:other_kmax_dependence} show periodical oscillations, with the period coinciding with the frequency of the time profile. 
We show the dependence of the pattern of oscillations on the background field in Fig.~\ref{fig:other_period}, in a larger time scale. In each panel, there is an oscillation in the probability with the period of the time profile ($T_{p}$), and in addition the amplitude of the oscillation is also oscillating with a longer period ($T_{a}$). In the top $3$ rows, as we move from the top to the bottom row, in each column, we observe that the amplitude of the oscillation changes but $T_{p}$ barely changes, and as we move from the left to the right column, in each row, we observe that all the $T_{p}$ change. We thus conclude that $T_{p}$ depends on the frequency of the time profile and is independent of $a_{0}$ and $l$. In the top $3$ rows, where $l=2\rm \,MeV$, we find that $T_{a}$ in all the panels are almost the same, which are different from the $T_{a}$ in the bottom $2$ panels, where $l=1\rm \, MeV$. We thus conclude that $T_{a}$ only depends on $l$, and is independent of $a_{0}$ and the frequency of the background field. In the limit of constant time profile, $T_{a}$ is related to the period induced by the energy difference between the vacuum and the lowest one-pair state; see the discussion of Fig.~\ref{fig:Schwinger} and Appendix~\ref{perturbation_theory}.

\section{Conclusions and Outlooks}\label{Conclusions}
We have investigated the vacuum pair production in inhomogeneous electric fields in the framework of tBLFQ. We truncated the Fock space up to the sector containing $4$-pairs of electrons and positrons, which allows us to investigate the emissions of multiple pairs of electrons and positrons. We calculated the time evolution of the exclusive probabilities of the production of up to $4$ pairs. We studied the longitudinal momentum distribution of the produced fermions, and the invariant mass of the system, in background fields with various intensities and momenta, as well as different time profiles. The results suggest that as the longitudinal momentum $l$ increases the electric field strength increases correspondingly, which leads to a more rapid acceleration of the produced particles and a faster increase of the invariant mass of the system.
We find a critical intensity $a_{0}$ around $0.1$, above which the production rate no longer exhibits periodic structure in time and thus the average production rate increases dramatically, which is reminiscent of Schwinger's result. Our results suggest that for inhomogeneous background fields even their strengths are below the critical intensity, at certain instants the resulting production rate may be much larger than Schwinger's results in certain time intervals during the exposure.

This paper can be improved in several directions. First, for more comprehensive descriptions of the pair production, we plan to increase the basis size by, e.g.~reinstating the degrees of freedom in the transverse directions, and adding Fock sectors with more pairs of electrons and positrons. Secondly, we plan to introduce the QED interaction between the produced particles, which would allow us to investigate, for example, the process of photon emission. Finally, we will adopt more complex background fields, possibly with nonzero transverse components, for more faithful modeling of the lasers in modern laser facilities.

\clearpage
\section{Acknowledgment}
The authors thank Anton Ilderton and James P.~Vary for their careful readings and important suggestions for the manuscript, and Yangguang Yang for his careful examination of the equations. Z.~L.~thanks Yunfei Zhang for his help in the development of numerical codes. X.~Z.~is supported by new faculty startup funding by the Institute of Modern Physics, Chinese Academy of Sciences, by Key Research Program of Frontier Sciences, Chinese Academy of Sciences, Grant No. ZDB-SLY-7020, by the Natural Science Foundation of Gansu Province, China, Grant No. 20JR10RA067 and by the Strategic Priority Research Program of the Chinese Academy of Sciences, Grant No. XDB34000000.
\appendix

\section{Quantum fields in discrete basis}\label{fields}
We define the light-front time and the longitudinal coordinate as $x^{\LCp}:=x^{0}+x^{3}$ and $x^{\LCm}:=x^{0}-x^{3}$, respectively, so $\ud x^{\LCp}\ud x^{\LCm}=2\ud x^{0}\ud x^{3}$~\cite{Hu:2019hjx,Zhao:2013cma}. The remaining $2$ spatial coordinates $x^{\LCperp}:=\left( x^{1},x^{2} \right)$ are the transverse directions. The bold-font letters $\mathbf{x}$ and $\mathbf{p}$ represent the $3$-coordinate $\left( x^{\LCm},x^{\LCperp} \right)$ and the $3$-momentum $\left( p^{\LCp},p^{\LCperp} \right)$, respectively.
The mode expansion of the fermion field is
\begin{equation}
   \Psi(\mathbf{x})=\sum_{k^{\LCp}, \lambda}\int \ud^{2} p^{\LCperp}\frac{1}{(2\pi)^{2} \sqrt{2 L}}\left(b(\mathbf{p}, \lambda) u(\mathbf{p}, \lambda) e^{-i \mathbf{p} \cdot \mathbf{x}}+d^{\dagger}(\mathbf{p}, \lambda) v(\mathbf{p}, \lambda) e^{+i \mathbf{p} \cdot \mathbf{x}}\right)\;,
   \label{field}
\end{equation}
in which
\begin{equation}
   \begin{split}
      u(\mathbf{p}, \uparrow)=\left(\begin{array}{c}{1} \\ {0} \\ {\frac{i m_{e}}{p^{\LCp}}} \\ {\frac{\left(i p^{1}-p^{2}\right)}{p^{\LCp}}}\end{array}\right)\;\;,\; u(\mathbf{p}, \downarrow)=\left(\begin{array}{c}{0} \\ {1} \\ {\frac{\left(-i p^{1}-p^{2}\right)}{p^{\LCp}}} \\ {\frac{i m_{e}}{p^{\LCp}}}\end{array}\right)\;\;,\\
      v(\mathbf{p}, \uparrow)=\left(\begin{array}{c}{0} \\ {1} \\ {\frac{\left(-i p^{1}-p^{2}\right)}{p^{\LCp}}} \\ {\frac{-i m_{e}}{p^{\LCp}}}\end{array}\right)\;\;,\; v(\mathbf{p}, \downarrow)=\left(\begin{array}{c}{1} \\ {0} \\ {\frac{-i m_{e}}{p^{\LCp}}} \\ {\frac{\left(i p^{1}-p^{2}\right)}{p^{\LCp}}}\end{array}\right)\;\;.
   \end{split}
\end{equation}
The anti-commutators of the creation and annihilation operators are 
\begin{equation}
   \left\{b(\mathbf{p}, \lambda), b^{\dagger}\left(\mathbf{p}^{\prime}, \lambda^{\prime}\right)\right\}=\left\{d(\mathbf{p}, \lambda), d^{\dagger}\left(\mathbf{p}^{\prime}, \lambda^{\prime}\right)\right\}=\left( 2\pi \right)^{2}\delta_{p^{\LCp}}^{p^{\prime+}} \delta^{\left( 2 \right)}\left( p^{\LCperp}-p'^{\LCperp} \right) \delta_{\lambda}^{\lambda^{\prime}}\;.
\end{equation}
The single-particle momentum eigenstates are
\begin{equation}
   |\,e^{-};\mathbf{p}, \lambda\,\rangle := b^{\dagger}(\mathbf{p}, \lambda)|\,0\,\rangle \text { and }|\,e^{+};\mathbf{p}, \lambda\,\rangle := d^{\dagger}(\mathbf{p}, \lambda)|\,0\,\rangle\;.
   \label{state}
\end{equation}
Our basis state is a Gaussian wave packet in the transverse plane, and can be obtained by taking the superposition of the momentum eigenstates above
\begin{equation}
   |\,e^{\pm},p^{\LCp},\Phi,\lambda\rangle=\int \frac{\ud^{2}p^{\LCperp}}{\left( 2\pi \right)^{2}}\Phi^{b}\left( p^{\LCperp} \right)|\,e^{\pm};\mathbf{p}, \lambda\,\rangle\;,
   \label{gaussianstate}
\end{equation}
where
\begin{equation}
   \Phi^{b}\left( p^{\LCperp} \right)=\frac{2\sqrt{\pi}}{b}\exp\left( -\frac{ p_{1}^{2}+p_{2}^{2} }{2b^{2}} \right)\;
   \label{}
\end{equation}
is the Gaussian profile, in which $b$ is the width.
\section{Matrix elements of the Hamiltonian}\label{melements}
The matrix element of the Hamiltonian~\eqref{eqn:fullhamiltonian} in our basis can be obtained by using the mode expansion of the field operator~\eqref{field} and the definition of the single-particle states~\eqref{gaussianstate}. The kinetic energy term $P^{\LCm}_{0}$ is diagonal in our basis, and its matrix element in the $n$-pair Fock sector is
\begin{equation}
   \begin{aligned}
      \left\langle\,\left(e^{+} e^{-}\right)^{n}\,\left|\,P_{0}^{\LCm}\right|\,\left(e^{+} e^{-}\right)^{n}\,\right\rangle:&=\int\prod_{i=1}^{2n}\frac{\ud^{2}p_{i}^{\LCperp}\ud^{2}p_{i}'^{\LCperp}}{\left( 2\pi \right)^{4n}}\Phi^{b}\left( p_{i}^{\LCperp}\right)\Phi^{b\ast}\left( p_{i}'^{\LCperp} \right)\\&\quad\times\langle\, \mathbf{p}_{1}',\lambda_{1}';\cdots;\mathbf{p}_{2n}',\lambda_{2n}'\,|\,P_{0}^{\LCm}\,|\,\mathbf{p}_{1},\lambda_{1};\cdots;\mathbf{p}_{2n},\lambda_{2n}\,\rangle\\
      &=\sum_{i=1}^{2 n} \frac{b^{2}+m_{e}^{2}}{p_{i}^{\LCp}} \prod_{i=1}^{2 n} \delta_{p_{i}^{\LCp}}^{p_{i}^{\prime\LCp}} \delta_{\lambda_{i}}^{\lambda_{i}^{\prime}}\;,
   \end{aligned}
\end{equation}
in which $i$ ranging from $1$ to $2n$ is the index of particles, with odd $i$ labelling the positrons and even $i$ labelling the electrons. 

The matrix element of the interaction Hamiltonian can be divided into two parts, i.e.~the pair production part which changes the particle number by $2$, and the acceleration part which does not change the particle number. The matrix element of the pair production part is
\begin{equation}
\begin{aligned}
   \left\langle\, e^{+} e^{-}\,|\,V\left( x^{\LCp} \right)|\, 0\,\right\rangle&=\int\frac{\ud^{2}p_{1}^{\LCperp}\ud^{2}p_{2}^{\LCperp}}{\left( 2\pi \right)^{4}}\Phi^{b\ast}\left( p_{1}^{\LCperp}\right)\Phi^{b\ast}\left( p_{2}^{\LCperp} \right)\left\langle\, \mathbf{p}_{1}, \lambda_{1} ; \mathbf{p}_{2}, \lambda_{2}\,\left|\,\int \ud^{2} x^{\LCperp} \ud x^{\LCm} \frac{e}{2} \mathcal{A}^{\LCm}\left( x^{\LCp},x^{\LCm} \right) \bar{\Psi}(\mathbf{x}) \gamma^{\LCp} \Psi(\mathbf{x})\,\right|\, 0\,\right\rangle\\
   &= \sum_{\substack{q^{\LCp}, \lambda_{q},r^{\LCp}, \lambda_{r}}}\int  \frac{\ud^{2} x^{\LCperp} \ud x^{\LCm}\ud^{2}p_{1}^{\LCperp}\ud^{2}p_{2}^{\LCperp}\ud^{2}r^{\LCperp}\ud^{2}q^{\LCperp}}{\left(2 \pi\right)^{6} 2 L} \bar{u}\left(\mathbf{q}, \lambda_{q}\right) \gamma^{\LCp} v\left(\mathbf{r}, \lambda_{r}\right)\mathrm e^{-i\left[(q^{\LCperp}+r^{\LCperp}) x^{\LCperp}-\frac{1}{2}(q^{\LCp}+r^{\LCp}) x^{\LCm}\right]} \\
   &\quad\times\Phi^{b}\left( p_{1}^{\LCperp}\right)\Phi^{b}\left( p_{2}^{\LCperp} \right)\left[-\delta_{q^{\LCp}}^{p^{\LCp}_{2}} \delta_{r^{\LCp}}^{p^{\LCp}_{1}}\left( 2\pi \right)^{4}\delta^{\left( 2 \right)}\left( p^{\LCperp}_{1}-r^{\LCperp} \right)\delta^{\left( 2 \right)}\left( p^{\LCperp}_{2}-q^{\LCperp} \right)\right] m_{e} a_{0} f\left( x^{\LCp} \right)\frac{1}{2}\left(\mathrm e^{\frac{i}{2} l x^{\LCm}}+\mathrm e^{-\frac{i}{2} l x^{\LCm}}\right) \\
&= \int  \frac{\ud^{2} x^{\LCperp} \ud x^{\LCm}\ud^{2}p_{1}^{\LCperp}\ud^{2}p_{2}^{\LCperp}}{\left(2\pi\right)^{2} 2 L}\Phi^{b}\left( p_{1}^{\LCperp}\right)\Phi^{b}\left( p_{2}^{\LCperp} \right)\mathrm e^{-i\left[\left(p_{2}^{\LCperp}+p_{1}^{\LCperp}\right) x^{\LCperp}-\frac{1}{2}\left(p_{2}^{\LCp}+p_{1}^{\LCp}-l\right) x^{\LCm}\right]} \frac{1}{2} m_{e} a_{0} f\left( x^{\LCp} \right)\cdot 2 \delta_{\lambda_{1}}^{-\lambda_{2}} \\
&=-m_{e} a_{0}f\left( x^{\LCp} \right) \delta_{l}^{p_{1}^{\LCp}+p_{2}^{\LCp}} \delta_{\lambda_{1}}^{-\lambda_{2}}\;,
\end{aligned}
\end{equation}
and matrix element for the acceleration part can be obtained similarly
\begin{equation}
\begin{aligned}
   \left\langle\, e^{+} e^{-}\,|\,V\left( x^{\LCp} \right)\,|\, e^{+} e^{-}\,\right\rangle& =\int\frac{\ud^{2}p_{1}^{\LCperp}\ud^{2}p_{2}^{\LCperp}\ud^{2}p_{1}'^{\LCperp}\ud^{2}p_{2}'^{\LCperp}}{\left( 2\pi \right)^{8}}\Phi^{b}\left( p_{1}^{\LCperp}\right)\Phi^{b}\left( p_{2}^{\LCperp} \right)\Phi^{b\ast}\left( p_{1}'^{\LCperp}\right)\Phi^{b\ast}\left( p_{2}'^{\LCperp} \right)\\&\quad\times\left\langle\, \mathbf{p}_{1}^{\prime}, \lambda_{1}^{\prime} ; \mathbf{p}_{2}^{\prime}, \lambda_{2}^{\prime}\,\left|\,V\left( x^{\LCp} \right)\,\right|\, \mathbf{p}_{1}, \lambda_{1} ; \mathbf{p}_{2}, \lambda_{2}\,\right\rangle\\
   &=m_{e} a_{0}f\left( x^{\LCp} \right)\Bigg[\left(  \delta^{p_{2}'^{\LCp}+l}_{p_{2}^{\LCp}}+\delta^{p_{2}'^{\LCp}-l}_{p_{2}^{\LCp}}\right) \delta_{\lambda_{2}}^{\lambda_{2}'}\delta_{p^{\LCp}_{1}}^{p'^{\LCp}_{1}}\delta_{\lambda_{1}}^{\lambda_{1}'}-\left(  \delta^{p_{1}'^{\LCp}+l}_{p_{1}^{\LCp}}+\delta^{p_{1}'^{\LCp}-l}_{p_{1}^{\LCp}}\right) \delta_{\lambda_{1}}^{\lambda_{1}'}\delta_{p_{2}^{\LCp}}^{p_{2}'^{\LCp}}\delta_{\lambda_{2}}^{\lambda_{2}'}\Bigg]\;.
\end{aligned}
\end{equation}
The treatment of the identical particles becomes important when considering the matrix elements in higher Fock sectors. In our calculation, the interaction $V$ changes the particle number by $0$ or $2$, and therefore most of the particles are spectators and give Kronecker $\delta$'s. These $\delta$'s can be compactly written by using the definition of the determinant. As a result, the pair production part of the interaction matrix element from the $(n-1)$-pairs sector to the $n$-pairs sector can be written as
\begin{equation}
	\begin{aligned}
           \langle\, (e^{+}e^{-})^n\,|\,V\left( x^{\LCp} \right)\,|\,(e^{+}e^{-})^{n-1}\,\rangle:&=\int\prod_{i=1}^{2n}\prod_{j=1}^{2n-2}\frac{\ud^{2}p_{j}^{\LCperp}\ud^{2}p_{i}'^{\LCperp}}{\left( 2\pi \right)^{4n-4}}\Phi^{b}\left( p_{j}^{\LCperp}\right)\Phi^{b\ast}\left( p_{i}'^{\LCperp} \right)\\&\quad\times\langle\, \mathbf{p}_{1}',\lambda_{1}';\cdots;\mathbf{p}_{2n}',\lambda_{2n}'\,|\,V\left( x^{\LCp} \right)\,|\,\mathbf{p}_{1},\lambda_{1};\cdots;\mathbf{p}_{2n-2},\lambda_{2n-2}\,\rangle\\
                &={}-m_{e}a_{0}f\left( x^{\LCp} \right) \sum_{i,j=1}^{n}\delta_{l}^{p_{2i-1}'^{\LCp}+p_{2j}'^{\LCp}}\delta_{\lambda_{2i-1}'}^{-\lambda_{2j}'}(B^{*}_{n})_{1j}(D_{n}^{*})_{1i}\;, 
	\end{aligned}
\end{equation}
in which 
\begin{equation}
   \begin{aligned}
      &\quad\quad B_{n}=\left(\begin{array}{lllll}
            \delta_{q^{\LCp}}^{p_{2n}'^{\LCp}}   \delta_{\lambda_{q}}^{\lambda_{2n}'}   & \delta_{p_{2}^{\LCp}}^{p_{2n}'^{\LCp}} \delta_{\lambda_{2}}^{\lambda_{2n}'} & \delta_{p_{4}^{\LCp}}^{p_{2n}'^{\LCp}}     \delta_{\lambda_{4}}^{\lambda_{2n}'}     & \cdots      &\delta_{p_{2n-2}^{\LCp}}^{p_{2n}'^{\LCp}}\delta_{\lambda_{2n-2}}^{\lambda_{2n}'}\\
            \vdots                 & \vdots                   & \vdots                       & \ddots      &\vdots\\
            \delta_{q^{\LCp}}^{p_{4}'^{\LCp}}    \delta_{\lambda_{q}}^{\lambda_{4}'}    & \delta_{p_{2}^{\LCp}}^{p_{4}'^{\LCp}}  \delta_{\lambda_{2}}^{\lambda_{4}'}  & \delta_{p_{4}^{\LCp}}^{p_{4}'^{\LCp}}      \delta_{\lambda_{4}}^{\lambda_{4}'}      & \cdots      &\delta_{p_{2n-2}^{\LCp}}^{p_{4}'^{\LCp}}\delta_{\lambda_{2n-2}}^{\lambda_{4}'}\\
            \delta_{q^{\LCp}}^{p_{2}'^{\LCp}}    \delta_{\lambda_{q}}^{\lambda_{2}'}    & \delta_{p_{2}^{\LCp}}^{p_{2}'^{\LCp}}  \delta_{\lambda_{2}}^{\lambda_{2}'}  & \delta_{p_{4}^{\LCp}}^{p_{2}'^{\LCp}}      \delta_{\lambda_{4}}^{\lambda_{2}'}      & \cdots      &\delta_{p_{2n-2}^{\LCp}}^{p_{2}'^{\LCp}}
\delta_{\lambda_{2n-2}}^{\lambda_{2}'}
      \end{array}\right),\\
\\
      &D_{n}=\left(\begin{array}{lllll}
            \delta_{r^{\LCp}}^{p_{2n-1}'^{\LCp}} \delta_{\lambda_{r}}^{\lambda_{2n-1}'} & \delta_{p_{1}^{\LCp}}^{p_{2n-1}'^{\LCp}} \delta_{\lambda_{1}}^{\lambda_{2n-1}'} & \delta_{p_{3}^{\LCp}}^{p_{2n-1}'}  \delta_{\lambda_{3}}^{\lambda_{2n-1}'}  & \cdots      & \delta_{p_{2n-3}^{\LCp}}^{p_{2n-1}'^{\LCp}}\delta_{\lambda_{2n-3}}^{\lambda_{2n-1}'}\\
            \vdots                 & \vdots                     & \vdots                      & \ddots      &\vdots\\
            \delta_{r^{\LCp}}^{p_{3}'^{\LCp}}    \delta_{\lambda_{r}}^{\lambda_{3}'}    & \delta_{p_{1}^{\LCp}}^{p_{3}'^{\LCp}}    \delta_{\lambda_{1}}^{\lambda_{3}'}    & \delta_{p_{3}^{\LCp}}^{p_{3}'^{\LCp}}     \delta_{\lambda_{3}}^{\lambda_{3}'}     & \cdots      &\delta_{p_{2n-3}^{\LCp}}^{p_{3}'^{\LCp}} \delta_{\lambda_{2n-3}}^{\lambda_{3}'}\\
            \delta_{r^{\LCp}}^{p_{1}'^{\LCp}}    \delta_{\lambda{r}}^{\lambda_{1}'}    & \delta_{p_{1}^{\LCp}}^{p_{1}'^{\LCp}}    \delta_{\lambda_{1}}^{\lambda_{1}'}    & \delta_{p_{3}^{\LCp}}^{p_{1}'^{\LCp}}     \delta_{\lambda_{3}}^{\lambda_{1}'}     & \cdots      & \delta_{p_{2n-3}^{\LCp}}^{p_{1}'^{\LCp}}\delta_{\lambda_{2n-3}}^{\lambda_{1}'}
      \end{array}\right)\;,
      \label{4to6}
   \end{aligned}
\end{equation}
and $B^{*}_{ji}$ is the $\left( j,i \right)$ minor of the matrix $B$.
The acceleration part of the interaction matrix element is
\begin{equation}
	\begin{aligned}
           \langle\, (e^{+}e^{-})^n\,|\,V\left( x^{\LCp} \right)\,|\,(e^{+}e^{-})^{n}\,\rangle:&=\int\prod_{i=1}^{2n}\frac{\ud^{2}p_{i}^{\LCperp}\ud^{2}p_{i}'^{\LCperp}}{\left( 2\pi \right)^{4n}}\Phi^{b}\left( p_{i}^{\LCperp}\right)\Phi^{b\ast}\left( p_{i}'^{\LCperp} \right)\\&\quad\times\langle\, \mathbf{p}_{1}',\lambda_{1}';\cdots;\mathbf{p}_{2n}',\lambda_{2n}'\,|\,V\left( x^{\LCp} \right)\,|\,\mathbf{p}_{1},\lambda_{1};\cdots;\mathbf{p}_{2n},\lambda_{2n}\,\rangle\\
           &={}m_{e}a_{0}f\left( x^{\LCp} \right) \sum_{i,j=1}^{n}\left(\delta^{p_{2n-2i+2}'^{\LCp}+l}_{p_{2j}^{\LCp}}+\delta^{p_{2n-2i+2}'^{\LCp}-l}_{p_{2j}^{\LCp}}\right)\delta^{\lambda_{2n-2i+2}'}_{\lambda_{2j}}\left(B_{n}'^{*}\right)_{ji}\times\left|\,D_{n}'\,\right|\,\\&\quad\quad\quad-\left(\delta^{p_{2n-2i+1}'^{\LCp}+l}_{p_{2j-1}^{\LCp}}+\delta^{p_{2n-2i+1}'^{\LCp}-l}_{p_{2j-1}^{\LCp}}\right)\delta^{\lambda_{2n-2i+1}'}_{\lambda_{2j-1}}\left|\,B_{n}'\,\right|\,\times \left(D_{n}'^{*}\right)_{ji}\;,
	\end{aligned}
\end{equation}
in which
\begin{equation}
   \begin{aligned}
      &B'_{n}=\left(\begin{array}{llll}
            \delta_{p_{2}^{\LCp}}^{p_{2n}'^{\LCp}}\delta_{\lambda_{2}}^{\lambda_{2n}'}   & \delta_{p_{4}^{\LCp}}^{p_{2n}'^{\LCp}}     \delta_{\lambda_{4}}^{\lambda_{2n}'}     & \cdots      &\delta_{p_{2n}^{\LCp}}^{p_{2n}'^{\LCp}}\delta_{\lambda_{2n}}^{\lambda_{2n}'}\\
            \vdots                 & \vdots                       & \ddots      &\vdots\\
            \delta_{p_{2}^{\LCp}}^{p_{4}'^{\LCp}}    \delta_{\lambda_{2}}^{\lambda_{4}'}    & \delta_{p_{4}^{\LCp}}^{p_{4}'^{\LCp}}      \delta_{\lambda_{4}}^{\lambda_{4}'}      & \cdots      &\delta_{p_{2n}^{\LCp}}^{p_{4}'^{\LCp}}\delta_{\lambda_{2n}}^{\lambda_{4}'}\\
            \delta_{p_{2}^{\LCp}}^{p_{2}'^{\LCp}}    \delta_{\lambda_{2}}^{\lambda_{2}'}    & \delta_{p_{4}^{\LCp}}^{p_{2}'^{\LCp}}      \delta_{\lambda_{4}}^{\lambda_{2}'}      & \cdots      &\delta_{p_{2n}^{\LCp}}^{p_{2}'^{\LCp}}\delta_{\lambda_{2n}}^{\lambda_{2}'}
      \end{array}\right),\quad
      D'_{n}=\left(\begin{array}{llll}
            \delta_{p_{1}^{\LCp}}^{p_{2n-1}'^{\LCp}}   \delta_{\lambda_{1}}^{\lambda_{2n-1}'}   & \delta_{p_{3}^{\LCp}}^{p_{2n-1}'^{\LCp}}     \delta_{\lambda_{3}}^{\lambda_{2n-1}'}     & \cdots      &\delta_{p_{2n-1}^{\LCp}}^{p_{2n-1}'^{\LCp}}\delta_{\lambda_{2n-1}}^{\lambda_{2n-1}'}\\
            \vdots                 & \vdots                       & \ddots      &\vdots\\
            \delta_{p_{1}^{\LCp}}^{p_{3}'^{\LCp}}    \delta_{\lambda_{1}}^{\lambda_{3}'}    & \delta_{p_{3}^{\LCp}}^{p_{3}'^{\LCp}}      \delta_{\lambda_{3}}^{\lambda_{3}'}      & \cdots      &\delta_{p_{2n-1}^{\LCp}}^{p_{3}'^{\LCp}}\delta_{\lambda_{2n-1}}^{\lambda_{3}'}\\
            \delta_{p_{1}^{\LCp}}^{p_{1}'^{\LCp}}    \delta_{\lambda_{1}}^{\lambda_{1}'}    & \delta_{p_{3}^{\LCp}}^{p_{1}'^{\LCp}}      \delta_{\lambda_{3}}^{\lambda_{1}'}      & \cdots      &\delta_{p_{2n-1}^{\LCp}}^{p_{1}'^{\LCp}}\delta_{\lambda_{2n-1}}^{\lambda_{1}'}
      \end{array}\right)\;,
      \label{4to6}
   \end{aligned}
\end{equation}
and $|\,B'\,|\,$ is the determinant of $B'$.
\section{Comparison with the time-dependent perturbation theory}\label{perturbation_theory}
\begin{figure*}[t!]
	\centering
	\begin{center}
		\begin{tabular}{@{}cccc@{}}
			\includegraphics[width=.50\textwidth]{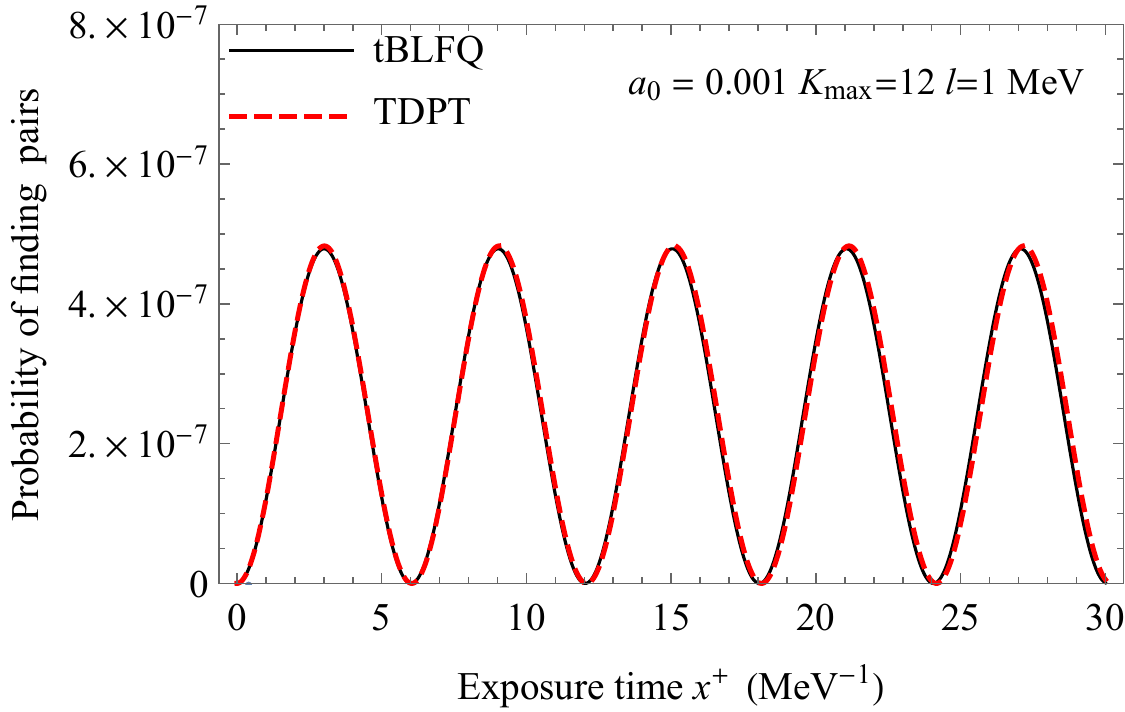} 
		\end{tabular}
                \caption{The comparison of the probability of finding electron-positron pairs between tBLFQ and time-dependent perturbation theory. Parameters: $a_{0}=0.001$, $K_{\max}=12$, $l=1 \rm{\,MeV}$, $b=m_{e}$ and $f\left(x^{\LCp} \right)=1$.
                }
		\label{fig:compare}
	\end{center}
\end{figure*}

In this section, we compare the total probability of finding electron-positron pairs obtained in tBLFQ and time-dependent perturbation theory, in a weak background field. For example, we take $a_{0}=0.001$, $l=1\rm \,MeV$, $b=m_{e} $ and $f\left( x^{\LCp} \right)=1$. In the limit of small $a_{0}$, we expect the dominance of the states directly connected to the vacuum by the background field, which are the $2$ states both containing a pair of electron and positron, with the following quantum numbers
\begin{equation}
   p_{1}^{\LCp}=\frac{1}{2}{\rm\, MeV},\lambda_{1}=\uparrow,p_{3}^{\LCp}=\frac{1}{2}{\rm\, MeV},\lambda_{3}=\downarrow\quad\text{and} \quad p_{1}^{\LCp}=\frac{1}{2}{\rm\, MeV},\lambda_{1}=\downarrow,p_{3}^{\LCp}=\frac{1}{2}{\rm\, MeV},\lambda_{3}=\uparrow\,.
   \label{}
\end{equation}
According to the time-dependent perturbation theory, the amplitude of finding the system in the $n$-th state to the leading order is  
\begin{equation}
   c_{n}^{(1)}(x^{\LCp})=-i \int_{0}^{x^{\LCp}} \ud t\frac{1}{2}V_{ni} \mathrm{ e}^{\frac{i}{2} \omega_{ni} t} =-\frac{V_{ni}}{2\omega_{ni}}\left(\mathrm e^{\frac{i}{2} \omega_{ni} t}-1\right)\;,
   \label{tpert}
\end{equation}
in which $i$ represents the initial state, namely the vacuum, and $n$ represents the final $2$ states, $V_{ni}=\langle \,n\,|\,V\,|\, i\,\rangle$ and $\omega_{ni}=\left(P_{0}^{\LCm}\right)_{n}-\left(P_{0}^{\LCm}\right)_{i}$, therefore, 
\begin{equation}
   \omega_{21}=\omega_{31}=\frac{1}{2}\left(\frac{2m_{e}^{2}}{p^{\LCp}_{1}}+\frac{2m_{e}^{2}}{p^{\LCp}_{3}}  \right)\approx1.04\mathrm{\,MeV}\quad\text{and}\quad V_{21}=V_{31}=-m_{e}a_{0}\approx0.000511 \mathrm{\,MeV}\;.
   \label{omega}
\end{equation}
Plugging~\eqref{omega} into~\eqref{tpert}, we obtain the total probability of finding electron-positron pairs by
\begin{equation}
   P_{\rm Pert}=\left|c_{2}^{(1)}\left( x^{\LCp} \right)\right|^{2}+\left|c_{3}^{(1)}\left( x^{\LCp} \right)\right|^{2} \;.
   \label{}
\end{equation}
Fig.~\ref{fig:compare} compares the results obtained in tBLFQ and time-dependent perturbation theory. As presented in the figure, both the period and the amplitude of the oscillation agree well in the two approaches.
\bibliography{article} 

\begin{thebibliography}{50}%
\makeatletter
\providecommand \@ifxundefined [1]{%
 \@ifx{#1\undefined}
}%
\providecommand \@ifnum [1]{%
 \ifnum #1\expandafter \@firstoftwo
 \else \expandafter \@secondoftwo
 \fi
}%
\providecommand \@ifx [1]{%
 \ifx #1\expandafter \@firstoftwo
 \else \expandafter \@secondoftwo
 \fi
}%
\providecommand \natexlab [1]{#1}%
\providecommand \enquote  [1]{``#1''}%
\providecommand \bibnamefont  [1]{#1}%
\providecommand \bibfnamefont [1]{#1}%
\providecommand \citenamefont [1]{#1}%
\providecommand \href@noop [0]{\@secondoftwo}%
\providecommand \href [0]{\begingroup \@sanitize@url \@href}%
\providecommand \@href[1]{\@@startlink{#1}\@@href}%
\providecommand \@@href[1]{\endgroup#1\@@endlink}%
\providecommand \@sanitize@url [0]{\catcode `\\12\catcode `\$12\catcode
  `\&12\catcode `\#12\catcode `\^12\catcode `\_12\catcode `\%12\relax}%
\providecommand \@@startlink[1]{}%
\providecommand \@@endlink[0]{}%
\providecommand \url  [0]{\begingroup\@sanitize@url \@url }%
\providecommand \@url [1]{\endgroup\@href {#1}{\urlprefix }}%
\providecommand \urlprefix  [0]{URL }%
\providecommand \Eprint [0]{\href }%
\providecommand \doibase [0]{http://dx.doi.org/}%
\providecommand \selectlanguage [0]{\@gobble}%
\providecommand \bibinfo  [0]{\@secondoftwo}%
\providecommand \bibfield  [0]{\@secondoftwo}%
\providecommand \translation [1]{[#1]}%
\providecommand \BibitemOpen [0]{}%
\providecommand \bibitemStop [0]{}%
\providecommand \bibitemNoStop [0]{.\EOS\space}%
\providecommand \EOS [0]{\spacefactor3000\relax}%
\providecommand \BibitemShut  [1]{\csname bibitem#1\endcsname}%
\let\auto@bib@innerbib\@empty
\bibitem [{\citenamefont {Vary}\ \emph {et~al.}(2010)\citenamefont {Vary},
  \citenamefont {Honkanen}, \citenamefont {Li}, \citenamefont {Maris},
  \citenamefont {Brodsky}, \citenamefont {Harindranath}, \citenamefont
  {de~Teramond}, \citenamefont {Sternberg}, \citenamefont {Ng},\ and\
  \citenamefont {Yang}}]{Vary:2009gt}%
  \BibitemOpen
  \bibfield  {author} {\bibinfo {author} {\bibfnamefont {J.~P.}\ \bibnamefont
  {Vary}}, \bibinfo {author} {\bibfnamefont {H.}~\bibnamefont {Honkanen}},
  \bibinfo {author} {\bibfnamefont {J.}~\bibnamefont {Li}}, \bibinfo {author}
  {\bibfnamefont {P.}~\bibnamefont {Maris}}, \bibinfo {author} {\bibfnamefont
  {S.~J.}\ \bibnamefont {Brodsky}}, \bibinfo {author} {\bibfnamefont
  {A.}~\bibnamefont {Harindranath}}, \bibinfo {author} {\bibfnamefont {G.~F.}\
  \bibnamefont {de~Teramond}}, \bibinfo {author} {\bibfnamefont
  {P.}~\bibnamefont {Sternberg}}, \bibinfo {author} {\bibfnamefont {E.~G.}\
  \bibnamefont {Ng}}, \ and\ \bibinfo {author} {\bibfnamefont {C.}~\bibnamefont
  {Yang}},\ }\href {\doibase 10.1103/PhysRevC.81.035205} {\bibfield  {journal}
  {\bibinfo  {journal} {Phys. Rev.}\ }\textbf {\bibinfo {volume} {C81}},\
  \bibinfo {pages} {035205} (\bibinfo {year} {2010})},\ \Eprint
  {http://arxiv.org/abs/0905.1411} {arXiv:0905.1411 [nucl-th]} \BibitemShut
  {NoStop}%
\bibitem [{\citenamefont {Li}\ \emph {et~al.}(2015)\citenamefont {Li},
  \citenamefont {Karmanov}, \citenamefont {Maris},\ and\ \citenamefont
  {Vary}}]{Li:2015iaw}%
  \BibitemOpen
  \bibfield  {author} {\bibinfo {author} {\bibfnamefont {Y.}~\bibnamefont
  {Li}}, \bibinfo {author} {\bibfnamefont {V.~A.}\ \bibnamefont {Karmanov}},
  \bibinfo {author} {\bibfnamefont {P.}~\bibnamefont {Maris}}, \ and\ \bibinfo
  {author} {\bibfnamefont {J.~P.}\ \bibnamefont {Vary}},\ }\href {\doibase
  10.1016/j.physletb.2015.07.014} {\bibfield  {journal} {\bibinfo  {journal}
  {Phys. Lett.}\ }\textbf {\bibinfo {volume} {B748}},\ \bibinfo {pages} {278}
  (\bibinfo {year} {2015})},\ \Eprint {http://arxiv.org/abs/1504.05233}
  {arXiv:1504.05233 [nucl-th]} \BibitemShut {NoStop}%
\bibitem [{\citenamefont {Li}\ \emph {et~al.}(2016)\citenamefont {Li},
  \citenamefont {Maris}, \citenamefont {Zhao},\ and\ \citenamefont
  {Vary}}]{Li:2015zda}%
  \BibitemOpen
  \bibfield  {author} {\bibinfo {author} {\bibfnamefont {Y.}~\bibnamefont
  {Li}}, \bibinfo {author} {\bibfnamefont {P.}~\bibnamefont {Maris}}, \bibinfo
  {author} {\bibfnamefont {X.}~\bibnamefont {Zhao}}, \ and\ \bibinfo {author}
  {\bibfnamefont {J.~P.}\ \bibnamefont {Vary}},\ }\href {\doibase
  10.1016/j.physletb.2016.04.065} {\bibfield  {journal} {\bibinfo  {journal}
  {Phys. Lett.}\ }\textbf {\bibinfo {volume} {B758}},\ \bibinfo {pages} {118}
  (\bibinfo {year} {2016})},\ \Eprint {http://arxiv.org/abs/1509.07212}
  {arXiv:1509.07212 [hep-ph]} \BibitemShut {NoStop}%
\bibitem [{\citenamefont {Karmanov}\ \emph {et~al.}(2016)\citenamefont
  {Karmanov}, \citenamefont {Li}, \citenamefont {Smirnov},\ and\ \citenamefont
  {Vary}}]{Karmanov:2016yzu}%
  \BibitemOpen
  \bibfield  {author} {\bibinfo {author} {\bibfnamefont {V.~A.}\ \bibnamefont
  {Karmanov}}, \bibinfo {author} {\bibfnamefont {Y.}~\bibnamefont {Li}},
  \bibinfo {author} {\bibfnamefont {A.~V.}\ \bibnamefont {Smirnov}}, \ and\
  \bibinfo {author} {\bibfnamefont {J.~P.}\ \bibnamefont {Vary}},\ }\href
  {\doibase 10.1103/PhysRevD.94.096008} {\bibfield  {journal} {\bibinfo
  {journal} {Phys. Rev.}\ }\textbf {\bibinfo {volume} {D94}},\ \bibinfo {pages}
  {096008} (\bibinfo {year} {2016})},\ \Eprint
  {http://arxiv.org/abs/1610.03559} {arXiv:1610.03559 [hep-th]} \BibitemShut
  {NoStop}%
\bibitem [{\citenamefont {Chen}\ \emph {et~al.}(2017)\citenamefont {Chen},
  \citenamefont {Li}, \citenamefont {Maris}, \citenamefont {Tuchin},\ and\
  \citenamefont {Vary}}]{Chen:2016dlk}%
  \BibitemOpen
  \bibfield  {author} {\bibinfo {author} {\bibfnamefont {G.}~\bibnamefont
  {Chen}}, \bibinfo {author} {\bibfnamefont {Y.}~\bibnamefont {Li}}, \bibinfo
  {author} {\bibfnamefont {P.}~\bibnamefont {Maris}}, \bibinfo {author}
  {\bibfnamefont {K.}~\bibnamefont {Tuchin}}, \ and\ \bibinfo {author}
  {\bibfnamefont {J.~P.}\ \bibnamefont {Vary}},\ }\href {\doibase
  10.1016/j.physletb.2017.04.024} {\bibfield  {journal} {\bibinfo  {journal}
  {Phys. Lett.}\ }\textbf {\bibinfo {volume} {B769}},\ \bibinfo {pages} {477}
  (\bibinfo {year} {2017})},\ \Eprint {http://arxiv.org/abs/1610.04945}
  {arXiv:1610.04945 [nucl-th]} \BibitemShut {NoStop}%
\bibitem [{\citenamefont {Li}\ \emph {et~al.}(2017)\citenamefont {Li},
  \citenamefont {Maris},\ and\ \citenamefont {Vary}}]{Li:2017mlw}%
  \BibitemOpen
  \bibfield  {author} {\bibinfo {author} {\bibfnamefont {Y.}~\bibnamefont
  {Li}}, \bibinfo {author} {\bibfnamefont {P.}~\bibnamefont {Maris}}, \ and\
  \bibinfo {author} {\bibfnamefont {J.~P.}\ \bibnamefont {Vary}},\ }\href
  {\doibase 10.1103/PhysRevD.96.016022} {\bibfield  {journal} {\bibinfo
  {journal} {Phys. Rev.}\ }\textbf {\bibinfo {volume} {D96}},\ \bibinfo {pages}
  {016022} (\bibinfo {year} {2017})},\ \Eprint
  {http://arxiv.org/abs/1704.06968} {arXiv:1704.06968 [hep-ph]} \BibitemShut
  {NoStop}%
\bibitem [{\citenamefont {Adhikari}\ \emph {et~al.}(2019)\citenamefont
  {Adhikari}, \citenamefont {Li}, \citenamefont {Li},\ and\ \citenamefont
  {Vary}}]{Adhikari:2018umb}%
  \BibitemOpen
  \bibfield  {author} {\bibinfo {author} {\bibfnamefont {L.}~\bibnamefont
  {Adhikari}}, \bibinfo {author} {\bibfnamefont {Y.}~\bibnamefont {Li}},
  \bibinfo {author} {\bibfnamefont {M.}~\bibnamefont {Li}}, \ and\ \bibinfo
  {author} {\bibfnamefont {J.~P.}\ \bibnamefont {Vary}},\ }\href {\doibase
  10.1103/PhysRevC.99.035208} {\bibfield  {journal} {\bibinfo  {journal} {Phys.
  Rev.}\ }\textbf {\bibinfo {volume} {C99}},\ \bibinfo {pages} {035208}
  (\bibinfo {year} {2019})},\ \Eprint {http://arxiv.org/abs/1809.06475}
  {arXiv:1809.06475 [hep-ph]} \BibitemShut {NoStop}%
\bibitem [{\citenamefont {Tang}\ \emph {et~al.}(2018)\citenamefont {Tang},
  \citenamefont {Li}, \citenamefont {Maris},\ and\ \citenamefont
  {Vary}}]{Tang:2018myz}%
  \BibitemOpen
  \bibfield  {author} {\bibinfo {author} {\bibfnamefont {S.}~\bibnamefont
  {Tang}}, \bibinfo {author} {\bibfnamefont {Y.}~\bibnamefont {Li}}, \bibinfo
  {author} {\bibfnamefont {P.}~\bibnamefont {Maris}}, \ and\ \bibinfo {author}
  {\bibfnamefont {J.~P.}\ \bibnamefont {Vary}},\ }\href {\doibase
  10.1103/PhysRevD.98.114038} {\bibfield  {journal} {\bibinfo  {journal} {Phys.
  Rev.}\ }\textbf {\bibinfo {volume} {D98}},\ \bibinfo {pages} {114038}
  (\bibinfo {year} {2018})},\ \Eprint {http://arxiv.org/abs/1810.05971}
  {arXiv:1810.05971 [nucl-th]} \BibitemShut {NoStop}%
\bibitem [{\citenamefont {Jia}\ and\ \citenamefont {Vary}(2019)}]{Jia:2018ary}%
  \BibitemOpen
  \bibfield  {author} {\bibinfo {author} {\bibfnamefont {S.}~\bibnamefont
  {Jia}}\ and\ \bibinfo {author} {\bibfnamefont {J.~P.}\ \bibnamefont {Vary}},\
  }\href {\doibase 10.1103/PhysRevC.99.035206} {\bibfield  {journal} {\bibinfo
  {journal} {Phys. Rev.}\ }\textbf {\bibinfo {volume} {C99}},\ \bibinfo {pages}
  {035206} (\bibinfo {year} {2019})},\ \Eprint
  {http://arxiv.org/abs/1811.08512} {arXiv:1811.08512 [nucl-th]} \BibitemShut
  {NoStop}%
\bibitem [{\citenamefont {Jia}\ and\ \citenamefont {Vary}(2018)}]{Jia:2018hxd}%
  \BibitemOpen
  \bibfield  {author} {\bibinfo {author} {\bibfnamefont {S.}~\bibnamefont
  {Jia}}\ and\ \bibinfo {author} {\bibfnamefont {J.~P.}\ \bibnamefont {Vary}},\
  }\href@noop {} {\  (\bibinfo {year} {2018})},\ \Eprint
  {http://arxiv.org/abs/1812.09340} {arXiv:1812.09340 [nucl-th]} \BibitemShut
  {NoStop}%
\bibitem [{\citenamefont {Lan}\ \emph {et~al.}(2019{\natexlab{a}})\citenamefont
  {Lan}, \citenamefont {Mondal}, \citenamefont {Jia}, \citenamefont {Zhao},\
  and\ \citenamefont {Vary}}]{Lan:2019vui}%
  \BibitemOpen
  \bibfield  {author} {\bibinfo {author} {\bibfnamefont {J.}~\bibnamefont
  {Lan}}, \bibinfo {author} {\bibfnamefont {C.}~\bibnamefont {Mondal}},
  \bibinfo {author} {\bibfnamefont {S.}~\bibnamefont {Jia}}, \bibinfo {author}
  {\bibfnamefont {X.}~\bibnamefont {Zhao}}, \ and\ \bibinfo {author}
  {\bibfnamefont {J.~P.}\ \bibnamefont {Vary}},\ }\href {\doibase
  10.1103/PhysRevLett.122.172001} {\bibfield  {journal} {\bibinfo  {journal}
  {Phys. Rev. Lett.}\ }\textbf {\bibinfo {volume} {122}},\ \bibinfo {pages}
  {172001} (\bibinfo {year} {2019}{\natexlab{a}})},\ \Eprint
  {http://arxiv.org/abs/1901.11430} {arXiv:1901.11430 [nucl-th]} \BibitemShut
  {NoStop}%
\bibitem [{\citenamefont {Lan}\ \emph {et~al.}(2019{\natexlab{b}})\citenamefont
  {Lan}, \citenamefont {Mondal}, \citenamefont {Jia}, \citenamefont {Zhao},\
  and\ \citenamefont {Vary}}]{Lan:2019rba}%
  \BibitemOpen
  \bibfield  {author} {\bibinfo {author} {\bibfnamefont {J.}~\bibnamefont
  {Lan}}, \bibinfo {author} {\bibfnamefont {C.}~\bibnamefont {Mondal}},
  \bibinfo {author} {\bibfnamefont {S.}~\bibnamefont {Jia}}, \bibinfo {author}
  {\bibfnamefont {X.}~\bibnamefont {Zhao}}, \ and\ \bibinfo {author}
  {\bibfnamefont {J.~P.}\ \bibnamefont {Vary}},\ }\href@noop {} {\  (\bibinfo
  {year} {2019}{\natexlab{b}})},\ \Eprint {http://arxiv.org/abs/1907.01509}
  {arXiv:1907.01509 [nucl-th]} \BibitemShut {NoStop}%
\bibitem [{\citenamefont {Du}\ \emph {et~al.}(2019{\natexlab{a}})\citenamefont
  {Du}, \citenamefont {Li}, \citenamefont {Zhao},\ and\ \citenamefont
  {Vary}}]{Du:2019qsz}%
  \BibitemOpen
  \bibfield  {author} {\bibinfo {author} {\bibfnamefont {W.}~\bibnamefont
  {Du}}, \bibinfo {author} {\bibfnamefont {Y.}~\bibnamefont {Li}}, \bibinfo
  {author} {\bibfnamefont {X.}~\bibnamefont {Zhao}}, \ and\ \bibinfo {author}
  {\bibfnamefont {J.~P.}\ \bibnamefont {Vary}},\ }in\ \href@noop {} {\emph
  {\bibinfo {booktitle} {{6th International Conference Nuclear Theory in the
  Supercomputing Era (NTSE-2018) Daejeon, Korea, October 29-November 2,
  2018}}}}\ (\bibinfo {year} {2019})\ \Eprint {http://arxiv.org/abs/1908.02237}
  {arXiv:1908.02237 [nucl-th]} \BibitemShut {NoStop}%
\bibitem [{\citenamefont {Mondal}\ \emph {et~al.}(2019)\citenamefont {Mondal},
  \citenamefont {Xu}, \citenamefont {Lan}, \citenamefont {Zhao}, \citenamefont
  {Li}, \citenamefont {Lamm},\ and\ \citenamefont {Vary}}]{Mondal:2019yph}%
  \BibitemOpen
  \bibfield  {author} {\bibinfo {author} {\bibfnamefont {C.}~\bibnamefont
  {Mondal}}, \bibinfo {author} {\bibfnamefont {S.}~\bibnamefont {Xu}}, \bibinfo
  {author} {\bibfnamefont {J.}~\bibnamefont {Lan}}, \bibinfo {author}
  {\bibfnamefont {X.}~\bibnamefont {Zhao}}, \bibinfo {author} {\bibfnamefont
  {Y.}~\bibnamefont {Li}}, \bibinfo {author} {\bibfnamefont {H.}~\bibnamefont
  {Lamm}}, \ and\ \bibinfo {author} {\bibfnamefont {J.~P.}\ \bibnamefont
  {Vary}},\ }\bibfield  {booktitle} {\emph {\bibinfo {booktitle} {{Proceedings,
  27th International Workshop on Deep Inelastic Scattering and Related Subjects
  (DIS 2019): Torino, Italy, April 8-12, 2019}}},\ }\href {\doibase
  10.22323/1.352.0190} {\bibfield  {journal} {\bibinfo  {journal} {PoS}\
  }\textbf {\bibinfo {volume} {DIS2019}},\ \bibinfo {pages} {190} (\bibinfo
  {year} {2019})}\BibitemShut {NoStop}%
\bibitem [{\citenamefont {Lan}\ \emph {et~al.}(2019{\natexlab{c}})\citenamefont
  {Lan}, \citenamefont {Mondal}, \citenamefont {Li}, \citenamefont {Li},
  \citenamefont {Tang}, \citenamefont {Zhao},\ and\ \citenamefont
  {Vary}}]{Lan:2019img}%
  \BibitemOpen
  \bibfield  {author} {\bibinfo {author} {\bibfnamefont {J.}~\bibnamefont
  {Lan}}, \bibinfo {author} {\bibfnamefont {C.}~\bibnamefont {Mondal}},
  \bibinfo {author} {\bibfnamefont {M.}~\bibnamefont {Li}}, \bibinfo {author}
  {\bibfnamefont {Y.}~\bibnamefont {Li}}, \bibinfo {author} {\bibfnamefont
  {S.}~\bibnamefont {Tang}}, \bibinfo {author} {\bibfnamefont {X.}~\bibnamefont
  {Zhao}}, \ and\ \bibinfo {author} {\bibfnamefont {J.~P.}\ \bibnamefont
  {Vary}},\ }\href@noop {} {\  (\bibinfo {year} {2019}{\natexlab{c}})},\
  \Eprint {http://arxiv.org/abs/1911.11676} {arXiv:1911.11676 [nucl-th]}
  \BibitemShut {NoStop}%
\bibitem [{\citenamefont {Mondal}\ \emph {et~al.}(2020)\citenamefont {Mondal},
  \citenamefont {Xu}, \citenamefont {Lan}, \citenamefont {Zhao}, \citenamefont
  {Li}, \citenamefont {Chakrabarti},\ and\ \citenamefont
  {Vary}}]{Mondal:2019jdg}%
  \BibitemOpen
  \bibfield  {author} {\bibinfo {author} {\bibfnamefont {C.}~\bibnamefont
  {Mondal}}, \bibinfo {author} {\bibfnamefont {S.}~\bibnamefont {Xu}}, \bibinfo
  {author} {\bibfnamefont {J.}~\bibnamefont {Lan}}, \bibinfo {author}
  {\bibfnamefont {X.}~\bibnamefont {Zhao}}, \bibinfo {author} {\bibfnamefont
  {Y.}~\bibnamefont {Li}}, \bibinfo {author} {\bibfnamefont {D.}~\bibnamefont
  {Chakrabarti}}, \ and\ \bibinfo {author} {\bibfnamefont {J.~P.}\ \bibnamefont
  {Vary}},\ }\href {\doibase 10.1103/PhysRevD.102.016008} {\bibfield  {journal}
  {\bibinfo  {journal} {Phys. Rev. D}\ }\textbf {\bibinfo {volume} {102}},\
  \bibinfo {pages} {016008} (\bibinfo {year} {2020})},\ \Eprint
  {http://arxiv.org/abs/1911.10913} {arXiv:1911.10913 [hep-ph]} \BibitemShut
  {NoStop}%
\bibitem [{\citenamefont {Du}\ \emph {et~al.}(2019{\natexlab{b}})\citenamefont
  {Du}, \citenamefont {Li}, \citenamefont {Zhao}, \citenamefont {Miller},\ and\
  \citenamefont {Vary}}]{Du:2019ips}%
  \BibitemOpen
  \bibfield  {author} {\bibinfo {author} {\bibfnamefont {W.}~\bibnamefont
  {Du}}, \bibinfo {author} {\bibfnamefont {Y.}~\bibnamefont {Li}}, \bibinfo
  {author} {\bibfnamefont {X.}~\bibnamefont {Zhao}}, \bibinfo {author}
  {\bibfnamefont {G.~A.}\ \bibnamefont {Miller}}, \ and\ \bibinfo {author}
  {\bibfnamefont {J.~P.}\ \bibnamefont {Vary}},\ }\href@noop {} {\  (\bibinfo
  {year} {2019}{\natexlab{b}})},\ \Eprint {http://arxiv.org/abs/1911.10762}
  {arXiv:1911.10762 [nucl-th]} \BibitemShut {NoStop}%
\bibitem [{\citenamefont {Xu}\ \emph {et~al.}(2020)\citenamefont {Xu},
  \citenamefont {Mondal}, \citenamefont {Lan}, \citenamefont {Zhao},
  \citenamefont {Li},\ and\ \citenamefont {Vary}}]{Xu:2020xbt}%
  \BibitemOpen
  \bibfield  {author} {\bibinfo {author} {\bibfnamefont {S.}~\bibnamefont
  {Xu}}, \bibinfo {author} {\bibfnamefont {C.}~\bibnamefont {Mondal}}, \bibinfo
  {author} {\bibfnamefont {J.}~\bibnamefont {Lan}}, \bibinfo {author}
  {\bibfnamefont {X.}~\bibnamefont {Zhao}}, \bibinfo {author} {\bibfnamefont
  {Y.}~\bibnamefont {Li}}, \ and\ \bibinfo {author} {\bibfnamefont {J.~P.}\
  \bibnamefont {Vary}} (\bibinfo {collaboration} {BLFQ}),\ }in\ \href {\doibase
  10.1142/9789811219313_0104} {\emph {\bibinfo {booktitle} {{18th International
  Conference on Hadron Spectroscopy and Structure}}}}\ (\bibinfo {year}
  {2020})\ pp.\ \bibinfo {pages} {607--611},\ \Eprint
  {http://arxiv.org/abs/2004.02464} {arXiv:2004.02464 [hep-ph]} \BibitemShut
  {NoStop}%
\bibitem [{\citenamefont {Xu}\ \emph {et~al.}(2021)\citenamefont {Xu},
  \citenamefont {Mondal}, \citenamefont {Lan}, \citenamefont {Zhao},
  \citenamefont {Li},\ and\ \citenamefont {Vary}}]{Xu:2021wwj}%
  \BibitemOpen
  \bibfield  {author} {\bibinfo {author} {\bibfnamefont {S.}~\bibnamefont
  {Xu}}, \bibinfo {author} {\bibfnamefont {C.}~\bibnamefont {Mondal}}, \bibinfo
  {author} {\bibfnamefont {J.}~\bibnamefont {Lan}}, \bibinfo {author}
  {\bibfnamefont {X.}~\bibnamefont {Zhao}}, \bibinfo {author} {\bibfnamefont
  {Y.}~\bibnamefont {Li}}, \ and\ \bibinfo {author} {\bibfnamefont {J.~P.}\
  \bibnamefont {Vary}} (\bibinfo {collaboration} {BLFQ}),\ }\href {\doibase
  10.1103/PhysRevD.104.094036} {\bibfield  {journal} {\bibinfo  {journal}
  {Phys. Rev. D}\ }\textbf {\bibinfo {volume} {104}},\ \bibinfo {pages}
  {094036} (\bibinfo {year} {2021})},\ \Eprint
  {http://arxiv.org/abs/2108.03909} {arXiv:2108.03909 [hep-ph]} \BibitemShut
  {NoStop}%
\bibitem [{\citenamefont {Wiecki}\ \emph {et~al.}(2015)\citenamefont {Wiecki},
  \citenamefont {Li}, \citenamefont {Zhao}, \citenamefont {Maris},\ and\
  \citenamefont {Vary}}]{Wiecki:2015xxa}%
  \BibitemOpen
  \bibfield  {author} {\bibinfo {author} {\bibfnamefont {P.}~\bibnamefont
  {Wiecki}}, \bibinfo {author} {\bibfnamefont {Y.}~\bibnamefont {Li}}, \bibinfo
  {author} {\bibfnamefont {X.}~\bibnamefont {Zhao}}, \bibinfo {author}
  {\bibfnamefont {P.}~\bibnamefont {Maris}}, \ and\ \bibinfo {author}
  {\bibfnamefont {J.~P.}\ \bibnamefont {Vary}},\ }\bibfield  {booktitle} {\emph
  {\bibinfo {booktitle} {{Proceedings, Theory and Experiment for Hadrons on the
  Light-Front (Light Cone 2014): Raleigh, North Carolina, USA, May 26-30,
  2014}}},\ }\href {\doibase 10.1007/s00601-015-0962-3} {\bibfield  {journal}
  {\bibinfo  {journal} {Few Body Syst.}\ }\textbf {\bibinfo {volume} {56}},\
  \bibinfo {pages} {489} (\bibinfo {year} {2015})},\ \Eprint
  {http://arxiv.org/abs/1502.02993} {arXiv:1502.02993 [nucl-th]} \BibitemShut
  {NoStop}%
\bibitem [{\citenamefont {Zhao}\ \emph {et~al.}(2013)\citenamefont {Zhao},
  \citenamefont {Ilderton}, \citenamefont {Maris},\ and\ \citenamefont
  {Vary}}]{Zhao:2013cma}%
  \BibitemOpen
  \bibfield  {author} {\bibinfo {author} {\bibfnamefont {X.}~\bibnamefont
  {Zhao}}, \bibinfo {author} {\bibfnamefont {A.}~\bibnamefont {Ilderton}},
  \bibinfo {author} {\bibfnamefont {P.}~\bibnamefont {Maris}}, \ and\ \bibinfo
  {author} {\bibfnamefont {J.~P.}\ \bibnamefont {Vary}},\ }\href {\doibase
  10.1103/PhysRevD.88.065014} {\bibfield  {journal} {\bibinfo  {journal} {Phys.
  Rev.}\ }\textbf {\bibinfo {volume} {D88}},\ \bibinfo {pages} {065014}
  (\bibinfo {year} {2013})},\ \Eprint {http://arxiv.org/abs/1303.3273}
  {arXiv:1303.3273 [nucl-th]} \BibitemShut {NoStop}%
\bibitem [{\citenamefont {Hu}\ \emph {et~al.}(2020)\citenamefont {Hu},
  \citenamefont {Ilderton},\ and\ \citenamefont {Zhao}}]{Hu:2019hjx}%
  \BibitemOpen
  \bibfield  {author} {\bibinfo {author} {\bibfnamefont {B.}~\bibnamefont
  {Hu}}, \bibinfo {author} {\bibfnamefont {A.}~\bibnamefont {Ilderton}}, \ and\
  \bibinfo {author} {\bibfnamefont {X.}~\bibnamefont {Zhao}},\ }\href {\doibase
  10.1103/PhysRevD.102.016017} {\bibfield  {journal} {\bibinfo  {journal}
  {Phys. Rev. D}\ }\textbf {\bibinfo {volume} {102}},\ \bibinfo {pages}
  {016017} (\bibinfo {year} {2020})},\ \Eprint
  {http://arxiv.org/abs/1911.12307} {arXiv:1911.12307 [nucl-th]} \BibitemShut
  {NoStop}%
\bibitem [{\citenamefont {Chen}\ \emph {et~al.}(2019)\citenamefont {Chen},
  \citenamefont {Li}, \citenamefont {Tuchin},\ and\ \citenamefont
  {Vary}}]{Chen:2018vdw}%
  \BibitemOpen
  \bibfield  {author} {\bibinfo {author} {\bibfnamefont {G.}~\bibnamefont
  {Chen}}, \bibinfo {author} {\bibfnamefont {Y.}~\bibnamefont {Li}}, \bibinfo
  {author} {\bibfnamefont {K.}~\bibnamefont {Tuchin}}, \ and\ \bibinfo {author}
  {\bibfnamefont {J.~P.}\ \bibnamefont {Vary}},\ }\href {\doibase
  10.1103/PhysRevC.100.025208} {\bibfield  {journal} {\bibinfo  {journal}
  {Phys. Rev. C}\ }\textbf {\bibinfo {volume} {100}},\ \bibinfo {pages}
  {025208} (\bibinfo {year} {2019})},\ \Eprint
  {http://arxiv.org/abs/1811.01782} {arXiv:1811.01782 [nucl-th]} \BibitemShut
  {NoStop}%
\bibitem [{\citenamefont {Li}\ \emph {et~al.}(2020)\citenamefont {Li},
  \citenamefont {Zhao}, \citenamefont {Maris}, \citenamefont {Chen},
  \citenamefont {Li}, \citenamefont {Tuchin},\ and\ \citenamefont
  {Vary}}]{Li:2020uhl}%
  \BibitemOpen
  \bibfield  {author} {\bibinfo {author} {\bibfnamefont {M.}~\bibnamefont
  {Li}}, \bibinfo {author} {\bibfnamefont {X.}~\bibnamefont {Zhao}}, \bibinfo
  {author} {\bibfnamefont {P.}~\bibnamefont {Maris}}, \bibinfo {author}
  {\bibfnamefont {G.}~\bibnamefont {Chen}}, \bibinfo {author} {\bibfnamefont
  {Y.}~\bibnamefont {Li}}, \bibinfo {author} {\bibfnamefont {K.}~\bibnamefont
  {Tuchin}}, \ and\ \bibinfo {author} {\bibfnamefont {J.~P.}\ \bibnamefont
  {Vary}},\ }\href {\doibase 10.1103/PhysRevD.101.076016} {\bibfield  {journal}
  {\bibinfo  {journal} {Phys. Rev. D}\ }\textbf {\bibinfo {volume} {101}},\
  \bibinfo {pages} {076016} (\bibinfo {year} {2020})},\ \Eprint
  {http://arxiv.org/abs/2002.09757} {arXiv:2002.09757 [nucl-th]} \BibitemShut
  {NoStop}%
\bibitem [{\citenamefont {Li}(2021)}]{Li:2020bti}%
  \BibitemOpen
  \bibfield  {author} {\bibinfo {author} {\bibfnamefont {M.}~\bibnamefont
  {Li}},\ }\href {\doibase 10.22323/1.387.0105} {\bibfield  {journal} {\bibinfo
   {journal} {PoS}\ }\textbf {\bibinfo {volume} {HardProbes2020}},\ \bibinfo
  {pages} {105} (\bibinfo {year} {2021})},\ \Eprint
  {http://arxiv.org/abs/2012.04438} {arXiv:2012.04438 [nucl-th]} \BibitemShut
  {NoStop}%
\bibitem [{\citenamefont {Gelis}\ and\ \citenamefont
  {Tanji}(2016)}]{Gelis:2015kya}%
  \BibitemOpen
  \bibfield  {author} {\bibinfo {author} {\bibfnamefont {F.}~\bibnamefont
  {Gelis}}\ and\ \bibinfo {author} {\bibfnamefont {N.}~\bibnamefont {Tanji}},\
  }\href {\doibase 10.1016/j.ppnp.2015.11.001} {\bibfield  {journal} {\bibinfo
  {journal} {Prog. Part. Nucl. Phys.}\ }\textbf {\bibinfo {volume} {87}},\
  \bibinfo {pages} {1} (\bibinfo {year} {2016})},\ \Eprint
  {http://arxiv.org/abs/1510.05451} {arXiv:1510.05451 [hep-ph]} \BibitemShut
  {NoStop}%
\bibitem [{\citenamefont {Hu}(2020)}]{Hu:2019dij}%
  \BibitemOpen
  \bibfield  {author} {\bibinfo {author} {\bibfnamefont {H.}~\bibnamefont
  {Hu}},\ }\href {\doibase 10.1080/00107514.2020.1775415} {\bibfield  {journal}
  {\bibinfo  {journal} {Contemp. Phys.}\ }\textbf {\bibinfo {volume} {61}},\
  \bibinfo {pages} {12} (\bibinfo {year} {2020})},\ \Eprint
  {http://arxiv.org/abs/1907.03786} {arXiv:1907.03786 [physics.atom-ph]}
  \BibitemShut {NoStop}%
\bibitem [{\citenamefont {Abramowicz}\ \emph {et~al.}(2021)\citenamefont
  {Abramowicz} \emph {et~al.}}]{Abramowicz:2021zja}%
  \BibitemOpen
  \bibfield  {author} {\bibinfo {author} {\bibfnamefont {H.}~\bibnamefont
  {Abramowicz}} \emph {et~al.},\ }\href@noop {} {\  (\bibinfo {year} {2021})},\
  \Eprint {http://arxiv.org/abs/2102.02032} {arXiv:2102.02032 [hep-ex]}
  \BibitemShut {NoStop}%
\bibitem [{\citenamefont {Bamber}\ \emph {et~al.}(1999)\citenamefont {Bamber}
  \emph {et~al.}}]{Bamber:1999zt}%
  \BibitemOpen
  \bibfield  {author} {\bibinfo {author} {\bibfnamefont {C.}~\bibnamefont
  {Bamber}} \emph {et~al.},\ }\href {\doibase 10.1103/PhysRevD.60.092004}
  {\bibfield  {journal} {\bibinfo  {journal} {Phys. Rev. D}\ }\textbf {\bibinfo
  {volume} {60}},\ \bibinfo {pages} {092004} (\bibinfo {year}
  {1999})}\BibitemShut {NoStop}%
\bibitem [{\citenamefont {Schutzhold}\ \emph {et~al.}(2008)\citenamefont
  {Schutzhold}, \citenamefont {Gies},\ and\ \citenamefont
  {Dunne}}]{Schutzhold:2008pz}%
  \BibitemOpen
  \bibfield  {author} {\bibinfo {author} {\bibfnamefont {R.}~\bibnamefont
  {Schutzhold}}, \bibinfo {author} {\bibfnamefont {H.}~\bibnamefont {Gies}}, \
  and\ \bibinfo {author} {\bibfnamefont {G.}~\bibnamefont {Dunne}},\ }\href
  {\doibase 10.1103/PhysRevLett.101.130404} {\bibfield  {journal} {\bibinfo
  {journal} {Phys. Rev. Lett.}\ }\textbf {\bibinfo {volume} {101}},\ \bibinfo
  {pages} {130404} (\bibinfo {year} {2008})},\ \Eprint
  {http://arxiv.org/abs/0807.0754} {arXiv:0807.0754 [hep-th]} \BibitemShut
  {NoStop}%
\bibitem [{\citenamefont {Dunne}\ \emph {et~al.}(2009)\citenamefont {Dunne},
  \citenamefont {Gies},\ and\ \citenamefont {Schutzhold}}]{Dunne:2009gi}%
  \BibitemOpen
  \bibfield  {author} {\bibinfo {author} {\bibfnamefont {G.~V.}\ \bibnamefont
  {Dunne}}, \bibinfo {author} {\bibfnamefont {H.}~\bibnamefont {Gies}}, \ and\
  \bibinfo {author} {\bibfnamefont {R.}~\bibnamefont {Schutzhold}},\ }\href
  {\doibase 10.1103/PhysRevD.80.111301} {\bibfield  {journal} {\bibinfo
  {journal} {Phys. Rev. D}\ }\textbf {\bibinfo {volume} {80}},\ \bibinfo
  {pages} {111301} (\bibinfo {year} {2009})},\ \Eprint
  {http://arxiv.org/abs/0908.0948} {arXiv:0908.0948 [hep-ph]} \BibitemShut
  {NoStop}%
\bibitem [{\citenamefont {Baier}\ and\ \citenamefont
  {Katkov}(2010)}]{Baier:2009it}%
  \BibitemOpen
  \bibfield  {author} {\bibinfo {author} {\bibfnamefont {V.~N.}\ \bibnamefont
  {Baier}}\ and\ \bibinfo {author} {\bibfnamefont {V.~M.}\ \bibnamefont
  {Katkov}},\ }\href {\doibase 10.1016/j.physleta.2010.01.051} {\bibfield
  {journal} {\bibinfo  {journal} {Phys. Lett. A}\ }\textbf {\bibinfo {volume}
  {374}},\ \bibinfo {pages} {2201} (\bibinfo {year} {2010})},\ \Eprint
  {http://arxiv.org/abs/0912.5250} {arXiv:0912.5250 [hep-ph]} \BibitemShut
  {NoStop}%
\bibitem [{\citenamefont {Bulanov}\ \emph {et~al.}(2010)\citenamefont
  {Bulanov}, \citenamefont {Mur}, \citenamefont {Narozhny}, \citenamefont
  {Nees},\ and\ \citenamefont {Popov}}]{Bulanov:2010ei}%
  \BibitemOpen
  \bibfield  {author} {\bibinfo {author} {\bibfnamefont {S.~S.}\ \bibnamefont
  {Bulanov}}, \bibinfo {author} {\bibfnamefont {V.~D.}\ \bibnamefont {Mur}},
  \bibinfo {author} {\bibfnamefont {N.~B.}\ \bibnamefont {Narozhny}}, \bibinfo
  {author} {\bibfnamefont {J.}~\bibnamefont {Nees}}, \ and\ \bibinfo {author}
  {\bibfnamefont {V.~S.}\ \bibnamefont {Popov}},\ }\href {\doibase
  10.1103/PhysRevLett.104.220404} {\bibfield  {journal} {\bibinfo  {journal}
  {Phys. Rev. Lett.}\ }\textbf {\bibinfo {volume} {104}},\ \bibinfo {pages}
  {220404} (\bibinfo {year} {2010})},\ \Eprint {http://arxiv.org/abs/1003.2623}
  {arXiv:1003.2623 [hep-ph]} \BibitemShut {NoStop}%
\bibitem [{\citenamefont {Gonoskov}\ \emph {et~al.}(2015)\citenamefont
  {Gonoskov}, \citenamefont {Bastrakov}, \citenamefont {Efimenko},
  \citenamefont {Ilderton}, \citenamefont {Marklund}, \citenamefont {Meyerov},
  \citenamefont {Muraviev}, \citenamefont {Sergeev}, \citenamefont {Surmin},\
  and\ \citenamefont {Wallin}}]{Gonoskov:2014mda}%
  \BibitemOpen
  \bibfield  {author} {\bibinfo {author} {\bibfnamefont {A.}~\bibnamefont
  {Gonoskov}}, \bibinfo {author} {\bibfnamefont {S.}~\bibnamefont {Bastrakov}},
  \bibinfo {author} {\bibfnamefont {E.}~\bibnamefont {Efimenko}}, \bibinfo
  {author} {\bibfnamefont {A.}~\bibnamefont {Ilderton}}, \bibinfo {author}
  {\bibfnamefont {M.}~\bibnamefont {Marklund}}, \bibinfo {author}
  {\bibfnamefont {I.}~\bibnamefont {Meyerov}}, \bibinfo {author} {\bibfnamefont
  {A.}~\bibnamefont {Muraviev}}, \bibinfo {author} {\bibfnamefont
  {A.}~\bibnamefont {Sergeev}}, \bibinfo {author} {\bibfnamefont
  {I.}~\bibnamefont {Surmin}}, \ and\ \bibinfo {author} {\bibfnamefont
  {E.}~\bibnamefont {Wallin}},\ }\href {\doibase 10.1103/PhysRevE.92.023305}
  {\bibfield  {journal} {\bibinfo  {journal} {Phys. Rev.}\ }\textbf {\bibinfo
  {volume} {E92}},\ \bibinfo {pages} {023305} (\bibinfo {year} {2015})},\
  \Eprint {http://arxiv.org/abs/1412.6426} {arXiv:1412.6426 [physics.plasm-ph]}
  \BibitemShut {NoStop}%
\bibitem [{\citenamefont {Hebenstreit}\ \emph {et~al.}(2013)\citenamefont
  {Hebenstreit}, \citenamefont {Berges},\ and\ \citenamefont
  {Gelfand}}]{Hebenstreit:2013qxa}%
  \BibitemOpen
  \bibfield  {author} {\bibinfo {author} {\bibfnamefont {F.}~\bibnamefont
  {Hebenstreit}}, \bibinfo {author} {\bibfnamefont {J.}~\bibnamefont {Berges}},
  \ and\ \bibinfo {author} {\bibfnamefont {D.}~\bibnamefont {Gelfand}},\ }\href
  {\doibase 10.1103/PhysRevD.87.105006} {\bibfield  {journal} {\bibinfo
  {journal} {Phys. Rev. D}\ }\textbf {\bibinfo {volume} {87}},\ \bibinfo
  {pages} {105006} (\bibinfo {year} {2013})},\ \Eprint
  {http://arxiv.org/abs/1302.5537} {arXiv:1302.5537 [hep-ph]} \BibitemShut
  {NoStop}%
\bibitem [{\citenamefont {Kasper}\ \emph {et~al.}(2014)\citenamefont {Kasper},
  \citenamefont {Hebenstreit},\ and\ \citenamefont {Berges}}]{Kasper:2014uaa}%
  \BibitemOpen
  \bibfield  {author} {\bibinfo {author} {\bibfnamefont {V.}~\bibnamefont
  {Kasper}}, \bibinfo {author} {\bibfnamefont {F.}~\bibnamefont {Hebenstreit}},
  \ and\ \bibinfo {author} {\bibfnamefont {J.}~\bibnamefont {Berges}},\ }\href
  {\doibase 10.1103/PhysRevD.90.025016} {\bibfield  {journal} {\bibinfo
  {journal} {Phys. Rev. D}\ }\textbf {\bibinfo {volume} {90}},\ \bibinfo
  {pages} {025016} (\bibinfo {year} {2014})},\ \Eprint
  {http://arxiv.org/abs/1403.4849} {arXiv:1403.4849 [hep-ph]} \BibitemShut
  {NoStop}%
\bibitem [{\citenamefont {Taya}(2017)}]{Taya:2017pdp}%
  \BibitemOpen
  \bibfield  {author} {\bibinfo {author} {\bibfnamefont {H.}~\bibnamefont
  {Taya}},\ }\emph {\bibinfo {title} {{Schwinger Mechanism in QCD and its
  Applications to Ultra-relativistic Heavy Ion Collisions}}},\ \href {\doibase
  10.15083/00075576} {Ph.D. thesis},\ \bibinfo  {school} {Tokyo U.} (\bibinfo
  {year} {2017})\BibitemShut {NoStop}%
\bibitem [{\citenamefont {Otto}\ \emph {et~al.}(2019)\citenamefont {Otto},
  \citenamefont {Graeveling},\ and\ \citenamefont {K\"ampfer}}]{Otto:2018hya}%
  \BibitemOpen
  \bibfield  {author} {\bibinfo {author} {\bibfnamefont {A.}~\bibnamefont
  {Otto}}, \bibinfo {author} {\bibfnamefont {D.}~\bibnamefont {Graeveling}}, \
  and\ \bibinfo {author} {\bibfnamefont {B.}~\bibnamefont {K\"ampfer}},\ }\href
  {\doibase 10.1088/1361-6587/ab1a21} {\bibfield  {journal} {\bibinfo
  {journal} {Plasma Phys. Control. Fusion}\ }\textbf {\bibinfo {volume} {61}},\
  \bibinfo {pages} {074002} (\bibinfo {year} {2019})},\ \Eprint
  {http://arxiv.org/abs/1812.10832} {arXiv:1812.10832 [hep-ph]} \BibitemShut
  {NoStop}%
\bibitem [{\citenamefont {Tang}\ \emph {et~al.}(1991)\citenamefont {Tang},
  \citenamefont {Brodsky},\ and\ \citenamefont {Pauli}}]{Tang:1991rc}%
  \BibitemOpen
  \bibfield  {author} {\bibinfo {author} {\bibfnamefont {A.~C.}\ \bibnamefont
  {Tang}}, \bibinfo {author} {\bibfnamefont {S.~J.}\ \bibnamefont {Brodsky}}, \
  and\ \bibinfo {author} {\bibfnamefont {H.~C.}\ \bibnamefont {Pauli}},\ }\href
  {\doibase 10.1103/PhysRevD.44.1842} {\bibfield  {journal} {\bibinfo
  {journal} {Phys. Rev. D}\ }\textbf {\bibinfo {volume} {44}},\ \bibinfo
  {pages} {1842} (\bibinfo {year} {1991})}\BibitemShut {NoStop}%
\bibitem [{\citenamefont {Askar}(1978)}]{askar1978askar}%
  \BibitemOpen
  \bibfield  {author} {\bibinfo {author} {\bibfnamefont {A.}~\bibnamefont
  {Askar}},\ }\href {\doibase 10.1063/1.436072} {\bibfield  {journal} {\bibinfo
   {journal} {J. Chem. Phys.}\ }\textbf {\bibinfo {volume} {68}},\ \bibinfo
  {pages} {2794} (\bibinfo {year} {1978})}\BibitemShut {NoStop}%
\bibitem [{\citenamefont {Parker}(1968)}]{Parker:1968mv}%
  \BibitemOpen
  \bibfield  {author} {\bibinfo {author} {\bibfnamefont {L.}~\bibnamefont
  {Parker}},\ }\href {\doibase 10.1103/PhysRevLett.21.562} {\bibfield
  {journal} {\bibinfo  {journal} {Phys. Rev. Lett.}\ }\textbf {\bibinfo
  {volume} {21}},\ \bibinfo {pages} {562} (\bibinfo {year} {1968})}\BibitemShut
  {NoStop}%
\bibitem [{\citenamefont {Tanji}(2009)}]{Tanji:2008ku}%
  \BibitemOpen
  \bibfield  {author} {\bibinfo {author} {\bibfnamefont {N.}~\bibnamefont
  {Tanji}},\ }\href {\doibase 10.1016/j.aop.2009.03.012} {\bibfield  {journal}
  {\bibinfo  {journal} {Annals Phys.}\ }\textbf {\bibinfo {volume} {324}},\
  \bibinfo {pages} {1691} (\bibinfo {year} {2009})},\ \Eprint
  {http://arxiv.org/abs/0810.4429} {arXiv:0810.4429 [hep-ph]} \BibitemShut
  {NoStop}%
\bibitem [{\citenamefont {Hebenstreit}\ \emph {et~al.}(2009)\citenamefont
  {Hebenstreit}, \citenamefont {Alkofer}, \citenamefont {Dunne},\ and\
  \citenamefont {Gies}}]{Hebenstreit:2009km}%
  \BibitemOpen
  \bibfield  {author} {\bibinfo {author} {\bibfnamefont {F.}~\bibnamefont
  {Hebenstreit}}, \bibinfo {author} {\bibfnamefont {R.}~\bibnamefont
  {Alkofer}}, \bibinfo {author} {\bibfnamefont {G.~V.}\ \bibnamefont {Dunne}},
  \ and\ \bibinfo {author} {\bibfnamefont {H.}~\bibnamefont {Gies}},\ }\href
  {\doibase 10.1103/PhysRevLett.102.150404} {\bibfield  {journal} {\bibinfo
  {journal} {Phys. Rev. Lett.}\ }\textbf {\bibinfo {volume} {102}},\ \bibinfo
  {pages} {150404} (\bibinfo {year} {2009})},\ \Eprint
  {http://arxiv.org/abs/0901.2631} {arXiv:0901.2631 [hep-ph]} \BibitemShut
  {NoStop}%
\bibitem [{\citenamefont {Kim}\ and\ \citenamefont
  {Schubert}(2011)}]{Kim:2011jw}%
  \BibitemOpen
  \bibfield  {author} {\bibinfo {author} {\bibfnamefont {S.~P.}\ \bibnamefont
  {Kim}}\ and\ \bibinfo {author} {\bibfnamefont {C.}~\bibnamefont {Schubert}},\
  }\href {\doibase 10.1103/PhysRevD.84.125028} {\bibfield  {journal} {\bibinfo
  {journal} {Phys. Rev. D}\ }\textbf {\bibinfo {volume} {84}},\ \bibinfo
  {pages} {125028} (\bibinfo {year} {2011})},\ \Eprint
  {http://arxiv.org/abs/1110.0900} {arXiv:1110.0900 [hep-th]} \BibitemShut
  {NoStop}%
\bibitem [{\citenamefont {Dabrowski}\ and\ \citenamefont
  {Dunne}(2016)}]{Dabrowski:2016tsx}%
  \BibitemOpen
  \bibfield  {author} {\bibinfo {author} {\bibfnamefont {R.}~\bibnamefont
  {Dabrowski}}\ and\ \bibinfo {author} {\bibfnamefont {G.~V.}\ \bibnamefont
  {Dunne}},\ }\href {\doibase 10.1103/PhysRevD.94.065005} {\bibfield  {journal}
  {\bibinfo  {journal} {Phys. Rev. D}\ }\textbf {\bibinfo {volume} {94}},\
  \bibinfo {pages} {065005} (\bibinfo {year} {2016})},\ \Eprint
  {http://arxiv.org/abs/1606.00902} {arXiv:1606.00902 [hep-th]} \BibitemShut
  {NoStop}%
\bibitem [{\citenamefont {Ilderton}(2021)}]{Ilderton:2021zej}%
  \BibitemOpen
  \bibfield  {author} {\bibinfo {author} {\bibfnamefont {A.}~\bibnamefont
  {Ilderton}},\ }\href@noop {} {\  (\bibinfo {year} {2021})},\ \Eprint
  {http://arxiv.org/abs/2108.13885} {arXiv:2108.13885 [hep-ph]} \BibitemShut
  {NoStop}%
\bibitem [{\citenamefont {Schwinger}(1951)}]{Schwinger:1951nm}%
  \BibitemOpen
  \bibfield  {author} {\bibinfo {author} {\bibfnamefont {J.~S.}\ \bibnamefont
  {Schwinger}},\ }\href {\doibase 10.1103/PhysRev.82.664} {\bibfield  {journal}
  {\bibinfo  {journal} {Phys. Rev.}\ }\textbf {\bibinfo {volume} {82}},\
  \bibinfo {pages} {664} (\bibinfo {year} {1951})},\ \bibinfo {note}
  {[,116(1951)]}\BibitemShut {NoStop}%
\bibitem [{\citenamefont {Cohen}\ and\ \citenamefont
  {McGady}(2008)}]{Cohen:2008wz}%
  \BibitemOpen
  \bibfield  {author} {\bibinfo {author} {\bibfnamefont {T.~D.}\ \bibnamefont
  {Cohen}}\ and\ \bibinfo {author} {\bibfnamefont {D.~A.}\ \bibnamefont
  {McGady}},\ }\href {\doibase 10.1103/PhysRevD.78.036008} {\bibfield
  {journal} {\bibinfo  {journal} {Phys. Rev. D}\ }\textbf {\bibinfo {volume}
  {78}},\ \bibinfo {pages} {036008} (\bibinfo {year} {2008})},\ \Eprint
  {http://arxiv.org/abs/0807.1117} {arXiv:0807.1117 [hep-ph]} \BibitemShut
  {NoStop}%
\bibitem [{\citenamefont {Tomaras}\ \emph {et~al.}(2001)\citenamefont
  {Tomaras}, \citenamefont {Tsamis},\ and\ \citenamefont
  {Woodard}}]{Tomaras:2001vs}%
  \BibitemOpen
  \bibfield  {author} {\bibinfo {author} {\bibfnamefont {T.~N.}\ \bibnamefont
  {Tomaras}}, \bibinfo {author} {\bibfnamefont {N.~C.}\ \bibnamefont {Tsamis}},
  \ and\ \bibinfo {author} {\bibfnamefont {R.~P.}\ \bibnamefont {Woodard}},\
  }\href {\doibase 10.1088/1126-6708/2001/11/008} {\bibfield  {journal}
  {\bibinfo  {journal} {JHEP}\ }\textbf {\bibinfo {volume} {11}},\ \bibinfo
  {pages} {008} (\bibinfo {year} {2001})},\ \Eprint
  {http://arxiv.org/abs/hep-th/0108090} {arXiv:hep-th/0108090} \BibitemShut
  {NoStop}%
\bibitem [{\citenamefont {Fried}\ and\ \citenamefont
  {Woodard}(2002)}]{Fried:2001ur}%
  \BibitemOpen
  \bibfield  {author} {\bibinfo {author} {\bibfnamefont {H.~M.}\ \bibnamefont
  {Fried}}\ and\ \bibinfo {author} {\bibfnamefont {R.~P.}\ \bibnamefont
  {Woodard}},\ }\href {\doibase 10.1016/S0370-2693(01)01384-3} {\bibfield
  {journal} {\bibinfo  {journal} {Phys. Lett. B}\ }\textbf {\bibinfo {volume}
  {524}},\ \bibinfo {pages} {233} (\bibinfo {year} {2002})},\ \Eprint
  {http://arxiv.org/abs/hep-th/0110180} {arXiv:hep-th/0110180} \BibitemShut
  {NoStop}%
\end{thebibliography}%

\end{document}